\documentclass[12pt]{report}
\usepackage{uidis}
\usepackage[dvipdfm]{color}
\usepackage[dvipdfm]{graphicx}
\usepackage[square,sort&compress]{natbib}
\usepackage{amsmath,amssymb}
\usepackage{dcolumn,array}
\newcolumntype{.}{D{.}{.}{-1}}
\newcolumntype{d}[1]{D{.}{.}{#1}}
\usepackage{epic}
\usepackage{eepic}
\usepackage{ifthen}
\usepackage{mydefs}
\doublespace
\newcommand\mach{\mathcal{M}}
\newcommand\cs{c_{\rm s 1}}
\newcommand\ed{\tilde{\delta}} 
\newcommand\msol{\hbox{${\rm M_{\odot}}$}}
\newcommand\dcs{C_{\rm s}}

\begin{document}
\pagenumbering{roman}
\title{An Analytic Model of Environmental Effects \\ on Cosmic
Structure Formation \\ and an Application to Cosmic Accretion Shocks}
\author{Vasiliki Pavlidou}
\advisor{Brian D. Fields}
\prevdegrees{ B.S., Aristotle University of Thesaloniki, 1999\\
              M.S., University of Illinois at Urbana-Champaign, 2001 }
\gradyear{2005} 
\maketitle
%
\stepcounter{page}
%
\begin{abstract}

We present a new analytic tool for the study of cosmic structure
formation, a {\em double distribution } of the number density 
of dark matter halos
with respect to both halo mass and local over- (or under-) density.
The double distribution provides a statistical treatment of the
properties of matter {\em surrounding} collapsed objects, and can be
used to provide analytical insight into 
environmental effects on hierarchical structure formation.

We apply this new tool to the case of cosmic accretion shocks. 
We investigate and quantify the effect of environmental factors
on the statistical properties of these shocks.
For this purpose, we explore two different models. The
first ``control'' model
 uses a Press-Schechter mass function to describe the population
of collapsed structures, and assumes that all objects accrete gas of
the same density and temperature. The second model treats the accreted
material as a multi-temperature, multi-density medium with densities and
temperatures derived from the double distribution. 
We find that the shock environment significantly
alters the physical impact of cosmic accretion
shocks on the intergalactic medium, as well as the cosmic history of
their properties. 

\end{abstract}

%
\chapter*{Acknowledgements}

\indent

I thank my advisor, Brian Fields, for his guidance and support
throughout my graduate studies, and especially during the necessary,
long, and not visibly productive period of development of the
analytical tools used in this work. I thank the members of my
doctoral  committee, Icko Iben, Telemachos Mouschovias, Laird Thompson,
and Ben Wandelt, for all their help and continued interest in my
work. 

I gratefully acknowledge enlightening scientific discussions which
enhanced the content of this dissertation with Rich Cyburt, Dimitris
Galanakis, Francesco Miniati, Telemachos Mouschovias,  Tijana
Prodanovi\'{c}, Kostas Tassis, and Ben Wandelt.

Parts of this work received support from: the Greek State
Scholarships Foundation in the form of an Award for Graduate Studies
Abroad;
Zonta International in the form of an Amelia
Earhart Fellowship; the University of Illinois in the form a Graduate
College Fellowship; and the National Science Foundation grant AST-0092939.

%
\singlespace
%
\tableofcontents
\clearpage
%
\addcontentsline{toc}{chapter}{\listtablename}
\listoftables
\clearpage
%
\addcontentsline{toc}{chapter}{\listfigurename}
\listoffigures
%
%
\doublespace
%
\begin{pqabstract}

We present a new analytic tool for the study of cosmic structure
formation, a {\em double distribution } function of dark matter halos,
with respect to both object mass and local over- (or under-) density.
The double distribution provides a statistical treatment of the
properties of matter {\em surrounding} collapsed objects, and can be
used to study environmental effects on hierarchical structure
formation.

We apply this new tool to the case of cosmic accretion shocks. 
We investigate and quantify the effect of environmental factors
on the statistical properties of these shocks.
For this purpose, we explore two different models. The
first ``control'' model
 uses a Press-Schechter mass function to describe the population
of collapsed structures, and assumes that all objects accrete gas of
the same density and temperature. The second model treats the accreted
material as a multi-temperature, multi-density medium with densities and
temperatures derived from the double distribution. 
We find that the shock environment significantly
alters the physical impact of cosmic accretion
shocks on the intergalactic medium, as well as the cosmic history of
their properties. 

\end{pqabstract}

%
\clearpage
\pagenumbering{arabic}
%
\chapter{Introduction}
\label{chapter:introduction}

\indent

The large-scale structure of the universe has been investigated 
with increasing intensity for almost a century. Cosmologists have 
been using multi-wavelength observations, analytical models, 
and massive numerical simulations to study the properties of 
galaxies, galaxy clusters, and superclusters, their formation, 
and their evolution. 

Structure in the universe is believed to be seeded by small, random 
density fluctuations in the early universe, which grow gravitationally
in a process further complicated by the interplay between 
baryons, radiation, dark matter, and dark energy. 
In this context, the observationally relevant predictions of 
any theory of structure formation are the {\em statistical properties}
of the resulting, observable population of cosmic structures.

\section{The Double Distribution}

\indent

Cosmological distributions have long 
been used for the study of cosmic structure formation 
with great success as an analytical tool complementary to 
numerical simulations. They have been used to constrain the cosmological 
parameters; interpret results of cosmological simulations; study
regions of the parameter space which cannot be approached by simulations 
due to prohibitive computational
cost; explore the effects of various physical 
processes in an efficient if approximate way.
The analytical tool used most widely in cosmology is
the mass function of dark matter halos (distribution of the number
density of halos with respect to halo mass). Analytical descriptions
of dark matter halos are usually based on the Press-Schechter
formalism (\cite{ps74,b91}) 
and its extensions (e.g., \cite{lc93,ph90,j95,bow91}). 
The Press-Schechter mass function has been shown to agree well
with results of N-body
simulations (e.g., \cite{wef,lc94,ecf}).
More sophisticated approaches taking into
account deviations from spherical symmetry (e.g., \cite{ls98,smt01,st02}) 
have improved this agreement even further.

To derive the Press-Schechter mass function, one begins with a density
perturbation field still in its linear regime. Regions in
space are then smoothed on successively smaller scales.
The mass of a collapsed object is then taken to be the 
largest smoothing mass scale
for which the average linear overdensity exceeds some threshold. 
In this way, matter in the universe is
distributed among collapsed structures of different masses, which all
share the same value of average overdensity (the threshold value). 
Information about the local environment of 
collapsed objects (whether they live in underdensities or
overdensities) is thus erased. 

For this reason, and despite its wide applicability, 
the expression for the mass function cannot
be used to address environment-related questions: Does the mass function of
structures in superclusters differ from the mass function 
inside  voids? Are structures of some particular mass more likely to reside
inside underdense or overdense regions in space? How does such a 
preference evolve with redshift, and  how sensitively does it depend
on the cosmological parameters? How does the state of the material 
surrounding and accreted by a collapsed object depend on the
mass of the object and the cosmic epoch? 

To address such questions, we seek a {\em double distribution} of the number 
density of structures with 
respect to mass but also to local overdensity (or underdensity). In
order to extract information about the surroundings of collapsed
structures, we use the same random walk formalism  which rigorously yields 
the Press-Schechter mass function (\cite{b91,lc93}). 
Integration of this distribution over density contrast should
return the Press-Schechter mass function so that the successes of the
Press-Schechter formalism be retained. 

Two complications arise in the effort to expand the Press-Schechter
mass function to incorporate a description of the local overdensity.
First, the concept of the ``local environment''
is somewhat vague and needs to be defined in a more rigorous way. The size
of the ``local environment'' cannot be the same for all structures. If
this was the case, very small structures would represent only a tiny
fraction of the ``environment'', while very large structures could
even exceed the size of the ``environment'', which would be an
unphysical situation. This problem is not exclusive to analytical
tools, but also needs to be addressed when analyzing the results of 
numerical simulations. 
Second, the Press-Schechter treatment of the density field uses linear 
theory, and ways of converting this information to a more physical 
non-linear result need to be determined. 

We address the first problem 
by introducing a {\em clustering scale parameter}, $\beta$, 
which allows us to define
the size of the ``environment'' of each structure as a function of its
mass. We address the second concern by calculating conversion
relations between the linear-theory overdensities (or underdensities)
and those predicted by the spherical-evolution model. 

Environment-related questions in cosmological structure
formation have also been addressed using analytical models for the
clustering properties of dark halos which evaluate quantities such as  
the cross-correlation function between dark halos and matter, and the
biasing factor (e.g. \cite{mw96,kns97,s98,tp98,lk99,st99,sj00}, also
see review by \cite{cs02} and references therein). These analyses are
based on  the same random walk formalism which we use here to derive our double
distribution (also see \cite{j98} for fitting formulae from N-body
simulation results). However, the information content 
of the double distribution, 
which treats the ``environment'' in a mass-dependent fashion, is
complementary to that of correlation functions, which 
describe the clustering properties of the dark halo
population at some {\em fixed spatial scale}. 
The double distribution is ideally fitted for population studies of
cosmological objects. If the properties of a single object can be
parametrized as a function of its mass and its environment, then the
double distribution can be used to predict the statistical properties
of such objects, as well as their evolution with time, for any
cosmological model.

We present such an application of the double distribution for the case
of cosmic accretion shocks. The properties of cosmic accretion shocks
depend
sensitively on their environment, and for this reason the double
distribution is the tool of choice to study this population of cosmic
structures.

\section{Cosmic Accretion Shocks}

\indent

The formation of shocks in the baryonic component of matter in the universe is
an inevitable and integral part of the process of cosmological structure
formation. 
The longstanding question of the  dark 
baryons\footnote{a fraction of the baryonic content of the universe which, 
until recently, has eluded detection, presumably because its temperature 
($10^5-10^7$ K) is too high to produce Ly$\alpha$ absorption  
while its density is too to be detected in emission with currently
available equipment}
\cite{HO86, Fuk98, Sper03, Cyb03}
is likely on its way to resolution with
the first evidence  \cite{Nic02, Mat03, finog03, nicastro} 
for a large component of
diffuse, low-density intergalactic gas
distributed in the filaments that comprise
the ``cosmic web,'' and
detected in the X-ray forest 
\cite{Hel98, PL98, FBC02} via absorption lines from
highly-ionized metals.
This warm-hot intergalactic medium 
is thought to arise in structure formation shocks \cite{Hel98, CO99,
D01, furlanetto, KRCS05}.
Moreover, structure formation shocks heat the intergalactic medium and 
are likely to act as acceleration sites for nonthermal particles, the 
``structure formation cosmic rays'' (e.g. \cite{min01, min01b, BSFG01,
BerrD03,  GB03p, BBCG04, KJ05}
and references therein).  Such a cosmic-ray population 
would have distinct $\gamma$-ray and light-element signatures, 
which are currently subjects of intense investigation
\cite{lw, tk, min02, SchMuk, SuIn02, ti, BerrD03, GB03g, kesh, min03,
RPSM, GB04, ProdF04, ProdF04b, SuIn04, KBH}.

Cosmic shocks occur
during different facets of structure formation and  in a variety of
environments, hence 
there are two possible ways to 
categorize them: according to the physical processes
causing them, and according to the state of the medium in 
which they form.

There are three principal processes associated with cosmic structure
formation which result in the formation of large-scale shocks.
\begin{enumerate}
\item Accretion of intergalactic gas by a collapsed, virialized
      structure. In this case, an {\em accretion shock} is formed at
      the interface between virialized and diffuse gas \cite{bert85,
      ryu97, min_shock}. The shock is driven by the gravitational
      attraction exerted on the diffuse gas by the accretor.
\item Merger of two collapsed structures. In this case, a {\em merger
      shock} is formed at the interface between the gas components of
      the merging objects \cite{min_shock, GB03p}. The shock is driven
      by the mutual gravitational attraction between the objects.
\item Blast-like expansion of a void (an underdensity in its
      non-linear regime). The regions compressed between expanding
      voids form large-scale filaments, and {\em filament shocks} are
      formed at the interface between the expanding void and the
      compressed gas \cite{bertV}. In this case, the shocks are driven
      by the expansion of the void rather than the gravity of the filaments.
\end{enumerate}

Shocks can also be divided according
to the state of the gas passing through them, into {\em external} and
{\em internal} shocks \cite{RKJ03}. 
External shocks process pristine material, which has never been
shocked before by any of the processes described above. External
shocks are mostly filament shocks, since the process of formation of
individual virialized structures (associated with the other two types of
shocks) occurs principally within filaments, and therefore in most
cases involves gas which has already been processed at least by filament
shocks. Internal shocks process gas which has already been shock-heated in the
past. All merger shocks, as well as many accretion shocks, are internal shocks.

Since external shocks process colder material of 
lower sound speed, their Mach numbers are generally higher
than those of internal shocks.
However, because the gas passing through internal shocks has 
already been compressed, internal shocks
process more mass and kinetic energy than external shocks.

In the second part of this thesis, 
we present an analytical study of the population 
of accretion shocks, in a concordance $\Omega_{\rm
m}+\Omega_\Lambda=1$ universe. We use the double distribution of
cosmic structures 
 to describe the accreting structures as well as their
environment at a given redshift. 
Other Press-Schechter extensions have been used to model different
``families'' of large-scale cosmic shocks by \cite{GB03p} and
\cite{inoue05}, who model merger shocks, and by \cite{furlanetto}, who 
describe large-scale
shocks that may appear when overdense perturbations reach and exceed
their turnaround point. Here, we present the first analytic model for
{\em accretion} shocks, and we include, for the first time, a detailed
treatment of the environment in which accreting structures
reside.

This thesis is organized as follows. 
In chapter 2 we derive the double distribution of
cosmic structures, and we explore its information content by plotting
the distribution itself as well as interesting derivative quantities
for different cosmological models.
In chapter 3 we apply the double distribution to study the effect of
environment on the statistical
properties of the population of cosmic accretion shocks around collapsed
structures. We discuss our findings in chapter 4.

\chapter{The Double Distribution Of Cosmic Structures}
\label{chapter:dd}

\section{Overview}

\indent

The double distribution of cosmic structures with respect to mass and
local overdensity is a cosmological statistical distribution which
describes how the number density of collapsed and virialized dark
matter objects is distributed among different masses and among
different local density contrasts with respect to the cosmic mean
density. In this chapter we derive the double distribution of cosmic
structures and we explore its information content. 

The derivation of the double distribution is based on the 
``random walk'' formalism, which  was introduced for the derivation of
cosmological mass functions from an early (linear) field of density 
fluctuations by \cite{b91} and by \cite{lc93}. 
The basic simplifying assumptions behind building a mass function
starting from a linear overdensity field are:
\begin{enumerate}
\item All matter in the universe can be distributed among collapsed
      objects, and there is no diffuse matter in the universe.
\item At any given cosmic epoch, all structures can be viewed as
      though they have just virialized. 
\item All collapsed objects at a given cosmic epoch have the same mean
      matter density.
\item All collapsed objects at a given cosmic epoch are spherically 
  symmetric.
\item The mean density of a certain region at a very early cosmic
      epoch (while all density perturbations
 are still growing linearly) contains
      adequate information to predict the mean density of the same
      region at much later times.
\end{enumerate}
The conceptual idea behind the process necessary to produce a mass
function is the following. 
\begin{itemize}
\item We focus at a infinitesimal mass element centered on a fixed point
      in space.
\item Starting at an infinite mass scale and gradually proceeding to
      smaller mass scales, we smooth the linear density field and 
      evaluate the mean overdensity at the scale under consideration
      (note that at an infinite mass scale the mean overdensity is
      always zero, since the smoothing identically returns the mean
      cosmic density).
\item We convert the smoothed linear overdensity to a nonlinear
      overdensity using the spherical evolution model. If the
      result is equal to the mean nonlinear overdensity of virialized
      objects, then the mass scale at which the smoothing took place
      is the mass of the collapsed object which hosts the
      infinitesimal mass under consideration. If not, we repeat the
      smoothing at a smaller mass scale, until the desired mean
      overdensity is reached.
\item  We then repeat the process for all points in space, and thus
       calculate how much mass corresponds to every {\em collapsed
       object mass} interval.
\end{itemize}
The random walk formalism now recognizes that the smoothed overdensity
executes a ``random walk'' with changing smoothing mass
scale, and uses this fact to calculate the probability that any given
point in space belongs to a collapsed object with mass in a specific
interval. This probability can then be 
manipulated to give us the mass function. The mathematical details of
this process are briefly reviewed in section \ref{rwoverview}.  

In our case, we would like to additionally 
retain information about the density of the
region within which a collapsed structure is embedded. For this reason, we
need to keep track of the mean density evaluated at a random walk
``step'' (a smoothing mass scale) preceding the final step which 
returns the virial density of the collapsed object. The ``distance''
in mass between the two ``steps'' of interest will determine how far
from the collapsed object the ``local environment'' extends. In our
derivation of the double distribution the ratio of the mass scale
representing the environment of an object (including the object itself) 
over the mass of the object 
is defined as $\beta$, the ``clustering scale parameter''. It
is a free parameter in our model, and, for every realization of our
model, it is the same for all structures.  In section \ref{ddderiv} we
present the details of the derivation of the double distribution
through this extension of the random walk formalism.

As we discussed above, a critical step in deriving both the
Press-Schechter mass function as well as the double distribution of
cosmic structures is the conversion of a density contrast evaluated
using linear theory to a density contrast evaluated using an exact (if
idealized) model. Since we have assumed that all structures are
spherically symmetric, we will consider the ``true'' density contrast
to be the one given by the spherical evolution model. In section
\ref{secconv} we derive such conversion relations between the linear
and the spherical evolution models, by expanding the spherical model
around zero overdensity and demanding that at early times the growth
of density perturbations according to both models agree to
first order. We derive exact conversion relations for both 
Einstein-de Sitter (flat, matter-only) and  $\Lambda$CDM (flat, matter
$+$ cosmological constant) universes. In addition, we present a useful
and simple approximate conversion relation, which exhibits the correct
asymptotic behavior at early as well as late times, and has an
accuracy better than $2\%$ throughout its domain.

In section \ref{derivquants} we define several interesting integral moments
of the double distribution, describing the statistical properties of
the population of collapsed cosmic structures and their distribution
among different (overdense or underdense) environments. Finally, in
section \ref{ddres} we explore the information content of the double
distribution by plotting both the double distribution itself as well
as several of its integral moments, for different redshifts and
different cosmologies. Our principal finding is that for every cosmic
epoch and every collapsed object mass, there is a pronounced peak in
the double distribution, corresponding to a ``most probable''
overdensity (or underdensity) for the local environment of the
structure. The location of this peak
\begin{itemize}
\item moves towards higher overdensities with increasing redshift;
\item moves towards higher overdensities with increasing object mass;
\item depends only very mildly on the value of $\beta$ for low $\beta$
(while it eventually moves towards zero overdensity as $\beta
\rightarrow \infty$)
\item has a qualitatively similar behavior for both cosmological models
      we examined.
\end{itemize}
At the present cosmic epoch, most structures are located inside
underdensities. 

\section{Random Walks and the Press-Schechter Mass Function}\label{rwoverview}

\indent

The Press-Schechter mass function of collapsed structures is the
comoving number density of virialized objects per differential mass interval,
$dn/dm$, for every cosmic epoch $a$ \footnote{The {\em proper} mass 
function, i.e. the number of
collapsed structures per unit proper volume per differential mass
interval is simply related to the comoving mass function via 
$dn/dm|_{\rm proper} = a^{-3} dn/dm|_{\rm comoving}$}.
A related quantity is the mass fraction, $P(>m, a)$, which is 
the fraction of matter in the universe belonging to
collapsed structures with mass $>m$. If $P(>m,a)$ is known, then
$dn/dm$ can be calculated from 
\begin{equation}\label{funcfrac}
\frac{dn}{dm}(m,a)= \frac{\rho_{\rm m,0}}{m}
\left|\frac{d}{dm}
P(>m,a)
\right|\,,
\end{equation}
where $\rho_{\rm m,0}$ is the present-day matter density of the
universe.

$P(>m,a)$ is in turn calculated by assigning, at each epoch $a$, every
infinitesimal element $dm$ in the universe to a
collapsed structure of some mass $m$.
A structure is considered ``collapsed'' if its
mean overdensity 
\begin{equation}\label{contrastdef}
\langle\delta\rangle = 
\frac{\langle\rho_{\rm structure}\rangle-\rho_{{\rm m}, a}}{\rho_{{\rm m}, a}}
\end{equation}
exceeds a certain critical value, $\delta_{\rm c}(a)$.
In Eq.~(\ref{contrastdef}), $\rho_{{\rm m}, a}$ 
is the mean matter density of the universe
at epoch $a$. The critical overdensity  $\delta_{\rm c}(a)$
is the mean overdensity predicted by the
spherical evolution model for a
structure virializing at epoch $a$. 
For each point in space, one calculates the
mean local overdensity by smoothing the
overdensity field $\delta(\vec{x},a)$ with a spherically symmetric
filter function of varying mass scale, starting
from $m \rightarrow \infty$, where 
one averages over the whole universe and finds
identically $\langle \delta \rangle = 0$, and proceeding to
successively smaller scales. When a mass scale is found for which the
mean overdensity becomes equal to $\delta_{\rm c}(a)$, it is taken
to be the mass $m$ of the parent object of the infinitesimal mass
at the point under consideration. 
This way of assigning object masses circumvents
the structure-in-structure problem, since
the mass of the parent object is always
the {\em largest possible mass} satisfying the criterion for
collapse. All information on substructure within
collapsed structures is thus erased from the resulting mass function. 

The way the average overdensity $\langle \delta \rangle$ 
changes when the smoothing mass scale is varied
resembles, under certain conditions, a 1D random walk
\cite{b91}. For all
``particles''(points in space in our case), the walk begins 
at the ``spatial origin'' ($\langle \delta \rangle =0$), at ``time zero''
($m \rightarrow \infty$). As ``time progresses'' ($m$ decreases), each
``particle'' may move either to the ``left''
 (negative $\langle \delta \rangle$) or to the ``right'' (positive 
 $\langle \delta \rangle$). An ``absorbing wall'' exists at $\delta_{\rm c}(a)$.
If this ``wall'' is reached,
the ``particle'' is ``removed'' from the walk (the point is
assigned its parent object mass and
removed from further consideration at smaller values of $m$). $P(>m,a)$
is then the fraction of ``particles'' which
have been ``lost'' by  ``time''
$m$, and it can be calculated using random-walk
theory. 

However, we must first ensure that simple random-walk theory is indeed applicable. 
First, each ``step'' of the ``walk'' should
be completely independent from the previous step. This requires that
the $k$-modes producing an increase $\Delta \langle \delta \rangle$
in the space-like variable not appear in any of the previous steps in
$\langle \delta \rangle$. A smoothing window function sharp in
$k-$space, 
\begin{equation}\label{what}
\hat{W}_m(k)=\left\{
\begin{array}{ll}
1 & k<k_{\rm c}(m)\\
0 &k>k_{\rm c}(m)
\end{array}
\right. \,.
\end{equation}
(see \cite{b91} and \cite{lc93} for more extended discussions on the
consequences of such a choice) enforces this condition, since
\begin{equation}
\langle \delta \rangle _{m,\vec{x}_0} =
\int W_m\left(|\vec{x}_0-\vec{x}|\right)\delta(\vec{x})d^3\vec{x}
= \int_{k \le k_{\rm c}(m)} 
\!\!\!\!\! \!\!\!\!\!
\!\!\!\!\!\!\!\!
\delta_k {\rm e}^{i\vec{x}_0\cdot \vec{k}} d^3\vec{k}
\end{equation}
and
\begin{equation}
\Delta \langle \delta \rangle _{\vec{x}_0} =
\int_{_{k_{\rm c}(m) \le k \le k_{\rm c}(m-dm)}} \!\!\!\!\!\!\!\!\!\!\!\!\!\!\!\!\!\!\!\! 
\!\!\!\!\!\!\!\!\!\!
\delta_k {\rm e}^{i\vec{x}_0\cdot \vec{k}} d^3\vec{k}
\end{equation}
which only involves new $k$-modes corresponding to scales from
$m$ to $m-dm$.

Second, there must be an
equal probability for the system to ``move'' towards any one of the
two available ``directions''. A Gaussian overdensity field 
(which is the usual assumption for deriving analytic mass functions
and which we adopt here) guarantees that
this condition is satisfied.

Finally, the
appropriate ``time-like'' variable (which should depend on $m$)
needs to be selected, given that 
the ``space-like'' variable is $\langle \delta \rangle$.
By direct analogy to the 1D 
random walk theory result $\langle x^2\rangle = 2Dt$, and from the
definition of the variance of the overdensity field $S(m)$, 
\begin{equation}\label{sdef}
S(m) =\sigma^2(m) = \langle|\delta (m,\vec{x})|^2\rangle \propto 
\int_{k=0}^{k(m)}\!\!\!k^2dk |\delta_k|^2
\end{equation}
we can immediately identify $Dt \rightarrow S(m)/2$.

Three further complications need to be addressed. 
First, our knowledge of $\delta_k$ and subsequently $S(m)$ 
is limited at late times. 
In the early universe, right after matter-radiation equality, 
$\langle|\delta_k|^2\rangle$ can be simply described in terms of a 
power-law in $k$ modified by a transfer function,
$\langle|\delta_k|^2\rangle \propto T^2(k)k^n$. 
While all $\delta$ are still in their linear regime, 
they simply grow by the linear growth factor
(independent of $k$). However, at later times, when certain structures start
departing from the linear regime, 
we cannot use our simple early-universe expressions for $\delta_k$.  
Second, $\langle \delta \rangle$ is limited 
to be $\ge -1$, which introduces a second,
reflecting ``wall'' at a value of $\langle \delta \rangle = -1$,
further complicating the random-walk calculations. 
 Finally, the true overdensity field
loses its Gaussianity as it evolves past the linear regime.

To circumvent these problems, we define the 
{\em linearly extrapolated overdensity field},
$\ed(\vec{x},a)$, as the overdensity field that would result if all
structures continued to grow according to the linear theory until time
$a$. Now $\ed(\vec{x},a)$ 
is not limited to be $\ge -1$, since it does not 
represent real overdensities. In addition, we can always
calculate $S(m)$ for  $\ed(\vec{x},a)$, since
$\ed_k$ is modified from its simple early-universe expression 
only by the linear growth factor. Finally, the extrapolated field
remains Gaussian at all times.

The linearly extrapolated overdensity $\ed(\vec{x},a)$
and the associated variance $S(m)$,
are time-varying, but  
the time dependence is well-known (see chapter 3), 
and the same for both $S$ and $\ed^2$ \footnote{ as seen by
Eq.~(\ref{sdef}) re-written for the extrapolated rather than the true
overdensity field}. Thus, the time dependence drops out of ratios 
$\ed/\sqrt{S}$ which appear in the mass function. For this reason we may,  without loss
of generality, choose any single epoch to evaluate these quantities, 
with the stipulation
that $\ed$ and $S(m)$ must refer to the same epoch.
Given that $\sigma(m)$ is often normalized to the present
value of $\sigma_8$, a convenient choice of epoch
is the present.  Then, 
Eq.~(\ref{sdef}) gives for $S(m)$ 
\begin{equation}
S(m) = \sigma_8^2 
\frac{\int_{k=0}^{k(m)} T^2(k)k^{n+2}dk}
{\int_{k=0}^{k(m_8)} T^2(k)k^{n+2}dk} \,.
\end{equation}

Thus we only consider $\ed(\vec{x},a_0)$ (the overdensity field
linearly extrapolated to the present epoch), which we use instead of 
the true field $\delta(\vec{x},a)$ in our
random walk formalism \footnote{Physically, the substitution of the true field by the extrapolated
field in the ``random walk'' corresponds to
smoothing the extrapolated overdensity field, and then mapping the mean
extrapolated overdensity value to a true overdensity value. That true
overdensity value is then assumed to accurately represent the result of
a smoothing of the true field, which implies
$
\delta\left(\left\langle\ed\right\rangle\right) = 
\left\langle\delta\left(\ed\right)\right\rangle$.
This would be exactly true only if $\delta(\ed)$ was linear in $\ed$,
which is not the case (see chapter 3). 
This assumption introduces an inaccuracy 
inherent to all calculations which employ it, including the
Press-Schechter mass function as well as the 
double distribution. }. 
To find the mass function at a particular cosmic epoch
$a$, we calculate the location of the ``absorbing wall'', $\ed_{\rm
  c}(a)$.
If a structure is predicted to 
collapse at epoch $a$ according to the spherical evolution model,
then $\ed_{\rm 0,c}(a)$ is the overdensity this same structure would have had if, 
instead of turning around and collapsing, 
it had continued its linear evolution until the present. 
This $\ed_{\rm 0,c}(a)$ is then our ``absorbing wall''.

We can now derive the mass fraction and mass function 
using random walk theory. If a particle executes a one-dimensional
random walk with an absorbing boundary at a point $x_1$, then its 
probability ${\cal W}(x,t) $ to be 
between $x$ and
$x+dx$ at time $t$ is \cite{ch}
\begin{equation}\label{rweq}
{\cal W}(x,t,x_1)dx = \frac{
\exp\left[-\frac{x^2}{4Dt}\right]
-\exp \left[-\frac{(2x_1-x)^2}{4Dt}\right]
}{2\sqrt{\pi Dt}}dx\, ,
\end{equation}
where $x \le x_1$.
In our case, the probability that a
point in space will be assigned an average extrapolated overdensity 
between $\ed$ and $\ed+d\ed$ when filtered at a scale $m$
corresponding to a variance of $S(m)$ is
\begin{equation}\label{wforPS}
{\cal W}(\ed,S,\ed_{\rm 0,c})d\ed = \frac{
\exp\left[-\frac{\ed^2}{2S}\right]
-\exp \left[-\frac{(2\ed_{\rm 0,c}-\ed)^2}{2S}\right]
}{\sqrt{2\pi S}}d\ed\, ,
\end{equation}
with $\ed \le \ed_{\rm 0,c}$.
The mass fraction $P(>m,a)$ is then the fraction of points 
already ``lost'' from the walk when filtering at higher mass scales,
which is one minus the fraction of points remaining in the walk,
\begin{eqnarray}
P(>m,a) &=& P(>\ed_{\rm 0,c}) 
= 1-\int_{-\infty}^{\ed_{\rm 0, c}} {\cal W}(\ed,S, \ed_{\rm 0,c})d\ed 
\nonumber \\
&=& {\rm erfc} \left(\frac{\ed_{\rm 0,c}(a)}{\sqrt{2S(m)}}\right)\,.
\end{eqnarray}
Then, 
\begin{equation}\label{dPdm}
\frac{dP(>m,a)}{dm} = 
\frac{1}{\sqrt{2\pi}} \frac{\ed_{\rm 0,c}(a)}{S(m)^{3/2}}\frac{dS}{dm}
\exp \left[-\frac{\ed_{\rm 0,c}(a)^2}{2S(m)}\right]\,,
\end{equation}
and the Press-Schechter mass function can be found using Eq.~(\ref{funcfrac}), 
\begin{equation}
\frac{dn}{dm}(m,a) = \sqrt{\frac{2}{\pi}}\frac{\rho_{\rm m,0}}{m^2}
\frac{\ed_{\rm 0,c}(a)}{\sqrt{S(m)}} \left|\frac{d\ln \sqrt{S}}{d\ln m}\right|
\exp \left[-\frac{\ed_{\rm 0,c}(a)^2}{2S(m)}\right]\,.
\end{equation}

\section{Derivation of the Double Distribution}\label{ddderiv}

\indent

We now use the random walk formalism described in the previous
section to derive the double distribution of the comoving number
density of collapsed structures with respect to object mass $m$ and local
environment overdensity $\delta_{\ell}$, $dn/(dm \, d\delta_{\ell})$.

For the reasons described in the previous section, 
we replace the true overdensity
field, $\delta(\vec{x},a)$, with its linear extrapolation to the
present time, $\ed(\vec{x},a_0)$. Thus, we derive the double
distribution of comoving $n$ with respect to object mass $m$ and {\em
  extrapolated} local environment overdensity $\ed_{\ell}$,
$dn/(dm\,d\ed_{\ell})$. We then use linear theory and the spherical
evolution model to establish a conversion relation $\delta(\ed,a)$
and calculate $dn/(dm \, d\delta_{\ell})$ as
\begin{equation} \label{convtotrue}
\frac{dn}{dmd\delta_{\ell}}(\delta_{\ell},m,a)dm d\delta_{\ell} 
= \frac{dn}{dmd\ed_{\ell}} \left[\ed_{\ell}(\delta_{\ell},a),m,a\right]dm\frac{\partial
  \ed}{\partial \delta_{\ell}}d\delta_{\ell} \,.
\end{equation}

First of all, we need to define the local environment extrapolated 
overdensity $\ed_{\ell}$ in a precise way. We would like $\ed_{\ell}$ to be a
measure of the density contrast of the
medium in which a collapsed structure is embedded. Clearly, the value
of $\ed_{\ell}$ depends on how far from the structure itself its 
``environment''
extends. We quantify this notion by introducing the 
{\em clustering scale parameter}, $\beta$, which is defined in the
following way: the ``environment'' of an object of mass $m$ is a
surrounding region in space which encompasses mass $\beta m$
(including the mass of the object). Hence, 
the local environment
extrapolated overdensity $\ed_{\ell}$ is the result of a filtering 
of $\ed(\vec{x},a_0)$ with a filter of scale $\beta m$ centered on the
object. 

Formally, the above definition translates as follows. Consider 
the sharp in $k-$space filtering function 
$\hat{W}_m(k)$ discussed previously (Eq.~\ref{what}). 
The relation between the cutoff wavenumber and the filter mass,
$k_{\rm c}(m)$, is found by considering the form of the filter function in
configuration space, 
\begin{equation}
W_m(r) = \frac{\sin \left[k_{\rm c}(m)r\right] 
- k_{\rm c}(m) r \cos \left[k_{\rm c}(m)r\right]}{2\pi^2r^3}\,,
\end{equation}
and multiplying by $\rho_{\rm m,0}$ and integrating over all space, which
yields
\cite{lc93}, 
\begin{equation}
k_{\rm c}(m) = \left(\frac{6\pi^2\rho_{\rm m,0}}{m}\right)^{1/3}\,.
\end{equation}
For a collapsed structure at an epoch $a$ which has mass $m$ and is
centered at a point $\vec{x}_0$, we can write
\begin{equation}
\ed(m,\vec{x_0}) = 
\int W_m\left(|\vec{x}_0-\vec{x}|\right)\ed(\vec{x},a_0)d^3\vec{x}
= \ed_{\rm 0,c}(a)\,,
\end{equation}
since the mean extrapolated overdensity of the collapsed structure
itself is always the critical value for collapse, $\ed_{\rm 0,c}(a)$.
For that same object, the {\em local environment extrapolated
  overdensity}, $\ed_{\ell}$, is
\begin{equation} \label{defbet}
\ed_{\ell}(m,\vec{x_0}) = \int W_{\beta m}
\left(|\vec{x}_0-\vec{x}|\right)\ed(\vec{x},a_0)d^3\vec{x}\,.
\end{equation}
Equation (\ref{defbet}) is then the definition of $\ed_{\ell}$ for a given
$\beta$. In our double distribution, $\beta$ is  free parameter, 
which is however constrained to be
between 1 and a few on physical grounds. It cannot be $<1$ since the
mass of the object's environment always includes the mass of the
object itself. In fact, as $\beta$  approaches $1$, the averaging
which produces $\ed_{\ell}$ is taken {\em only} over the collapsed object
itself, and inevitably returns the critical overdensity for
collapse, $\ed_{\rm 0,c}$, for all objects.   
In the other extreme, $\beta \gg 1$, the average
$\ed$ on a scale $\beta m$ is no longer a local quantity with respect to the
central object. When $\beta$ grows without bound, $\ed_{\ell}$ 
approaches $0$ for all collapsed structures, since averaging over the whole
universe identically returns the background matter density,
which corresponds to a vanishing density contrast. In appendix \ref{ap_lim} we
show that  our double distribution becomes proportional to a 
Dirac delta-function around $\ed_{\ell}=0$ in the limit $\beta
\rightarrow \infty$ and proportional to a Dirac delta-function around
$\ed_{\ell}=\ed_{\rm 0,c}$ in the limit $\beta \rightarrow 1$.

We are now ready to use random walk theory results to 
calculate the double distribution. We first find 
the fraction of points in space which belong to
structures of mass between $m$ and $m+dm$, which in turn are embedded
in a medium of mean linearly extrapolated overdensity between $\ed_{\ell}$
and $\ed_{\ell}+d\ed_{\ell}$, $f(m, \ed_{\ell}, \beta)d\ed_{\ell}\,dm$. The double
distribution then is
\begin{equation}\label{prel1}
\frac{dn}{dm d\ed_{\ell}}= \frac{\rho_{\rm m,0}}{m}f(m,\ed_{\ell},\beta)dm d\ed_{\ell}.
\end{equation}

The quantity $f$ can be written as
\begin{equation}\label{prel2}
f dm \, d\ed_{\ell} = (f_1d\ed_{\ell})(f_2dm)
\end{equation}
where $f_1d\ed_{\ell}$ is the fraction
 of points in space which have an average overdensity between $\ed_{\ell}$
 and $\ed_{\ell}+d\ed_{\ell}$ on a smoothing scale $\beta m$, and $f_2 dm$ is the
 fraction of points satisfying the previous condition which belong to
 collapsed structures of mass between $m$ and $m+dm$.

The first of the two factors above is the fraction of points 
still in the walk which are found between
$\ed$ and $\ed+d\ed$ at a ``time'' $\beta m$.
This is the solution of the 
1D random walk problem of $\ed_{\ell}$ as a function of 
$S$, with an absorbing boundary at 
the critical collapse threshold $\ed_{\rm 0,c}$, as given by Eq.~(\ref{wforPS}) 
but for a smoothing scale $\beta m$, 
\begin{equation}
f_1d\ed_{\ell} = 
\frac{
\exp\left[-\frac{\ed_{\ell}^2}{2S(\beta m)}\right]
-\exp \left[\frac{(\ed_{\ell}-2\ed_{\rm 0,c}(a))^2}{2S(\beta m)}\right]}
{\sqrt{2\pi S(\beta m)}}
d\ed\,.
\end{equation}

The second factor ($f_2$) is the {\em conditional probability} that a point in
space originating from $(\beta m, \ed)$ in the mass - overdensity
plane,will reach the ``wall'' for the first time for  
a smoothing scale between $m$ and $m + dm$.
This is then the probability that 
a particular point in space is absorbed by the ``wall''
$\ed_{\rm 0,c}(a)$ at a particular ``time'' $S(m)$, 
provided that the origin of the walk is transferred from $(0,0)$ 
to $(S(\beta m),\ed_{\ell})$. This probability can then be found if,
in the expression for $dP(>m,a)/dm$ (Eq.~\ref{dPdm}), we perform the
substitutions $\ed_{\rm 0,c} \rightarrow \ed_{\rm 0,c}-\ed_\ell$ and 
$S(m) \rightarrow S(m)-S(\beta m)$.
Similar conditional probabilities were originally calculated by
\cite{b91} and \cite{lc93}
in the context of rates of mergers between halos. In our case, it is
\begin{equation}
f_2 dm = \frac{\left[\ed_{\rm 0,c}(a)-\ed_{\ell}\right]
\exp\left[-\frac{\left(\ed_{\rm 0,c}(a)-\ed_{\ell}\right)^2}{2\left[S(m)-S(\beta m)\right]}\right]}
{\sqrt{2\pi}\left[S(m)-S(\beta m)\right]^{3/2}}
\left|\frac{dS}{dm}\right|_m dm\,.
\end{equation}

Equations (\ref{prel1}) and (\ref{prel2}) then give
\begin{eqnarray} \label{dd}
\frac{dn}{dmd\ed_{\ell}}(m,\ed_{\ell},\beta,a)&=&
\frac{\rho_{\rm m,0}}{m} \,\,
\frac{\ed_{\rm 0,c}(a)-\ed_{\ell}}{2\pi }\,\,
\frac{
\exp \left[-\frac{\ed_{\ell}^2}{2S(\beta m)}\right]
- \exp\left[-\frac{\left(\ed_{\ell} - 2 \ed_{\rm 0,c}(a)\right)^2}{2S(\beta
    m)}\right]
}
{[S(\beta m)]^{1/2}
\left[S(m)-S(\beta m)\right]^{3/2}}
\left|\frac{dS}{dm}\right|_m  \nonumber \\
&& \times \exp\left[-\frac{\left(\ed_{\rm 0,c}(a)-\ed_{\ell}\right)^2}
{2\left[S(m)-S(\beta m)\right]}\right]
\end{eqnarray}

with $\ed_{\ell} \le \ed_{\rm 0,c}(a)$ and $\beta > 1$ so $S(m) > S(\beta m)$
\footnote{since $S(m)$ monotonically decreases with $m$ for all
physically interesting power spectra}.
Equation (\ref{dd}) is the double distribution we have sought and is the
central result of this chapter. 
Integrating $dn/(dmd\ed_{\ell})$ over $\ed_{\ell}$ yields the Press-Schechter 
mass function, as it should. The result is independent of the value of $\beta$.
We explicitly perform this integration in appendix \ref{ap_int}.

Note that the functional form of our
double distribution is similar with that of the integrand used by \cite{mw96}
in their calculation of the cross-correlation between dark halos and
mass using random walk theory, however the second variance of the
field (corresponding to our $S(\beta m)$) in their case refers to a
fixed clustering radius and is independent of object mass. 

\section{Converting Between Linear And Spherical Density
Contrasts}\label{secconv}

\indent

In this chapter, we derive and discuss
 conversion relations between the linearly
extrapolated density contrast entering the double distribution as we
derived it using random walk theory, and the more physical density
contrast predicted by the spherical evolution theory. In \S
\ref{ap1} and \ref{ap2} we derive exact expressions for 
$\ed_{\ell}(\delta_{\ell},a)$ in the case of the spherical evolution
model, for an $\Omega_{\rm m}=1$ (\S \ref{ap1}) and an
$\Omega_{\rm m}+\Omega_\Lambda=1$ (\S \ref{ap2}) universe 
(note however that all of the equations we have presented up to this point are
cosmology-independent, and can therefore be adapted for any cosmological
model). In \S \ref{aprel} we present a useful approximation, valid for
both types of cosmological models.

\subsection{Exact Conversion in an Einstein-de Sitter 
Universe}\label{ap1}

\indent

In this section we derive a conversion relation
$\ed_0(a,\delta)$ for an $\Omega_{\rm m}=1$ cosmology (here, $\delta$ is the
density contrast predicted for a density perturbation at cosmic epoch
$a$ by the spherical evolution model and $\ed_0$ is the extrapolation
of the density contrast to the present cosmic epoch using
linear theory). In order to do so, we first calculate $\delta(a)$
from the spherical evolution solution, then calculate $\ed_0$ using
linear theory, and finally require that $\delta(a)$ and $\ed_a$ (the 
linear-theory density contrast at epoch $a$) should
agree at early times.

\subsubsection{Spherical Evolution Model in an {\boldmath $\Omega_{\rm
      m}=1$ }Universe}

\indent

The evolution of a spherically symmetric, overdense perturbation in an
otherwise homogeneous $\Omega_{\rm m}=1$ universe is described by the
parametric equations 
\begin{equation}\label{par2}
a_{\rm p} = \frac{2a_{\rm coll}}{(12\pi)^{2/3}} 
(1-\cos \theta) \, {\rm ,\,\, and \,\,}
a =  a_{\rm coll}\left(\frac{\theta - \sin \theta}{2\pi}\right)^{2/3} \,\, ,
\end{equation}
where $a_{\rm coll}$ is the scale factor of the universe when the
perturbation formally collapses to a point, $a_{\rm p}$ is the scale
factor of the perturbation, and $\theta$ is the development
angle. Note that the perturbation will turn around (reach its maximum
size, $a_{\rm p,max} = 4a_{\rm coll}(12\pi)^{-2/3}$) when $\theta=\pi$, at
a time $a=a_{\rm coll}/2^{2/3}$.

The normalization of Eq.~(\ref{par2}) is such that the density
contrast $\delta$ can be expressed as 
\begin{equation}\label{gend}
\delta = \left(\frac{a}{a_{\rm p}}\right)^3-1\,.
\end{equation}
Hence, for any density contrast $\delta$, Eqs. (\ref{par2}) and
(\ref{gend}) can be combined to give a unique development angle
$\theta(\delta)$ which is the solution to the transcendental equation
\begin{equation}\label{thetaofdelta}
\frac{6^{2/3}(\theta-\sin\theta)^{2/3}}{2(1-\cos \theta)}
-(1+\delta)^{1/3} = 0\,.
\end{equation}

Similarly, the spherical evolution solution for an underdensity is
given by the parametric equations
\begin{equation}\label{upar2}
a_{\rm p}=A_{\rm p}(\cosh \eta -1)  \, {\rm ,\,\, and \,\,} a = A_{\rm
  p}
\frac{6^{2/3}}{2}(\sinh \eta -\eta)^{2/3}\,.
\end{equation}
where $\eta$ is the development angle in this case. Equation
(\ref{upar2}) together with Eq.~(\ref{gend}) can be combined as
before to give $\eta(\delta)$ as the solution to the transcendental
equation
\begin{equation}\label{etaofdelta}
\frac{6^{2/3}(\sinh\eta-\eta)^{2/3}}{2(\cosh \eta-1)}
-(1+\delta)^{1/3} = 0\,.
\end{equation}

\subsubsection{{\boldmath $\ed_0(a,\delta)$} 
according to the spherical evolution model}

\indent

The behavior 
of $\delta$ in the linear regime in this cosmology is
\begin{equation}\label{linom1}
\ed = \ed_0 a\,.
\end{equation}
This result should coincide with the linear expansion of the spherical
evolution result at early times. Expanding the 
parametric solution to second nonvanishing order in $\theta$ 
and eliminating $\theta$, we obtain
\begin{equation}\label{oper}
a_{\rm p}(a) = a \left[1-\frac{(12\pi)^{2/3}}{20}
\frac{a}{a_{\rm coll}}\right]\,.
\end{equation}
We then substite Eq.~(\ref{oper}) in the definition of
$\delta$ (Eq.~\ref{gend}) to get
\begin{equation}
\ed = \frac{3(12\pi)^{2/3}}{20 a_{\rm coll}}a
\end{equation}
which, by comparison to Eq.~(\ref{linom1}) gives
\begin{equation}\label{acollofed}
\ed_0 = \frac{3(12\pi)^{2/3}}{20 a_{\rm coll}}\,.
\end{equation}
Then, the conversion relation we seek is
\begin{equation}\label{otrue}
\ed_0(a,\delta) = \frac{6^{2/3}3}
{20a}\left[\theta(\delta)-\sin \theta(\delta)\right]^{2/3}
\end{equation}
where $\theta(\delta)$ is given by Eq.~(\ref{thetaofdelta}).

Equation (\ref{otrue}) has the undesirable property that it diverges as $\theta$
approaches $2\pi$. This is of course a consequence of the perturbation
formally collapsing to a singularity in the spherical evolution model instead
of reaching virial equilibrium. If we make the usual assumption that at
virialization the radius of the perturbation is $a_{\rm max}/2$ and we
additionally require that 
\begin{itemize}
\item $\ed_0(a,\delta)$ is continuous and smooth at $\theta=3\pi/2$
\item $a_{\rm p}=a_{\rm p,max}$ for all $a\ge a_{\rm coll}$
\end{itemize}
then for $\theta >3\pi/2$ (which corresponds to $\delta>9(3\pi+2)^2/8$)
we can replace Eq.~(\ref{otrue}) with
\begin{eqnarray}
\ed_0(a, \delta) &=& 
\ed_{\rm 0,v} + \ed_{\rm 0,v}'(\delta - \delta_{\rm v}) \nonumber \\ 
&+&\frac{3(\ed_{\rm0,c}-\ed_{\rm 0,v})-(\delta_{\rm c} - \delta_{\rm v})
(2\ed_{\rm 0,v}'+\ed_{0,c}')}{\left(\delta_{\rm c} - \delta_{\rm
    v}\right)^2}(\delta -\delta_{\rm v})^2 \nonumber\\
&+& \frac{(\ed_{\rm 0,c}'+\ed_{\rm 0,v}')(\delta_{\rm c}- \delta_{\rm
    v})
-2(\ed_{\rm 0,c}-\ed_{\rm 0,v})}{(\delta_{\rm c} -\delta_{\rm v})^3}
(\delta -\delta_{\rm v})^3 \nonumber\\
\label{patch}
\end{eqnarray}
(see \S \ref{ap2} for a discussion of the reasons for employing
this particular functional form, and \S \ref{ddres} for a
discussion on why the effect of such a choice on the double
distribution is negligible). In Eq.~(\ref{patch}), 
\begin{eqnarray}
\delta_{\rm v}&=&\left(\frac{a|_{\theta=3\pi/2}}{a_{\rm
    p}|_{\theta=3\pi/2}}\right)^3-1=
 \frac{9(3\pi+2)^2}{8}-1\nonumber \\
\delta_{\rm c}&=& \left(\frac{a|_{\theta=2\pi}}{a_{\rm
    p}|_{\theta=3\pi/2}}\right)^3-1= 18\pi^2-1 \nonumber\\
\ed_{\rm 0,v} &=& \ed_0(a,\delta_{\rm v}) = 
\frac{3^{5/3}}{20a} (3\pi+2)^{2/3} \nonumber \\
\ed_{\rm 0,c}' &=& \left.\frac{\partial \ed_0}{\partial \delta}\right|
_{\delta=\delta_{\rm c}} =
\frac{1}{10a(1+\delta_{\rm c})^{2/3}}\,,
\end{eqnarray}
the last equality coming from the fact that after $a_{\rm coll}$
the radius of a perturbation remains constant and equal to $a_{\rm
  p,max}/2$, while its density contrast $\delta$ changes only due to the
expansion of the background universe, $\delta = (2a/a_{\rm
  p,max})^3-1$ or $\delta = (10 a \ed_0/3)^3-1$. Finally, 
$\ed_{\rm 0,c}$ is given by Eq.~(\ref{matteredc}) while 
$\ed_{\rm 0,v}'$ is given by Eq.~(\ref{oder}) for $\delta=\delta_{\rm
    v}$ and $\theta=3\pi/2$.

To get the linear behavior of $\delta$ for an underdensity
we expand the parametric solution \ref{upar2} to
second nonvanishing order in $\eta$ and we eliminate $\eta$ to get
\begin{equation}\label{uper}
a_{\rm p} =  A_{\rm p}\frac{6^{2/3}}{2} \left[
1+\frac{1}{10}\frac{a}{A_{\rm p}}
\right]
\,.
\end{equation}
Substituting Eq.~(\ref{uper}) in the definition of
$\delta$ (Eq.~\ref{gend}), we get for the time dependence of 
$\delta$ at early times, 
\begin{equation}\label{ulin}
\ed =  -\frac{3}{10A_{\rm p}}a
\end{equation}
from which, by comparison to Eq.~(\ref{linom1}), we get
\begin{equation}
\ed_0 = -\frac{3}{10A_{\rm p}}\,.
\end{equation}
Then, $\ed_0(a,\delta)$ will be 
\begin{equation}\label{utrue}
\ed_0(a,\delta) = -\frac{6^{2/3}3}
{20a}\left[\sinh \eta(\delta) - \eta(\delta)\right]^{2/3}
\end{equation}
where $\eta(\delta)$ is given by Eq.~(\ref{etaofdelta}).

Equation 
(\ref{utrue}) is valid for all $\eta$ and its limit as $\ed \rightarrow
-\infty$ is $\delta(\ed) \rightarrow -1$. Thus, although the linearly
extrapolated field can become $<-1$, the corresponding value of the actual
$\delta$ is always $\ge -1$, as the physical requirement $\rho_{\rm p} \ge 0$
demands.

\subsubsection{Critical extrapolated overdensity for collapse, {\boldmath
    $\ed_{\rm 0,c}(a)$}}

\indent

The critical extrapolated overdensity for collapse can be found from
Eq.~(\ref{acollofed})
\begin{equation}\label{matteredc}
\ed_{\rm 0,c}(a_{\rm coll}) = \frac{3(12\pi)^{2/3}}{20}a_{\rm coll}^{-1}
\approx 1.69 a_{\rm coll}^{-1}\,.
\end{equation}
Note that the above equation has the functional form 
$\ed_{\rm 0,c}(a_{\rm coll})\propto 
1/D(a_{\rm coll})$, where $D(a)$ is the linear growth factor for this
cosmology. This is also true in the $\Omega_{\rm m}+\Omega _\Lambda = 1$ case.

\subsubsection{{\boldmath$\partial \ed_0 / \partial \delta |_a$}}

\indent

In addition to the relation between $\delta$ and $\ed_0$, we will also need the
derivative $\partial \ed_0 / \partial \delta |_a$ in order to convert between
true and extrapolated overdensity differentials in
Eq.~(\ref{convtotrue}). 
In the case of an overdense structure, $\delta>0$, Eq.~(\ref{otrue}) gives
\begin{equation}\label{oder}
\left.\frac{\partial \ed_0}{\partial \delta}\right|_{a} =
\frac{6^{2/3}}{10a}\frac{1-\cos \theta (\delta)}
{\left[\theta(\delta)-\sin \theta (\delta)\right]^{1/3}}
\frac{d\theta}{d\delta} \,.
\end{equation}
To evaluate $d\theta/d\delta$ we define the
auxiliary function
\begin{equation}\label{faux}
F_{\rm a}(\theta, \delta) = 6^{2/3}(\theta-\sin \theta)^{2/3}
-2(1-\cos\theta)(1+\delta)^{1/3}\,.
\end{equation}
From Eq.~(\ref{thetaofdelta}) we get immediately
$F_{\rm a}(\theta,\delta) = 0$, and differentiating we get
$dF_{\rm a} = 0 = \frac{\partial F_{\rm a}}{\partial \theta}d\theta+
\frac{\partial F_{\rm a}}{\partial \delta}d\delta$. Hence, 
\begin{equation}
\frac{d \theta}{d\delta}
= -\frac{\partial F_{\rm a}}{\partial \delta}\left(\frac{\partial
  F_{\rm a}}{\partial \theta}\right)^{-1} \,,
\end{equation}
where
\begin{equation}
\frac{\partial F_{\rm a}}{\partial \delta} = 
-\frac{2}{3}\frac{1-\cos \theta}{(1+\delta)^{2/3}}
\end{equation}
and
\begin{equation}
\frac{\partial F_{\rm a}}{\partial \theta} = 
\frac{6^{2/3}2}{3}\frac{1-\cos \theta}{(\theta-\sin \theta)^{1/3}}
-2(1+\delta)^{1/3}\sin \theta \,.
\end{equation}

Equation (\ref{oder}) is valid only for $0<\delta<\delta_{\rm v}$. For
$\delta > \delta_{\rm v}$ Eq.~(\ref{patch}) gives
\begin{eqnarray}
\left.\frac{\partial \ed_0}{\partial \delta}\right|_{a} &=&
\ed_{\rm 0,v}'  \nonumber \\ 
&&\!\!\!\!\! +
2\frac{3(\ed_{\rm0,c}-\ed_{\rm 0,v})-(\delta_{\rm c} - \delta_{\rm v})
(2\ed_{\rm 0,v}'+\ed_{0,c}')}{\left(\delta_{\rm c} - \delta_{\rm
    v}\right)^2}(\delta -\delta_{\rm v}) \nonumber\\
&& \!\!\!\!\! +
3\frac{(\ed_{\rm 0,c}'+\ed_{\rm 0,v}')(\delta_{\rm c}- \delta_{\rm
    v})
-2(\ed_{\rm 0,c}-\ed_{\rm 0,v})}{(\delta_{\rm c} -\delta_{\rm v})^3}
(\delta -\delta_{\rm v})^2\,. \nonumber\\
\end{eqnarray}

In the case of an underdense structure, $\delta<0$, Eq.~(\ref{utrue}) gives
\begin{equation}
\left.\frac{\partial \ed_0}{\partial \delta}\right|_{a} =
-\frac{6^{2/3}}{10a}\frac{\cosh \eta (\delta)-1}
{\left[\sinh \eta (\delta)-\eta(\delta)\right]^{1/3}}
\frac{d\eta}{d\delta} \,.
\end{equation}
As before, in order to evaluate $d\eta / d\delta$ we define the
auxiliary function
\begin{equation}\label{gaux}
G_{\rm a}(\eta, \delta) = 
6^{2/3}(\sinh \eta - \eta)^{2/3}
-2(\cosh\eta-1)(1+\delta)^{1/3}\,.
\end{equation}
Equation  (\ref{etaofdelta}) implies
$G_{\rm a}(\eta,\delta) = 0$ so
\begin{equation}
\frac{d \eta}{d \delta}
= -\frac{\partial G_{\rm a}}{\partial \delta}\left(\frac{\partial
  G_{\rm a}}{\partial \eta}\right)^{-1} \,,
\end{equation}
where
\begin{equation}
\frac{\partial G_{\rm a}}{\partial \delta} = 
-\frac{2}{3}
\frac{\cosh \eta-1}{(1+\delta)^{2/3}}\,,
\end{equation}
and
\begin{equation}
\frac{\partial G_{\rm a}}{\partial \eta} = 
\frac{6^{2/3}2}{3}\frac{\cosh \eta -1}{(\sinh \eta - \eta)^{1/3}}
-2(1+\delta)^{1/3}\sinh \eta\,.
\end{equation}

\subsection{Exact Conversion in
  an \boldmath{$\Omega_{\rm m}+\Omega_\Lambda=1$} Universe}\label{ap2}

\indent

In this section we will derive a conversion between true and
extrapolated overdensity, $\delta(\ed_0, a)$ for an 
$\Omega_{\rm m}+\Omega_\Lambda = 1$
cosmological model. We will do so by first
calculating the true density contrast $\delta(a)$ of a density
perturbation at cosmic epoch $a$ as predicted by the spherical
evolution model, then calculating $\ed_0$, which is the overdensity of the same
spherical perturbation if extrapolated according to the linear theory 
until the present cosmic epoch, and finally requiring that at early times
linear theory and the linear expansion of the spherical evolution
model should give the same result. 

\subsubsection{Spherical Evolution Model in an $\Omega_{\rm m} 
+ \Omega_{\rm \Lambda} = 1$ Cosmology: The Evolution Equation}

\indent

In the spherical evolution model, the spherical density perturbation
under consideration behaves as an independent non-flat sub-universe. Its evolution
is dictated by a Friedmann equation, 
\begin{equation}\label{friedt}
\left(\frac{da_{\rm p}}{dt} \right)^2= H_0^2 \Omega_{\rm m}a_{\rm p}^2
\left(a_{\rm p}^{-3}+\omega -\kappa a_{\rm p}^{-2}\right)
\end{equation}
where $a_{\rm p}$ is the  radius 
of such a spherical density perturbation in an otherwise homogeneous 
universe, $\omega=\Omega_\Lambda/\Omega_{\rm m}=\Omega_{\rm m}^{-1}-1$ (where
$\Omega_{\rm m}$ and $\Omega _{\rm \Lambda}$ are the matter and vacuum
density parameters of the background universe)
and $\kappa$ is a
constant characteristic of the amplitude and sign of the perturbation: the
larger the $|\kappa |$, the larger the deviation from homogeneity at a given
time, while a positive $\kappa$ corresponds to an overdensity and a negative
$\kappa$ to an underdensity. Clearly then in Eq.~(\ref{friedt}), the
first term in parentheses on the RHS is the matter term, the second is
the vacuum term and the third is the curvature term, which can have a
positive or negative sign depending on whether we are studying an
``open''(underdensity) or ``closed'' (overdensity) perturbation. 
The normalization of $a_{\rm p}$ 
is such that, had the specific spherical region begun
its evolution with no curvature ($\kappa=0$), $a_{\rm p}$ at the present
cosmic epoch would have been $a_{p(\kappa=0),0}=1$. For this
reason, the density contrast $\delta$ of the perturbation at epoch $a$
is given by Eq.~(\ref{gend})

The behavior of the perturbation radius $a_{p}$ as a function of the
universe scale factor $a$ can be found by taking the ratio of the
Friedmann equations of the perturbation and the background universe, 
thus obtaining \cite{peeb84}
\begin{equation}\label{genl}
\left(\frac{da_{\rm p}}{da}\right)^2 = 
\frac{a_{\rm p}^{-1}+\omega a_{\rm p}^2-\kappa}{a^{-1} + \omega a^2}
= \frac{a}{a_{\rm p}}\frac{\omega a_{\rm p}^2-\kappa a_{\rm
    p}+1}{\omega a^3 +1}
\,.
\end{equation}
Equation (\ref{genl}) implies that the smallest positive perturbation
which will turn around and collapse corresponds to the smallest positive
$\kappa$ for which the equation 
\begin{equation}\label{lf0}
\omega a_{\rm p}^3 -\kappa a_{\rm p} +1=0
\end{equation}
has a real positive solution \cite{ecf}. This gives
\begin{equation}\label{kmin}
\kappa_{\rm min, coll} = 3\omega^{1/3}/2^{2/3}\,.
\end{equation}
Equation (\ref{genl}) can then be re-written as
\begin{equation}\label{spel}
\frac{da_{\rm p}}{da} = \left\{
\begin{array}{ll}
\left(\frac{a_{\rm p}^{-1}+
\omega a_{\rm p}^2-\kappa}{a^{-1}+\omega a^2}\right)^{1/2},
& \kappa < \kappa_{\rm min,coll} \,\,\,\, {\it {\bf  \,\, or}}\\
& \kappa \ge \kappa_{\rm min,coll}, \,\,a<a_{\rm ta}\\ & \\ & \\
- \left(\frac{a_{\rm p}^{-1}+\omega a_{\rm p}^2-\kappa}{a^{-1}+
\omega a^2}\right)^{1/2},
& \kappa \ge \kappa_{\rm min,coll}, \,\,a>a_{\rm ta}\\
\end{array}
\right.
\end{equation}
where
$a_{\rm ta}$ is the scale factor of the universe when the perturbation
reaches its 
maximum (or {\it turnaround}) radius. The turnaround radius is the
smallest of the two positive solutions of Eq.~(\ref{lf0}), 
\begin{equation}\label{aptaofk}
a_{\rm p,ta} =
\omega^{-1/3}\sqrt{\frac{4}{3}\frac{\kappa}{\omega^{\frac{1}{3}}}}
\cos \frac{1}{3}\left(
\cos ^{-1}\sqrt{\frac{27}{4}\left(\frac{\kappa}{\omega^{\frac{1}{3}}}
\right)^{-3}} +\pi
\right)\,.
\end{equation}
Equation (\ref{aptaofk}) has an 
asymptotic behavior 
$a_{\rm p,ta} \approx
1/\kappa$ when $\kappa/\omega^{1/3} \gg 1$, 
as expected from Eq.~(\ref{lf0}). The maximum possible turnaround
radius, $a_{\rm p,ta,max}$ is achieved for $\kappa=\kappa_{\rm min,coll}$ and
is $a_{\rm p,ta,max}=(2\omega)^{-1/3}$. All other collapsing
overdensities will have $a_{\rm p,ta}<a_{\rm p,ta,max}$.

\subsubsection{Qualitative Description of the Evolution of Structures}

\indent

The introduction of the additional vacuum term in the Friedmann
equation considerably complicates the simple classification of density
perturbations to overdensities (all of which turn around and collapse
in an $\Omega_{\rm m}=1$ cosmology) and underdensities (all of which
expand forever). In 
the $\Omega_{\rm m}+\Omega_{\rm \Lambda} =1$
universe there exist overdensities which will continue to expand
forever. The behavior of a perturbation in such a cosmology is
parametrized by the quantity $\kappa/\omega^{1/3}$, and we can
identify the following cases.

{\bf Case I, \boldmath{$\kappa/\omega^{1/3} \leq -1$}: large
  underdensities, expanding forever}. The table below shows the 
  relative magnitude of the three terms in the Friedmann equation
  (matter, curvature and vacuum term) for different values of the
  scale factor of the perturbation. The first line in the table
  indicates the hierarchy of the three terms, from largest to smaller,
  for each range of the scale factor. The second line indicates the
  dominant term in each scale factor range.  The third line shows the
  approximate dependence of the radius of the perturbation, $a_{\rm p}$, on
  time, assuming that only the dominant term contributes to the
  Friedmann equation in each range.
\begin{center}
\begin{tabular}{|c|c|c|c|}
\hline
$a_{\rm p}<\frac{1}{|\kappa|}$ &
$\frac{1}{|\kappa|}<a_{\rm p}< \frac{1}{\sqrt[3]{\omega}} $&
$ \frac{1}{\sqrt[3]{\omega}} < a_{\rm p} < \sqrt{\frac{|\kappa|}{\omega}}$ 
& $a_{\rm p}>\sqrt{\frac{|\kappa|}{\omega}}$\\
\hline
\hline
MCV & CMV& CVM & VCM\\
\hline
matter & curvature & curvature & vacuum \\
\hline
$a_{\rm p} \sim t^{2/3}$ & $a_{\rm p} \sim t$ & $a_{\rm p} \sim t$ &
$a_{\rm p} \sim {\rm e}^t$\\
\hline 
\end{tabular}
\end{center}

{\bf Case II, \boldmath {$-1 < \kappa/\omega^{1/3} \leq 1$}: small
  perturbations, expanding forever}. 
These can be either underdensities ($\kappa <0$) or
  overdensities ($\kappa >0$). In both cases 
  the curvature term never becomes dominant. The following table shows
  their different evolutionary stages (as in Case I).
\begin{center}
\begin{tabular}{|c|c|c|c|}
\hline
$a_{\rm p}<\sqrt{\frac{|\kappa|}{\omega}}$ &
$\sqrt{\frac{|\kappa|}{\omega}}<a_{\rm p}< \frac{1}{\sqrt[3]{\omega}}$&
$ \frac{1}{\sqrt[3]{\omega}} < a_{\rm p} < \frac{1}{|\kappa|}$& 
$a_{\rm p}>\frac{1}{|\kappa|}$\\
\hline
\hline
MCV & MVC& VMC & VCM\\
\hline
matter & matter & vacuum & vacuum \\
\hline
$a_{\rm p} \sim t^{2/3}$ & $a_{\rm p} \sim t^{2/3}$ & $a_{\rm p} \sim
    {\rm e}^t$ &$a_{\rm p} \sim {\rm e}^t$ \\
\hline
\end{tabular}
\end{center}

{\bf Case III, \boldmath{$1 < \kappa/\omega^{1/3} < 3/2^{2/3}$}:
``coasting'' overdensities, expanding forever}.
These overdensities continue to expand forever despite the fact that
they go through a phase in their evolution when the curvature term
becomes dominant and their expansion slows down. 
During this phase, the contributions of the matter 
and vacuum terms, which are the ones driving the expansion, add up to 
a value always higher than the curvature term, although the curvature
term is larger than each one of them. When the
perturbation enters the curvature-dominated phase, the expansion rate 
decreases and the perturbation grows much more mildly than $t^{2/3}$.
The expansion rate reaches a minimum at $a_{\rm p}=(2\omega)^{-1/3}$, 
after which it increases again as the perturbation 
approaches the phase of exponential expansion. This phase between 
the matter-like expansion and the exponential expansion is denoted 
by $(*)$ in the table below.
\begin{center}
\begin{tabular}{|c|c|c|c|}
\hline
$a_{\rm p}<\frac{1}{\kappa}$ &
$\frac{1}{\kappa}<a_{\rm p}< \frac{1}{\sqrt[3]{\omega}} $&
$ \frac{1}{\sqrt[3]{\omega}} < a_{\rm p} < \sqrt{\frac{\kappa}{\omega}}$ 
& $a_{\rm p}>\sqrt{\frac{\kappa}{\omega}}$\\
\hline
\hline
MCV & CMV& CVM & VCM\\
\hline
matter & curvature & curvature & vacuum \\
\hline
$a_{\rm p} \sim t^{2/3}$ &  $(*)$ & $(*)$ &
$a_{\rm p} \sim {\rm e}^t$\\
\hline 
\end{tabular}
\end{center}

Cases I-III are all sub-cases of the Lema\^{\i}tre model (\cite{lem1}, \cite{lem2}), which
features an inflection point at $a_{\rm p,e}=(2\omega)^{-1/3}$ where 
$\ddot{a}_{\rm p}=0$ while $\dot{a}_{\rm p}>0$. The rate of expansion
initially decreases to achieve its minimum (positive) value when
$a_{\rm p} = a_{\rm p,e}$, after which point the expansion accelerates again.

{\bf Special Case, \boldmath{$\kappa/\omega^{1/3} = 3/2^{2/3}$}:
Eddington Overdensity}.
This overdensity is the lowest $\kappa$ overdensity which does not 
expand to an infinite radius. However, it does not turn around and 
collapse, but it approaches its (finite) turnaround radius, $a_{\rm
  p,max}=(2\omega)^{-1/3}$ (from Eq.~\ref{aptaofk}) as $t\rightarrow \infty$.
As seen by an observer inside this overdensity, as $t\rightarrow
\infty$ the part of the universe outside $a_{\rm p,max}$ will accelerate away
and eventually exit the horizon, and the observable universe (``local
Eddington bubble'') 
will asymptotically approach the Einstein static universe (as in the
Eddington model with a cosmological constant).

{\bf Case IV, \boldmath{$\kappa/\omega^{1/3} > 3/2^{2/3}$}:
large overdensities, eventually collapsing}. When such a structure
enters the dominant-curvature-term phase, its expansion rate 
starts to decrease ($a_{\rm p}\sim t^{\epsilon}$
with $\epsilon = \epsilon(t)$ monotonically decreasing from $2/3$ to
0), until the expansion halts, at $a_{\rm p}
= a_{\rm p,ta}$ which occurs at a time $t_{\rm ta}$, given in table
\ref{timescales}. 
After $t_{\rm ta}$ the perturbation turns around and contracts, 
its evolution being symmetrical in time about $t_{\rm ta}$, 
i.e. $a_{\rm p}(t) = a_{\rm p}(2t_{\rm ta}-t)$ for $t > t_{\rm ta}$
(this is a consequence of Eq.~(\ref{friedt}) and holds for any
cosmological model as long as the RHS of the Friedmann equation
involves no explicit time-dependence).
Eventually, the perturbation will formally collapse to a singularity 
at time $t_{\rm coll} = 2t_{\rm ta}$.
\begin{center}
\begin{tabular}{|c|c||c|c|}
\hline
$a_{\rm p}<\frac{1}{\kappa}$ &
$\frac{1}{\kappa}<a_{\rm p}< a_{\rm p,ta} $&
$a_{\rm p,ta}>a_{\rm p}>\frac{1}{\kappa}$&
$\frac{1}{\kappa}>a_{\rm p}$
\\
\hline
\hline
MCV & CMV & CMV & MCV\\
\hline
matter & curvature & curvature & matter\\
\hline
$a_{\rm p} \sim t^{2/3}$ & $a_{\rm p} \sim t^\epsilon$ &
$a_{\rm p} \sim (2t_{\rm ta}\!\!-t)^\epsilon$ & 
$a_{\rm p} \!\!\sim \!\!(2t_{\rm ta}\!\!-t)^{2/3}$
\\
\hline 
expansion & expansion & contraction & contraction \\
\hline
\end{tabular}
\end{center}

In all of the cases discussed above, the transitions between different
phases of their evolution occur at characteristic times, those of
matter-vacuum equality $t_{\rm MV}$, matter-curvature
equality $t_{\rm MC}$ and curvature-vacuum equality $t_{\rm CV}$. At
these times (shown in table \ref{timescales}), 
the corresponding terms in the Friedmann equation become
equally important. Note that in the case of the Eddington overdensity 
and of case IV
collapsing overdensities, matter-vacuum equality and curvature-vacuum
equality are never reached, and the vacuum term never dominates over
any of the other terms.

\begin{table}
\caption{Characteristic times of the spherical evolution model in an
  $\Omega_{\rm m}+\Omega_{\rm \Lambda} = 1$
  universe. \label{timescales}}
\begin{tabular}{lll}
\hline \hline
{\bf event} &$\,\,\,$ & {\bf time} \\
$\,$turnaround & &
$t_{\rm ta} = \frac{1}{H_0\sqrt{\Omega_{\rm m}}}
\int _{0}^{a_{\rm p, ta}} da_{\rm p}\sqrt{
\frac{a_{\rm p}}{\omega a_{\rm p}^3-\kappa a_{\rm p}+1}}$\\
\hline 
\begin{tabular}{l}
matter-vacuum \\equality \end{tabular} && 
$t_{\rm MV} = \frac{1}{H_0\sqrt{\Omega_{\rm m}}}
\int _{0}^{\omega^{-1/3}} da_{\rm p}\sqrt{
\frac{a_{\rm p}}{\omega a_{\rm p}^3-\kappa a_{\rm p}+1}}$ \\
\hline 
\begin{tabular}{l}
matter-curvature \\equality \end{tabular} &&
$t_{\rm MC} = \frac{1}{H_0\sqrt{\Omega_{\rm m}}}
\int _{0}^{|\kappa|^{-1}} da_{\rm p}\sqrt{
\frac{a_{\rm p}}{\omega a_{\rm p}^3-\kappa a_{\rm p}+1}}$ \\
\hline
\begin{tabular}{l}
curvature-vacuum \\equality \end{tabular} &&
$t_{\rm CV} = \frac{1}{H_0\sqrt{\Omega_{\rm m}}}
\int _{0}^{\sqrt{\frac{|\kappa|}{\omega}}} da_{\rm p}\sqrt{
\frac{a_{\rm p}}{\omega a_{\rm p}^3-\kappa a_{\rm p}+1}}$\\
\hline \hline
\end{tabular}
\end{table}

In the next section we derive exact solutions for the time-evolution
of $a_{\rm p}$ for perturbations of different curvature. 
However, surprisingly accurate approximate solutions can be derived
using only linear theory and 
Eq.~(\ref{magic}). Solving for the spherical collapse 
density contrast we get 
\begin{equation}
\delta_a \approx  \left(1-\frac{\ed_a}{\ed_{\rm c}}\right)^{-\ed_{\rm c}}-1\,.
\end{equation}
Since $a_{\rm p} = a(1+\delta_a)^{-1/3}$, we can write for collapsing
overdensities 
\begin{eqnarray}\label{apsc}
a_{\rm p} &\approx& a  \left[1-\frac{\ed_{\rm c}
  D(a)/D(a_{\rm c})}{\ed_{\rm c}}\right]^{\ed_{\rm c}/3}\nonumber \\
&=& a\left[1-\frac{D(a)}{D(a_{\rm coll})}\right]^{\ed_{\rm c}/3}\,.
\end{eqnarray}
where the initial conditions (curvature) of the
perturbation are parametrized by its collapse epoch, $a_{\rm coll}$, while the
cosmology enters through the functional form of the linear growth
factor and the linear collapse overdensity, $\ed_{\rm c}$. Similarly, for
perturbations which expand forever we can write
\begin{eqnarray}\label{apse}
a_{\rm p} &\approx& a  \left[1-\frac{\ed_0
  D(a)/D(a_{0})}{\ed_{\rm c}}\right]^{\ed_{\rm c}/3}\nonumber \\
&=& a\left[1-\frac{\ed_0}{\ed_c}\frac{D(a)}{D(a_0)}\right]^{\ed_{\rm c}/3}\,.
\end{eqnarray}
where the curvature of the perturbation is parametrized by its
extrapolated linear density contrast at the present epoch, $\ed_0$.
Note that for overdensities which expand forever, $\ed_0>0$ and
$a_{\rm p}<a$, while for underdensities $\ed_0<0$ and $a_{\rm p}>a$. Also, because
$D(a)$ asymptotes to a constant value for $a\rightarrow \infty$ (as we
will see in the next sections), 
$a_{\rm p}$ grows proportionally to $a$ at late
times. This is the exponential expansion phase, described in our
analysis above. 

\subsubsection{Solutions of the Evolution Equation}

\indent

For eventually collapsing structures ($\kappa \ge \kappa_{\rm min,coll}$), 
separation of
variables in Eq.~(\ref{spel}) and integration yields, 
\begin{equation}\label{sepvarint}
\int_0^a \frac{\sqrt{y}dy}{\sqrt{\omega y^3+1}} = 
\left\{
\begin{array}{lr}
\int_{_0}^{^{a_{\rm p}}}\!\!\!\!\!
\frac{\sqrt{x}dx}{\sqrt{\omega x^3 - \kappa x + 1}}
&  a<a_{\rm ta}
\\ \\
2\int_{_0}^{^{a_{\rm p, ta}}}\!\!\!\!\!\!\!\!\!\!\!
\frac{\sqrt{x}dx}{\sqrt{\omega x^3 - \kappa x + 1}}
\!  - \!\! \int_{_0}^{^{a_{\rm p}}} \!\!\!\!\!\!
\frac{\sqrt{x}dx}{\sqrt{\omega x^3 - \kappa x + 1}}
&  a\ge a_{\rm ta}
\end{array}
\right. \,,
\end{equation}
where $a_{\rm ta}$ is the cosmic epoch when $a_{\rm p} = a_{\rm p,ta}$.
Now the integral on the LHS of Eq.~(\ref{sepvarint}) can be calculated using
\cite{ecf}
\begin{equation}\label{lhsint}
\int \frac{\sqrt{y}dy}{\sqrt{\omega y^3+1}} = 
\frac{2}{3}\omega^{-1/2}\sinh^{-1}\sqrt{\omega y^3}\,\,.
\end{equation}
The integral of the RHS can be re-written as 
\begin{equation}\label{defv1}
\int_0^{a_{\rm p}} \frac{\sqrt{x}dx}{\sqrt{\omega x^3 -
    \kappa x +1}}=
\frac{2}{3}\omega^{-1/2}\mathcal{V}_1(r, \mu)
\end{equation}
where $\mathcal{V}_1$ is the {\em incomplete vacuum integral 
of the first kind}, defined in appendix \ref{vac_ints}, and 
\begin{equation}\label{rmu}
\begin{array}{lcccr}
r = a_{\rm p} / a_{\rm p, ta} \,,
& &&&\mu = (\omega a_{\rm p, ta}^3)^{-1}\\
\end{array} \,\,.
\end{equation}
Note that for this
range of curvature values, $\kappa/\omega^{1/3}$
(which is the quantity which parametrizes the behavior
of the perturbation with time) is a function of $\mu$ alone, with
 \begin{equation}\kappa/\omega^{1/3}=(1+\mu)/\mu^{2/3}\,.\end{equation}

Using Eqs. (\ref{lhsint}) and (\ref{defv1}), Eq.~(\ref{sepvarint})
can be rewritten as, 
\begin{equation} \label{eqfora}
a = \left\{
\begin{array}{lr}
\omega^{-1/3}\left\{\sinh
\left[\mathcal{V}_1(r,\mu)\right]\right\}^{2/3} 
\,, & a\le a_{\rm ta} \\ \\
\omega^{-1/3}\left\{\sinh \left[2 \mathcal{V}_1(1,\mu) - 
\mathcal{V}_1(r,\mu)\right]\right\}^{2/3} \,, & a >a_{\rm ta}
\end{array}
\right. \,.
\end{equation}
Equation (\ref{eqfora}) is the spherical evolution solution for a collapsing 
perturbation with
amplitude $\kappa$ (parametrized above by $\mu$) 
in an $\Omega_\Lambda + \Omega _{\rm m} =1$,
$\Omega_\Lambda/\Omega_{\rm m}=\omega$ universe, and gives $r$ (and hence
$a_{\rm p}$) as a function of
$a$ for this model. Note that $\mathcal{V}_1(r,\mu)$ is the
development angle for this cosmology.

The scale factor of the universe at turnaround for a given collapsing
overdensity can be found immediately from Eq.~(\ref{eqfora}), 
\begin{equation}
a_{\rm ta} = \omega^{-1/3} \left[\sinh \mathcal{V}_1(1,\mu)\right]^{2/3}\,.
\end{equation}
The scale factor of the universe at collapse, $a_{\rm
  coll}$ (when the scale factor of the
perturbation becomes formally zero, $a_{\rm p,c}= r_{\rm c} =0$) is, 
from Eq.~(\ref{eqfora}) and since $\mathcal{V}_1(0,\mu)=0$, 
\begin{equation} \label{acollapse}
a_{\rm coll} = \omega^{-1/3} \left[\sinh 2
  \mathcal{V}_1(1,\mu)\right]^{2/3} 
\,.
\end{equation}

Equation (\ref{eqfora}) should not be applied 
"literally" until the final collapse of the perturbation to a
singularity, since the physical picture for the late stages of the
evolution of a perturbation involves virialization at a finite radius.
It has been shown by \cite{lah}
that the analogous arguments which give $a_{p,v}=a_{p,ta}/2$ for the
$\Omega_{\rm m}=1$ cosmology give, for an $\Omega_{\rm m}+\Omega_\Lambda=1$ universe, 
\begin{equation}\label{3vofta}
4\omega a_{\rm p,v}^3 - \frac{2+2\omega a_{p,ta}^3}{a_{p,ta}}a_{p,v} + 1 =0\,.
\end{equation}
The physically meaningful solution of \ref{3vofta} which gives the correct
behavior for $\omega \rightarrow 0$ is (using Eq.~\ref{rmu})
\begin{eqnarray}\label{apvir}
a_{\rm p,v} 
&=& a_{\rm p,ta} \sqrt{\frac{2\mu+2}{3}} \cos \frac{1}{3}
\left(\cos^{-1}\sqrt{\frac{27\mu^2}{(2\mu+2)^3}}+\pi\right)\,.\nonumber \\
\end{eqnarray}
The scale factor $a_{\rm v}$ of the universe when the scale factor of the 
perturbation {\em past its turnaround} becomes equal to $a_{\rm p,v}$, will be
given by the second branch of Eq.~(\ref{eqfora}), for $a_{\rm p} 
= a_{\rm p,v}$. Then, the  
validity range for the second branch of Eq.~(\ref{eqfora}) is 
$a_{\rm ta}<a<a_{\rm v}$.

For $a>a_{\rm v}$, we can no longer use the spherical evolution
solution to describe the physical picture of interest (virialization). In the
next section we will present a simple recipe we will use to follow
the late stages of evolution of the perturbation which satisfies the
desired boundary conditions ($a_{\rm p} = a_{\rm p,v}$ at $a_{\rm
  coll}$ and constant thereafter).

For perpetually expanding structures ($\kappa < \kappa_{\rm min,coll}$), 
Eq.~(\ref{spel}) gives 
\begin{equation}\label{sepvarint_nc}
\int_0^a \frac{\sqrt{y}dy}{\sqrt{\omega y^3+1}} = 
\int_0^{a_{\rm p}}\frac{\sqrt{x}dx}{\sqrt{\omega x^3 - \kappa x + 1}}
\,.
\end{equation}
In this case, the integral on the RHS can be rewritten as
\begin{equation}\label{defh1}
\int_0^{a_{\rm p}} \frac{\sqrt{x}dx}{\sqrt{\omega x^3 -
    \kappa x +1}}=
\frac{2}{3}\omega^{-1/2}\mathcal{H}_1(r, \varpi)
\end{equation}
where $\mathcal{H}_1$ is the {\em hyperbolic vacuum integral of the first
kind}, defined in appendix \ref{vac_ints}, and
\begin{equation}\label{rvarpi}
\begin{array}{lcccr}
r = a_{\rm p} / |a_{\rm p, R}| \,,
& &&&\varpi = (\omega |a_{\rm p, R}|^3)^{-1}\\
\end{array} \,\,, 
\end{equation}
where $a_{\rm p,R}$ is the only real (and always negative) root of
Eq.~(\ref{lf0}) when $\kappa < \kappa_{\rm min,coll}$, 
\begin{equation}
 a_{\rm p,R} = 
\frac{-\omega^{\frac{1}{3}}}
{\!\!\!\!
\sqrt[3]{\left(\frac{1}{2}-\sqrt{\frac{1}{4}-\frac{\kappa^3}{27\omega}}
\right)^2}\!\!\!+\!
{\sqrt[3]{\left(\frac{1}{2}+\sqrt{\frac{1}{4}-\frac{\kappa^3}{27\omega}
}\right)^2}\!-\!\frac{\kappa}{3\omega^{\frac{1}{3}}}
}}
\end{equation}
As in the case of collapsing perturbations,
$\kappa/\omega^{1/3}$ is a function of $\varpi$ alone, with
\begin{equation}\kappa/\omega^{1/3} =
  (1-\varpi)/\varpi^{2/3}\,.\end{equation} Then, $\varpi=1$ is a 
flat subuniverse (not perturbed with respect to the background), and 
perturbations with $\varpi>1$ are underdensities while $1/4<\varpi<1$
correspond to non-collapsing overdensities.

Then, Eq.~(\ref{sepvarint_nc}) becomes
\begin{equation}\label{eqfora_nc}
a = 
\omega^{-1/3}\left\{\sinh
\left[\mathcal{H}_1(r,\varpi)\right]\right\}^{2/3} 
\end{equation}
which is the spherical evolution solution for a non-collapsing
perturbation with
amplitude $\kappa$ (parametrized above by $\varpi$) 
in an $\Omega_\Lambda + \Omega _{\rm m} =1$,
$\Omega_\Lambda/\Omega_{\rm m}=\omega$ universe.
As for collapsing perturbations, $\mathcal{H}_1(r,\varpi)$ is the
development angle.

\subsubsection{{\boldmath $\ed_0(\delta)$} 
according to the spherical evolution model}

\indent

The linear theory result for a growing-mode perturbation in an
$\Omega_{\rm m}+\Omega_\Lambda=1$ cosmology is \cite{peeb}
\begin{equation}\label{genlin}
\ed = \ed_0\frac{D(a)}{D(a_0)}
\end{equation}  
where $D$, the linear growth factor, is given by $D(a)=A[(2\omega)^{1/3}a]$
with
\begin{equation}
A(x) = \frac{(x^3+2)^{1/2}}{x^{3/2}} \int_0^x 
\left(\frac{u}{u^3+2}\right)^{3/2}du \,.
\end{equation}
To find the relation between $\ed_0$ and $\kappa$, we expand the exact
($\delta = a^3/a_{\rm p}^3-1$) and linear relations for the overdensity to first
order in $a$ and demand that the coefficients be equal. Thus we get
\cite{ecf}
\begin{equation}\label{kofdnot}
\kappa = \frac{(2\omega)^{1/3}}{3A\left[(2\omega)^{1/3}a_0\right]}\ed_0
= \frac{(2\omega)^{1/3}}{3A\left[(2\omega)^{1/3}\right]}\ed_0\,,
\end{equation}
since $a_0=1$ .
This is the value of the constant $\kappa$ for a perturbation which at a
cosmic epoch $a_0=1$ has a linearly extrapolated overdensity $\ed_0$.
Note that since the linear theory result is the same for both underdensities
and overdensities, Eq.~(\ref{kofdnot}) holds for both cases. For underdensities,
both $\kappa$ and $\ed$ will be negative, while for overdensities both will be
positive. 

At any given epoch $a$, there is a unique perturbation (parametrized
by $\kappa$ or, equivalently, by $\mu$ or $\varpi$) which will have
achieved a true density contrast $\delta$ at that time. Therefore, to
calculate the desired conversion relation $\ed_0(a, \delta)$ we first
calculate $\kappa$ (or, equivalently, $\mu$ or $\varpi$) from the
given $a$ and $\delta$ and the appropriate solution of the evolution
equation (\ref{eqfora} or \ref{eqfora_nc}). Then, we use Eq.~(\ref{kofdnot}) to evaluate $\ed_0$. 

\begin{table}
\caption{Applicability limits for different branches of the spherical
  evolution solution, where:  
$\delta_{\rm Ed}(a)$ is the density contrast of the
Eddington overdensity;  $\delta_{\rm ta}(a)$ is the density contrast
of an overdensity turning around at $a$; $\delta_{\rm v}(a)$ is the
density contrast of overdensity reaching its virial size at $a$; and 
 $\delta_{\rm c}(a)$ is the density contrast of virialized overdensity
formally collapsing to a point at $a$.\label{table_limits}}
\begin{tabular}{lll}
\hline
\hline
{\bf Limit} & {\bf Expression} 
&{\bf Auxiliary Relations} \\
\hline
$\delta_{\rm Ed}(a)$
 & $\delta_{\rm Ed}(a) = 2\omega \left(\frac{a}{r_{\rm
    Ed}(a)}\right)^3 -1$
&$\sinh ^{-1} \sqrt{\omega a^3} - \mathcal{V}_1(r_{\rm Ed},2) = 0$\\
\hline
$\delta_{\rm ta}(a)$ &
$\delta _{\rm ta}(a) = \omega a^3 \mu_{\rm ta}(a)-1$ &
$\sinh^{-1}\sqrt{\omega a^3} - \mathcal{V}_{1}(1,\mu_{\rm ta}) =0$ \\
\hline
$\delta_{\rm v}(a)$&
$\delta_{\rm v}(a) = \left(\frac{a}{a_{\rm p,v}
\left[\mu_{\rm v}(a)\right]}\right)^3 - 1$ &
$\sinh ^{-1} \sqrt{\omega a^3} - 2\mathcal{V}_1(1,\mu_{\rm v})
+ \mathcal{V}_1\left[
r_{\rm v}(\mu_{\rm v}), \mu_{\rm v}\right] =0$\\
&& $r_{\rm v}(\mu) = a_{\rm p,v}(\mu)/a_{\rm p,ta}(\mu)$ \\
\hline 
$\delta_{\rm c}(a)$&
$\delta_{\rm c}(a) = \left[ \frac{a}{a_{\rm p,v}\left[\mu_{\rm c}(a)\right]}
\right]^3-1$ &
$\sinh^{-1}\sqrt{\omega a^3}-2\mathcal{V}_{1}(1,\mu_{\rm c}) =0$ \\ 
\hline
\hline
\end{tabular}
\end{table}
To determine which is the appropriate solution of the evolution
equation we need to use for each $\delta$, we calculate the limits of
applicability of each equation in terms of $\delta$. 
\begin{itemize}
\item Equation (\ref{eqfora_nc}) is applicable for all forever expanding
  perturbations. For any given epoch $a$, the maximum density contrast
  of such perturbations is achieved by the Eddington perturbation and
  is equal to $\delta_{\rm Ed}(a)$, given in line 1 of Table
  \ref{table_limits}. 
Then, the applicability domain of Eq.~(\ref{eqfora_nc}) is $-1 <
\delta \le \delta_{\rm Ed}(a)$, and the conversion relation in this
case takes the form shown in the 1st line of Table \ref{mytable1}. 

\item The first branch of Eq.~(\ref{eqfora}) is applicable for
  eventually collapsing perturbations which, however, have not reached
  their turnaround radius yet. The maximum $\delta$ of all such
  perturbations at a given $a$ is achieved by the perturbation which
  is turning around at $a$, and is equal to $\delta_{\rm ta}(a)$,
  given in line 2 of Table \ref{table_limits}. 
The applicability domain of the first branch of Eq.~(\ref{eqfora})
is then $\delta_{\rm Ed}(a) < \delta \le \delta_{\rm ta}(a)$ and the
conversion relation in this case is shown in the 2nd line of table
\ref{mytable1}. 

\item The second branch of Eq.~(\ref{eqfora}) is applicable for
  eventually collapsing perturbations which are past their turnaround
  but which have not yet reached their virial radius. 
  The maximum $\delta$ of such
  perturbations at a given $a$ is achieved by the perturbation which
  is reaching its virial size at $a$, and is equal to $\delta_{\rm
  v}(a)$, given in line 3 of Table \ref{table_limits}.
The applicability domain of the second branch of Eq.~(\ref{eqfora})
is then $\delta_{\rm ta}(a) < \delta \le \delta_{\rm v}(a)$ and the
conversion relation in this case is shown in the 3rd line of table
\ref{mytable1}. 

\item Perturbations which have reached their virial size but have not
  yet reached their designated collapse time, $a_{\rm coll}(\mu)$,
  need to be treated separately, since the spherical collapse model
  fails (does not agree with the physical picture we would like to
  describe, although it is still formally applicable) for radii
  smaller than the virial radius. Since a realistic   
  description of the microphysical dissipation processes which lead to
  virialization is far beyond the scope of this analytical
  calculation, we will adopt a prescription which is driven by
  mathematical simplicity. We will assume that for $\delta_{\rm v}(a)
  < \delta \le \delta_{\rm c}(a)$ the conversion relation
  $\ed_0(a,\delta)$ has the simplest polynomial form which satisfies
  the following physically motivated boundary conditions: 
\begin{itemize}
\item The extrapolated overdensity is continuous and smooth at
  $\delta_{\rm v}$, so $\ed_0(a,\delta_{\rm v}) = \ed_{\rm 0,v}$
and $\left. \partial \ed_0 / \partial \delta \right|_{\delta_{\rm
    v}} = \ed'_{\rm 0,v}$ as given by the appropriate relations of the
  previous branch (3rd line of tables \ref{mytable1} and
  \ref{mytable2} correspondingly).
\item After time $a_{\rm coll}(\mu)$ the radius of the perturbation
  remains constant and equal to the virial radius so changes in the
  (true) overdensity are only due to the increase of the scale factor
  of the background universe. This then implies that
$\ed_0(a,\delta_{\rm c}) = 
\ed_{\rm 0,c}$ given by Eq.~(\ref{edc_lambda}) and 
$\left. \partial \ed_0 / \partial \delta \right|_{\delta_{\rm
    c}} = (\partial \ed_0 / \partial \mu|_{\mu_{\rm c}})
(\partial \mu / \partial \delta|
_{\delta_{\rm c}})
= \ed'_{\rm 0,c}$ given in Table \ref{mytable1} line 4.
\end{itemize}
The
conversion relation in this case is shown in Table \ref{mytable1} line
4,while its applicability domain is $\delta_{\rm v}(a) < \delta \le 
\delta_{\rm c}(a)$, with $\delta_{\rm c}(a)$ given in Table
\ref{table_limits} line 4.
Past their
collapse time, perturbations are treated as virialized objects
without substructure and are not relevant as ``local environment'' of
other objects for the purposes of our double distribution calculation,
hence it is not necessary to have a conversion relation of $\delta >
\delta_{\rm c}(a)$.
The calculation would be simplified (this last branch would be
unnecessary) if we chose to
regard perturbations as virialized objects after the moment they
reached their virial size after turnaround, at time $a_v$. However,
we will retain the usual assumption that objects virialize at time
$a_{\rm coll}$ for consistency with existing Press-Schechter calculations.
\end{itemize}

\begin{table*}
\caption{\label{mytable1} 
Different branches of conversion relation $\ed_0(a,\delta)$
  for an $\Omega_{\rm m}+\Omega_{\rm \Lambda}=1$ universe, where:
branch I is $-1 < \delta \le \delta_{\rm Ed}$; branch II is 
$\delta_{\rm Ed} < \delta \le \delta_{\rm ta}$; branch III is 
$\delta_{\rm ta} < \delta \le \delta_{\rm v}$; and branch IV is 
$\delta_{\rm v} < \delta \le \delta_{\rm c}$.}
\begin{tabular}{lll}
\hline
\hline
{\bf Branch}  
&{\boldmath $\ed_0(\delta,a) = $} &{\bf Auxiliary relations} \\ 
\hline 
I &
$
\frac{3A\left[(2\omega)^{1/3}\right]}{2^{1/3}}
\frac{1-\varpi(a,\delta)}{\left[\varpi(a,\delta)\right]^{2/3}}$
 & $\sinh^{-1} \sqrt{\omega a^3} - \mathcal{H}_1\left[
a\left(\frac{\varpi \omega}{1+\delta}\right)^{1/3}, \varpi 
\right]=0$\\
\hline 
II &
$
\frac{3A\left[(2\omega)^{1/3}\right]}{2^{1/3}}
\frac{1+\mu(a,\delta)}{\left[\mu(a,\delta)\right]^{2/3}}$
& $\sinh^{-1}\sqrt{\omega a^3} -
\mathcal{V}_1\left[a\left(\frac{\mu
    \omega}{1+\delta}\right)^{1/3},\mu\right] =0$\\
\hline
III &
$
\frac{3A\left[(2\omega)^{1/3}\right]}{2^{1/3}}
\frac{1+\mu(a,\delta)}{\left[\mu(a,\delta)\right]^{2/3}}$
&  $  \begin{array}{l}
\sinh^{-1} \sqrt{\omega a^3} -2\mathcal{V}_1(1,\mu)\\
\,\,\,\,\, +\mathcal{V}_1\left[a\left(\frac{\mu
    \omega}{1+\delta}\right)^{1/3},\mu\right]=0\end{array}$\\
\hline
IV &
$\begin{array}{l}
\ed_{\rm 0,v} + \ed_{\rm 0,v}'(\delta - \delta_{\rm v}) \\ 
+\frac{3(\ed_{\rm0,c}-\ed_{\rm 0,v})-(\delta_{\rm c} - \delta_{\rm v})
(2\ed_{\rm 0,v}'+\ed_{0,c}')}
{\left(\delta_{\rm c} - \delta_{\rm
    v}\right)^2} \\
\,\,\,\,\,\,\,\, \times (\delta -\delta_{\rm v})^2 \\
+ \frac{(\ed_{\rm 0,c}'+\ed_{\rm 0,v}')(\delta_{\rm c}- \delta_{\rm
    v})
-2(\ed_{\rm 0,c}-\ed_{\rm 0,v})}{(\delta_{\rm c} -\delta_{\rm v})^3}\\
\,\,\,\,\,\,\,\, \times (\delta -\delta_{\rm v})^3\\
\end{array}$
& \begin{tabular}{l}
$\ed_{\rm 0,v} = \ed_0(a,\delta_{\rm v})$ (this Table line 3)\\
$\ed'_{\rm 0,v} = \left. \frac{\partial \ed_0}{\partial \delta}
  \right|_{\delta_{\rm v}} $ (Table \ref{mytable2} line 3)\\
$ \ed_{\rm 0,c} = \ed_0(a,\delta_{\rm c})$
(Eq.~\ref{edc_lambda}) \\
$\begin{array}{lll}\ed'_{\rm 0,c} &=& \left.\frac{\partial \ed_0 }{\partial
  \delta}\right|_{\delta_{\rm c}}  \\
&=& - \frac{A\left[(2\omega)^{1/3}\right]
\left(
6\omega^{2/3}a_{\rm p,v}^2\mu_{\rm c}^{2/3} -1 -\mu_{\rm c}\right)}{
\left(2\mu_{\rm c}^{2}\right)^{1/3}(\delta_{\rm c}+1)} \end{array}$ \\
$\mu_{\rm c}(a)$ in Table \ref{table_limits} line 4
\end{tabular}
\\
\hline
\hline 
\end{tabular}
\end{table*}

\subsubsection{Critical extrapolated overdensity for collapse, {\boldmath
    $\ed_{\rm 0,c}(a)$}}

\indent

We need to find the critical $\ed_{\rm 0,c}(a)$ for collapse if the field is linearly
extrapolated to the present epoch, i.e. the value the linearly extrapolated to
the present overdensity must have, for a structure to have collapsed at
universe scale factor $a$. This, from Eq.~(\ref{kofdnot}), 
will be 
\begin{equation}\label{edc_lambda}
\ed_{\rm 0,c}(a) = \frac{3
  A\left[(2\omega)^{1/3}\right]}{(2)^{1/3}}\frac{1+\mu_{\rm c}(a)}
{\left[\mu_{\rm c}(a)\right]^{2/3}}
\end{equation}
where again $\mu_{\rm c}(a)$ is given by Table \ref{table_limits} line 4.

The dependence of $\ed_{\rm 0,c}$ on $a$ can also be expressed in terms
of the linear growth factor, $D(a)$, as was the case for the
$\Omega_{\rm m}=1$ universe. The conversion relation between
$\delta(a)$ and $\ed_a$ (the linear-theory result for the density
contrast at time $a$) is independent of $a$. In other words, as long
as $\delta$ and $\ed$ both refer to the same time, knowledge of the one
uniquely defines the other, independently of the actual time at which
they are both evaluated. As $\delta(a) \rightarrow \infty$,
$\ed_a\rightarrow \ed_{\rm c}$, the linear-theory density contrast at
the time of collapse (given by Eq.~\ref{edc_lambda} for $a=1$). 
Therefore $\ed_{\rm c}$ is the same for perturbations of all
curvatures, and, using Eq.~(\ref{genlin}), we can write 
\begin{equation}
\ed_{\rm 0,c} (a_{\rm coll })
= \ed_{\rm c} \frac{D(a_0)}{D(a_{\rm coll})}\,.
\end{equation}

\subsubsection{{\boldmath$\partial \ed_0 / \partial \delta |_a$}}
 
\indent

In addition to the relation between $\delta$ and $\ed_0$, we will also
need the 
derivative $\partial \ed_0 / \partial \delta |_a$ in order to convert
between 
true and extrapolated overdensity differentials in
Eq.~(\ref{convtotrue}). The calculation is similar as in the
case of an $\Omega_{\rm m}=1$ universe, and the results are summarized
in table \ref{mytable2}.

\begin{table*}
\caption{\label{mytable2} 
Different branches of derivative $\partial \ed_0/\partial \delta |_a$
  for an $\Omega_{\rm m}+\Omega_{\rm \Lambda}=1$ universe.
The roman numerals correspond to different branches as in Table
\ref{mytable1}}
\begin{tabular}{llll}
\hline
\hline
& {\bf Aux. Function}  &
{\boldmath $\left.\frac{\partial \ed_0}{\partial
  \delta}\right|_a\!\!\!\!(a,\delta)= $} 
&{\bf Additional Functions}\\ 
\hline 
I & $
\begin{array}{ll} \Phi_1
= &\sinh^{-1}\sqrt{\omega a^3} \\
& - \mathcal{H}_1(r,\varpi)\end{array}$ 
&$-\frac{\partial \Phi_1}{\partial \delta}\left(
\frac{\partial \Phi_1}{\partial \ed_0}\right)^{-1}$ &
$\frac{\partial \Phi_1}{\partial \delta} = 
\frac{a^{3/2}}{2(1+\delta)^{3/2}
\sqrt{\frac{a^3}{(1+\delta)}-\frac{a(\varpi-1)}{\varpi^{2/3}
(1+\delta)^{1/3}}
+ \frac{1}{\omega}}}$\\
&&& $\frac{\partial \Phi_1}{\partial \ed_0}= - \frac{(2\varpi^2)^{1/3} }
{3A[(2\omega)^{1/3}]}
\mathcal{H}_2 \left[
a\left(\frac{\varpi \omega}{1+\delta}\right)^{1/3}\!\!\!
,\varpi\right]$
\\
&&& $\sinh^{-1} \sqrt{\omega a^3} \!\! - \!\! \mathcal{H}_1\left[
a\left(\frac{\varpi \omega}{1+\delta}\right)^{1/3}\!\!\!\!\!, \varpi 
\right]=0$\\
\hline 
II  &
$\begin{array}{ll} 
\Phi_2
= &\sinh^{-1}\sqrt{\omega a^3} \\ &-\mathcal{V}_1(r,\mu)
\end{array}$&
$-\frac{\partial \Phi_2}{\partial \delta}\left(
\frac{\partial \Phi_2}{\partial \ed_0}\right)^{-1}$&
$\frac{\partial \Phi_2}{\partial \delta} = 
\frac{a^{3/2}}{2(1+\delta)^{3/2}
\sqrt{\frac{a^3}{(1+\delta)}-\frac{a(\mu+1)}{\mu^{2/3}
(1+\delta)^{1/3}}
+ \frac{1}{\omega}}}$
\\
&&&$\frac{\partial \Phi_2}{\partial \ed_0}=-\frac{(2\mu^2)^{1/3} }
{3A[(2\omega)^{1/3}]}
\mathcal{V}_2 \left[
a\left(\frac{\mu \omega}{1+\delta}\right)^{1/3},\mu\right]$\\
&&& $\sinh^{-1} \sqrt{\omega a^3} \!-\!\! \mathcal{V}_1\left[
a\left(\frac{\mu \omega}{1+\delta}\right)^{1/3}\!\!\!\!, \mu 
\right]=0$\\
\hline
III &
$\begin{array}{ll}
\Phi_3 =& \sinh^{-1}\sqrt{\omega a^3} \\ 
& -2\mathcal{V}_1(1,\mu) \\ &+ \mathcal{V}(r,\mu)\end{array}$
&
$-\frac{\partial \Phi_3}{\partial \delta}\left(
\frac{\partial \Phi_3}{\partial \ed_0}\right)^{-1}$&
$\frac{\partial \Phi_3}{\partial \delta} = 
- \frac{[a/(1+\delta)]^{3/2}}{2
\sqrt{\frac{a^3}{(1+\delta)}-\frac{a(\mu+1)}{\mu^{2/3}
(1+\delta)^{1/3}}
+ \frac{1}{\omega}}}$
\\
&&&$\begin{array}{ll} 
\frac{\partial \Phi_3}{\partial \ed_0}=&\frac{(2\mu^2)^{1/3} }
{3A[(2\omega)^{1/3}]}\left\{\mathcal{V}_2 \left[
a\left(\frac{\mu \omega}{1+\delta}\right)^{1/3}\!\!\!
\!\!\!,\mu\right]
\right. \\
& \left. \,\,\,- \frac{6\mu}{\mu-2}\frac{d\mathcal{V}_1(1,\mu)}{d\mu}
\right\}\end{array}$\\
&&&  $  \begin{array}{l} 
\sinh^{-1} \sqrt{\omega a^3} -2\mathcal{V}_1(1,\mu)\\
\,\,\,\,\,\,\,\, +\mathcal{V}_1\left[a\left(\frac{\mu
    \omega}{1+\delta}\right)^{1/3},\mu\right]=0\end{array}$\\
\\
\hline
IV & ----- & 
\multicolumn{2}{l}{
\begin{tabular}{l}
$\begin{array}{l}
\ed_{\rm 0,v}' 
+2\frac{3(\ed_{\rm0,c}-\ed_{\rm 0,v})-(\delta_{\rm c} - \delta_{\rm v})
(2\ed_{\rm 0,v}'+\ed_{0,c}')}{\left(\delta_{\rm c} - \delta_{\rm
    v}\right)^2}(\delta -\delta_{\rm v}) \\
+ 3\frac{(\ed_{\rm 0,c}'+\ed_{\rm 0,v}')(\delta_{\rm c}- \delta_{\rm
    v})
-2(\ed_{\rm 0,c}-\ed_{\rm 0,v})}{(\delta_{\rm c} -\delta_{\rm v})^3}
(\delta -\delta_{\rm v})^2\\
\end{array}$ \\  
 with $\ed_{\rm0,c}$, $\ed_{\rm 0,v}$ 
$\ed_{\rm 0,c}'$, $\ed_{\rm 0,v}'$,\\
as in Table \ref{mytable1} line 4\end{tabular}} 
\\
\hline
\hline 
\end{tabular}
\end{table*}

In table \ref{mytable2}, $\mathcal{H}_2(r, \varpi)$ 
is the {\em hyperbolic  vacuum
integral of the second kind}, and 
$\mathcal{V}_2(r, \mu)$ is the {\em incomplete vacuum
integral of the second kind}, defined in appendix \ref{vac_ints}.

\subsection{An Approximate Conversion Relation}\label{aprel}

\indent

An excellent approximation 
to these conversion relations can be derived from
the expression
\begin{equation}\label{magic}
\ed_a \approx \ed_{\rm c}\left[1-(1+\delta_a)^{-1/\ed_{\rm c}}\right]\,.
\end{equation}
Similar approximations were suggested by \cite{b94} and \cite{s98}.
Equation (\ref{magic}) relates the linear overdensity at a time $a$ to
the true overdensity at the same time, and its accuracy is {\em  better than
$2\%$  throughout its domain} for both $\Omega_{\rm m}=1$ and
$\Omega_{\rm m}+\Omega_\Lambda=1$ cosmologies. Its functional form is
much simpler and more intuitive than the more accurate fit of
\cite{mw96}. 
The cosmological model enters only through $\ed_{\rm c}$.
For the Einstein-deSitter universe,
$\ed_{\rm c}$ is given by Eq.~(\ref{matteredc}) for $a_{\rm coll}=1$, while
for the $\Omega_{\rm m}+\Omega_\Lambda=1$ universe it is given by 
Eq.~(\ref{edc_lambda}) for $a=1$. Note that $\ed_{\rm c}$ is related to the
quantity $\ed_{\rm 0,c}(a)$ (which appears explicitly in our double
distribution expression) through 
\begin{equation}
\ed_{\rm 0,c}(a) = \ed_{\rm c} \frac{D(a_0)}{D(a)}
\end{equation}
where $D(a)$ is the linear growth factor in the relevant cosmology.

The limits of Eq.~(\ref{magic}) are the same as the ones required
for the exact conversion relation. When
$|\delta_a| \ll 1$, $\ed_a \approx \delta_a$. In addition, 
$\ed_a \rightarrow -\infty$ as  $\delta_a \rightarrow -1$, and 
$\delta_a \rightarrow \infty$ as $\ed_a \rightarrow \ed_{\rm c}$.

Using Eq.~(\ref{magic}), 
\begin{equation}\label{apconv}
\ed_{\ell} \approx \frac{D(a_0)}{D(a)}\ed_{\rm
  c}\left[1-(1+\delta_{\ell})^{-1/\ed_{\rm c}}\right]\,,
\end{equation}
where $a$ is the time at which 
we want to evaluate the double distribution.
Note that close to virialization, Eq.~(\ref{apconv}) loses its
applicability (as does the spherical collapse model), and has to be 
replaced by a recipe which does not diverge in $\delta_{\ell}$. We 
have presented
such recipes in \S \ref{ap1} and \ref{ap2}, however the exact
functional form of the conversion relation in this regime cannot
affect any of the physically interesting results as the amplitude of
the double distribution decreases rapidly enough with $\delta_\ell$
that the contribution of the high-delta tail 
to the integrated mass function is negligible. We have
verified this fact by comparing the integral of our double distribution
over $\delta_\ell$ with the Press-Schechter mass function. When we extended
the integration up to $\delta_{\rm 0,c}$, the results agreed to the accuracy of
the numerical integration. When we extended our integration only up to
$\delta_{\rm 0,v}$ (just below the application of our virialization
recipe), the error relative to the Press-Schechter mass function was
less than $0.02\%$.

\subsection{Clustering Scale Lengths
and Correction for Central Object Contamination}

\indent

The definition of $\beta$ and $\ed_{\ell}$ described above was sufficient
for us to derive the double distribution from random walk
theory. However, from a physical point of view, the presence of a
collapsed structure at the center of the ``environment sphere''
contaminates the evaluation of the average ``environmental''
overdensity. If we want the double distribution to describe the
properties of matter {\em surrounding} collapsed objects, we need to
correct for the presence of the objects themselves. 

We will employ a simple, ``top-hat'' physical picture to calculate an
appropriate correction (see also \cite{s98}). Note however that our correction is 
approximate, since the filter we used to smooth the
overdensity field was $k-$sharp rather than top-hat in space. 

Let $\delta_{\rm c}$ be the (true) overdensity of a collapsed object of mass
m and radius $R_{\rm v}$, and $\delta_{\ell}$ be the overdensity of the
``environment sphere'' of radius $R_{\rm e}$. The ``environment sphere''
encompasses a mass $\beta m$, including the central collapsed
object. We want to find the average overdensity $\delta_{\rm ext}$ of
that part of the ``environment sphere'' which is {\em external} to the central object. For the
collapsed object we can write
\begin{equation}\label{cor1}
m = \frac{4}{3} \pi R_{\rm v}^3(1+\delta_{\rm c})\rho_{\rm m}\,,
\end{equation}
where $\rho_{\rm m}$ is the mean matter density of the universe at the
epoch of interest. For the environment sphere, including the central
object, we can write
\begin{equation}\label{cor2}
\beta m = \frac{4}{3} \pi R_{\rm e}^3(1+\delta_{\ell})\rho_{\rm m}\,.
\end{equation}
From Eqs. (\ref{cor1}) and (\ref{cor2}) we get $R_{\rm v}^3 = R_{\rm
  e}^3
(1+\delta_{\ell})/\beta(1+\delta_{\rm c})$. 
It thus 
follows that the length scale $R_{\rm e}$ associated
with the clustering parameter $\beta$ is
\begin{eqnarray}
\label{eq:size}
R_{\rm e} &=& \left( \frac{\beta(1+\delta_{\rm
 c})}{1+\delta_{\ell}} \right)^{1/3}  R_{\rm v} \nonumber \\
 &=& \left( \frac{3 \beta m}{4 \pi (1+\delta_{\ell}) \rho_{\rm m}} \right)^{1/3} \, .
\end{eqnarray}
We see that for a fixed clustering parameter $\beta$,
the length scale associated with an object of mass $m$
is mass-dependent, scaling linearly with the virial
radius but larger by a factor
$\left[\beta(1+\delta_{\rm c})/(1+\delta_{\ell})\right]^{1/3} > 1$.
Thus, we can roughly think of the clustering scale parameter
as a measure of how many virial radii
we include as the local environment around each
structure \footnote{Note however that for fixed $\beta$, the number of
virial radii included in the environment depends on $\delta_{\ell}$
and is larger for underdense environments.}.

Having identified the environmental length scale,
we can now isolate the environmental overdensity
from that of the collapsed object.
The volume of the environment sphere external to the
central object contains a mass
\begin{equation}\label{cor3}
(\beta-1) m = \frac{4}{3} \pi (R_{\rm e}^3-R_{\rm v}^3)
(1+\delta_{\rm ext})\rho_{\rm m}\,.
\end{equation}
Using Eq.~(\ref{eq:size}) to eliminate $R_{\rm v}$, and dividing by
Eq.~(\ref{cor2}) we obtain 
\begin{equation}
\delta_{\rm ext} = \frac{(\beta-1)(1+\delta_{\ell})(1+\delta_{\rm c})}
{\beta(1+\delta_{\rm c})-(1+\delta_{\ell})}-1\,,
\end{equation}
which is the contamination-corrected overdensity for an environment
sphere with uncorrected overdensity 
$\delta_{\ell}$. Then, the contamination-corrected double
distribution will be given by 
\begin{equation}
\frac{dn}{dmd\delta_{\rm ext}}(\delta_{\rm ext},m,a)=
\frac{dn}{dmd\delta_{\ell}} \left[\delta_{\ell}(\delta_{\rm
    ext},a),m,a\right]\frac{d
  \delta_{\ell}}{d\delta_{\rm ext}} \,, 
\end{equation}
where 
\begin{equation}
\delta_{\ell}(\delta_{\rm ext}) = \frac{\beta(1+\delta_{\rm
    ext})(1+\delta_{\rm c})} {(\beta-1)(1+\delta_{\rm c})+(1+\delta_{\rm ext})}-1
\end{equation}
and 
\begin{equation}
\frac{d\delta_{\ell}}{d\delta_{\rm ext}} = 
\frac{\beta (\beta -1)(1+\delta_{\rm c})^2}
{\left[(\beta-1)(1+\delta_{\rm c})+(1+\delta_{\rm ext})\right]^2}\,.
\end{equation}

\section{Derivative Quantities of the Double Distribution}\label{derivquants}

\indent

We now have enough tools to calculate derivative quantities of
interest. The number density of collapsed objects of mass greater than
some minimum $m_{\rm min}$ \footnote{The introduction of a finite
  minimum mass $m_{\rm min}$ is necessary for both physical and
  technical reasons. Physically, the mass of collapsed objects is
  strictly forced to have a lower bound, not only due to the finite
  mass of the dark matter particle, but also due to the existence of a
dark matter Jeans mass, however small this may be. In addition, the
dark matter perturbation transfer function imposes cutoffs at scales of order
of an earth mass \cite{hss01,ghs04,lz05,dms05}.
Practically, 
the Press-Schechter $dn/dm$ diverges as $m \rightarrow 0$ and setting
a minimum mass is required to extract interesting information. For the
purposes of this thesis, the selected mass cutoff will generally be such that
the population of interest will include super-galactic scales only.}
embedded 
in a medium of local overdensity between $\delta_{\rm ext}$ and 
$\delta_{\rm ext}+d \delta_{\rm ext}$ is 
\begin{equation}
\frac{dn}{d\delta_{\rm ext}}(>m_{\rm min})d\delta_{\rm ext} = 
d\delta_{\rm ext} \frac{\partial\ed_{\ell}}{\partial\delta_{\ell}}
\frac{d\delta_{\ell}}{d\delta_{\rm ext}}
\int_{m=m_{\rm min}}^{\infty} 
\!\!\!\frac{dn}{dmd\ed_{\ell}} dm\,,
\end{equation}
while the density of matter in collapsed objects of mass $ > m_{\rm min}$
embedded in a medium of local overdensity between $\delta_{\rm ext}$ 
and $\delta_{\rm ext}
+d\delta_{\rm ext}$ is
\begin{equation}\label{matdelta}
\frac{d\rho}{d\delta_{\rm ext}}(>m_{\rm min})d\delta_{\rm ext} = 
d\delta_{\rm ext} \frac{\partial\ed_{\ell}}{\partial\delta_{\ell}}
\frac{d\delta_{\ell}}{d\delta_{\rm ext}}
\int_{m=m_{\rm min}}^{\infty} \!\!\!m 
\frac{dn}{dmd\ed_{\ell}} dm\,.
\end{equation}

Of all the matter in the universe which belongs to collapsed objects
of mass $>m_{\rm min}$, the fraction by mass which lives in underdense
neighborhoods is
\begin{equation}
f_{\rm \rho,un} = \frac{
\int_{\delta_{\rm ext}=-1}^{0} \frac{d\rho}{d\delta_{\rm
    ext}}(>m_{\rm min})d\delta_{\rm ext}
}
{\int_{\delta_{\rm ext}=-1}^{\delta_{\rm c}} 
\frac{d\rho}{d\delta_{\rm ext}}
(>m_{\rm min})d\delta_{\rm ext}}\,.
\end{equation}
Then, the mass fraction of the matter defined above which 
lives in overdensities will be $f_{\rm \rho, ov} = 1 - f_{\rm \rho, un}$.

Similarly, of all the objects with mass $m>m_{\rm min}$, a fraction
by number which lives inside underdensities is
\begin{equation}
f_{\rm n,un} = \frac{
\int_{\delta_{\rm ext}=-1}^{0} \frac{dn}{d\delta_{\rm ext}}(>m_{\rm
  min})d\delta_{\rm ext}
}
{\int_{\delta_{\rm ext}=-1}^{\delta_{\rm c}} \frac{dn}{d\delta_{\rm
      ext}}(>m_{\rm min})d\delta_{\rm ext}}\,.
\end{equation}
The complementary number fraction of such structures living inside
overdensities will be $f_{\rm n, ov} = 1 - f_{\rm n,un}$.

The number-density--weighted mean $\delta_{\rm ext}$ for structures of mass
$>m_{\rm min}$ is 
\begin{equation}\label{parnmean}
\langle\delta\rangle_{\rm n} = 
\frac{\int_{\delta_{\rm ext}=-1}^{\delta_{\rm c}}\delta_{\rm ext}
  \frac{dn}{d\delta_{\rm ext}}(>m_{\rm
    min})d\delta_{\rm ext}}
{\int_{\delta_{\rm ext}=-1}^{\delta_{\rm c}}\frac{dn}{d\delta_{\rm ext}}(>m_{\rm
    min})d\delta_{\rm ext}}
\end{equation}
with a variance
\begin{equation}\label{parnvar}
\sigma^2_{\rm \delta,n} =  
\frac{\int_{\delta_{\rm ext}=-1}^{\delta_{\rm c}}(\delta_{\rm ext}
  -\langle\delta\rangle_{\rm n})^2 \frac{dn}{d\delta_{\rm ext}}(>m_{\rm
    min})d\delta_{\rm ext}}
{\int_{\delta_{\rm ext}=-1}^{\delta_{\rm c}}\frac{dn}{d\delta_{\rm ext}}(>m_{\rm
    min})d\delta_{\rm ext}}\,.
\end{equation}

Similarly, the matter-density--weighted mean $\delta$ for structures of mass
$>m_{\rm min}$ is 
\begin{equation}
\langle\delta\rangle_{\rm \rho}\label{parrhomean} = 
\frac{\int_{\delta_{\rm ext}=-1}^{\delta_{\rm c}}\delta_{\rm ext}
  \frac{d\rho}{d\delta_{\rm ext}}(>m_{\rm
    min})d\delta_{\rm ext}}
{\int_{\delta_{\rm ext}=-1}^{\delta_{\rm c}}\frac{d\rho}{d\delta_{\rm ext}}(>m_{\rm
    min})d\delta_{\rm ext}}
\end{equation}
with a variance
\begin{equation}\label{parrhovar}
\sigma^2_{\rm \delta,\rho}  =
\frac{\int_{\delta_{\rm ext}=-1}^{\delta_{\rm c}}\left(\delta_{\rm ext} -
  \langle\delta\rangle_{\rho}\right)^2 
\frac{d\rho}{d\delta_{\rm ext}}(>m_{\rm
    min})d\delta_{\rm ext}}
{\int_{\delta_{\rm ext}=-1}^{\delta_{\rm c}}\frac{d\rho}{d\delta_{\rm ext}}(>m_{\rm
    min})d\delta_{\rm ext}}\,.
\end{equation}

\section{Results}\label{ddres} 

\indent

In this section we present plots of the double distribution itself
as well as of its various physically interesting derivative
quantities. We compare results derived for a concordance, $\Omega_{\rm
m} +\Omega_\Lambda=1$ universe with WMAP parameters ($\sigma_8=0.84$, 
$h=0.71$, $\Omega_{\rm m}=0.27$, $\Omega_{\rm b} = 0.04$,
\cite{Sper03}), and for an Einstein-deSitter ($\Omega_{\rm m}=1$)
universe  with $h=0.71$, $\Omega_{\rm b} = 0.04$, but $\sigma_8=0.45$.
The different power-spectrum normalization in the Einstein-deSitter
case was selected so that the Press-Schechter mass function in this
case coincides with that of the concordance universe on a mass scale 
of $5.5\times 10^{14} {\rm M_\odot}$, which is between the values of
$m_8$ (mass included in a sphere of comoving radius $8h^{-1}$ Mpc) for
the two cosmologies ($m_8=2\times10^{14} {\,\rm M_\odot}$ 
for the concordance universe while $m_8=8\times10^{14} {\,\rm
  M_\odot}$ for the Einstein-de Sitter universe). 
This value of $\sigma_8$ is also consistent with
the fits of \cite{ecf} given the WMAP result for the concordance
universe. 
Finally, we use fitting formulae of \cite{bard86} for the adiabatic cold dark
matter transfer function to calculate the density field variance $S(m)$.
In this section, $\delta$ always refers to $\delta_{\rm ext}$, the
true overdensity of that part of the ``environment sphere'' which is
external to the central object. 

\begin{figure*}
\resizebox{2.9in}{!}{
\includegraphics{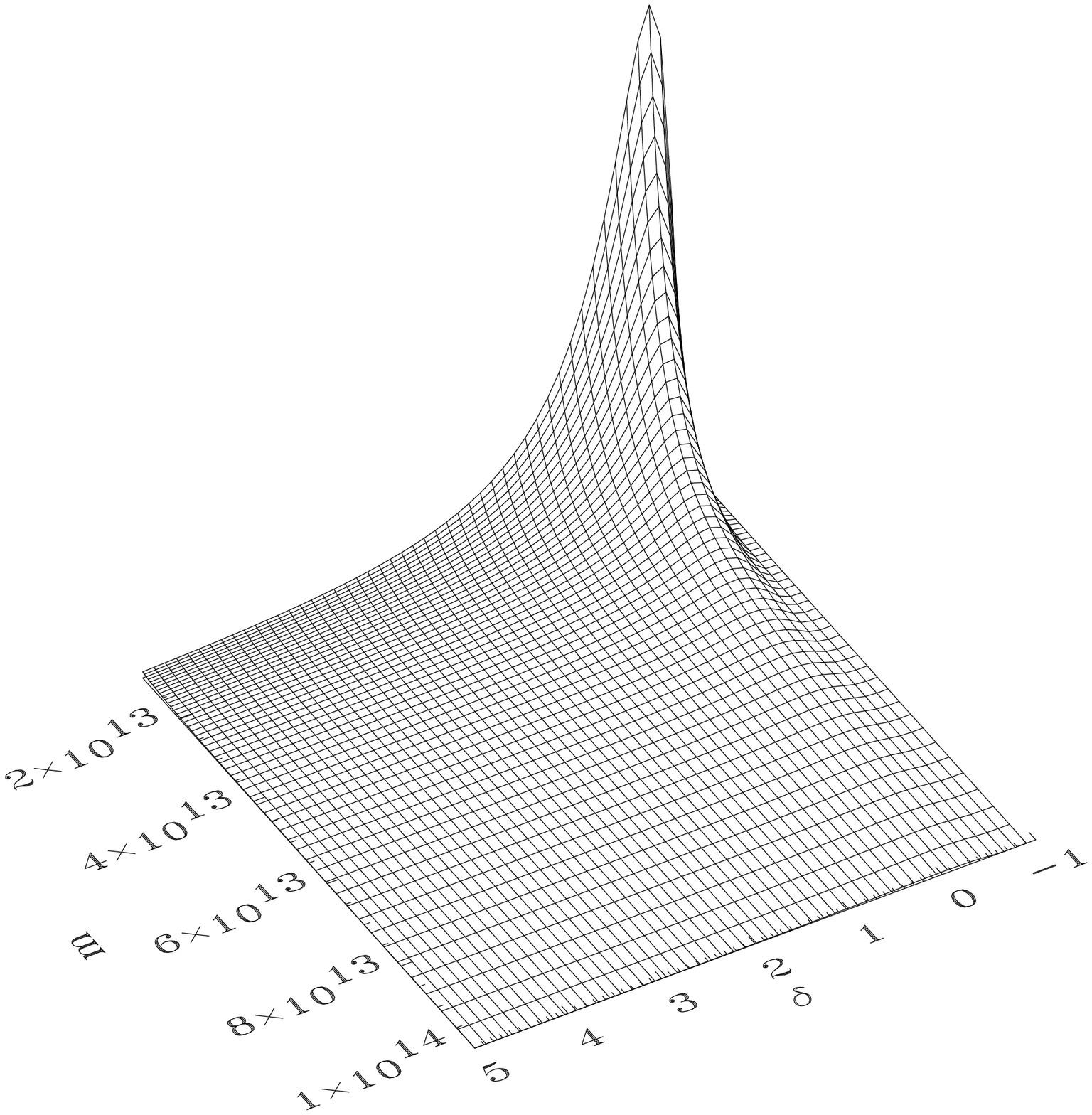}}
\resizebox{2.9in}{!}{
\includegraphics{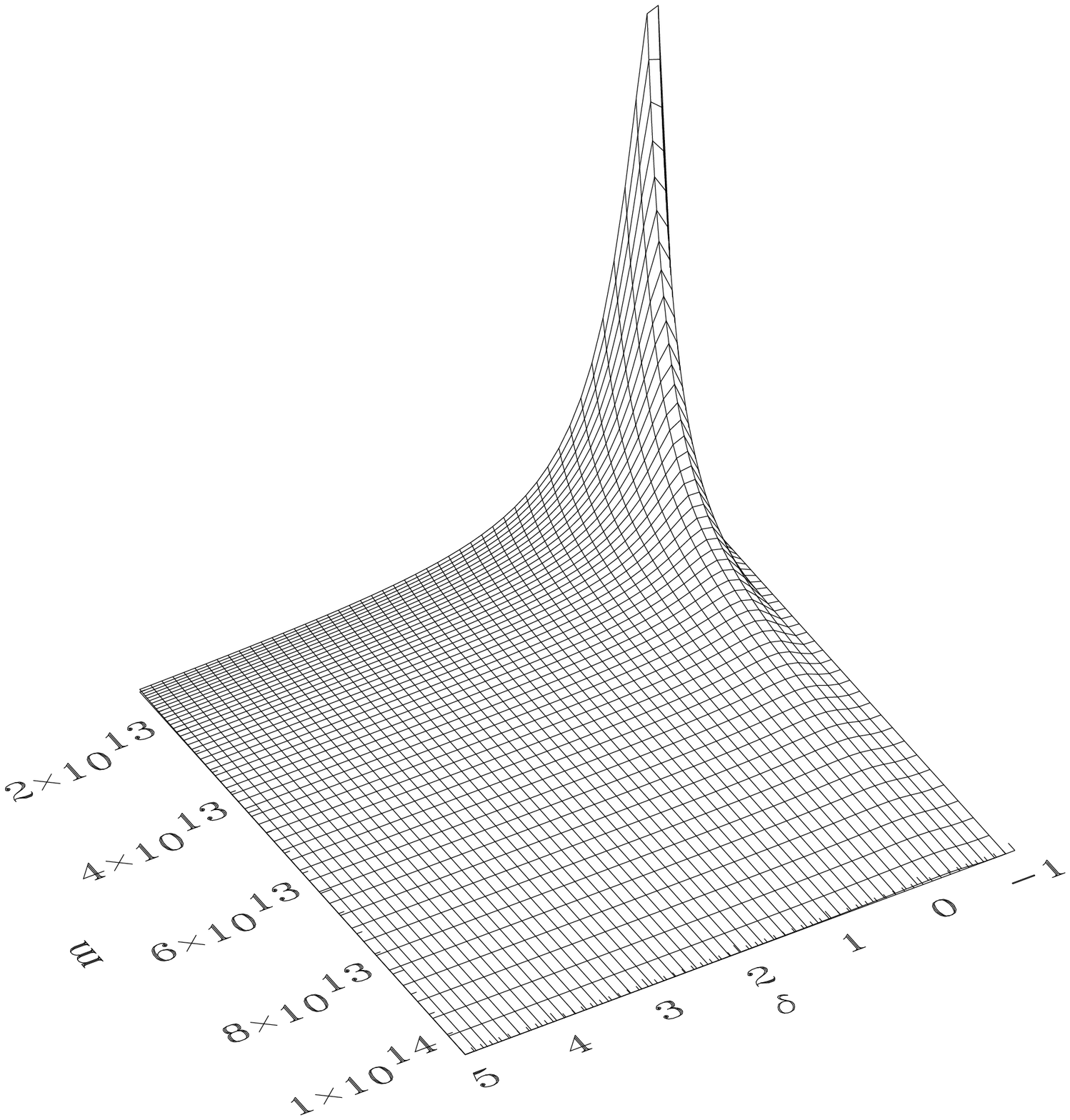}}
\caption{\label{fig:surfaces} Surface plots of the double distribution
for $z=0$ and $\beta=2$ in $\Omega_{\rm m}+\Omega_{\rm \Lambda} =
1$ (left panel) and Einstein-deSitter (right panel) universes. The mass
is measured in ${\rm M_\odot}$. The vertical axis is linear, with the
$m-\delta$ axes level corresponding to $dn/dmd\delta_{\ell} = 0$ 
and the highest
point corresponding to $dn/dmd\delta_{\ell} = 2.72\times 10^{-2}$ (left
panel)  and
$dn/dmd\delta_{\ell} =1.89\times10^{-1} $ (right panel) objects per 
${\rm Mpc ^3}$ per $10^{15} {\rm M_\odot}$.}
\end{figure*}

Figure \ref{fig:surfaces} shows a 3-dimensional rendering of our double
distribution as a function of mass and overdensity for fixed $\beta =
2$ and $z=0$.  The left panel corresponds to the concordance universe
while the right panel corresponds to the Einstein-deSitter
universe, and this arrangement is retained throughout this section.

\begin{figure*}
\resizebox{2.9in}{!}{
\includegraphics{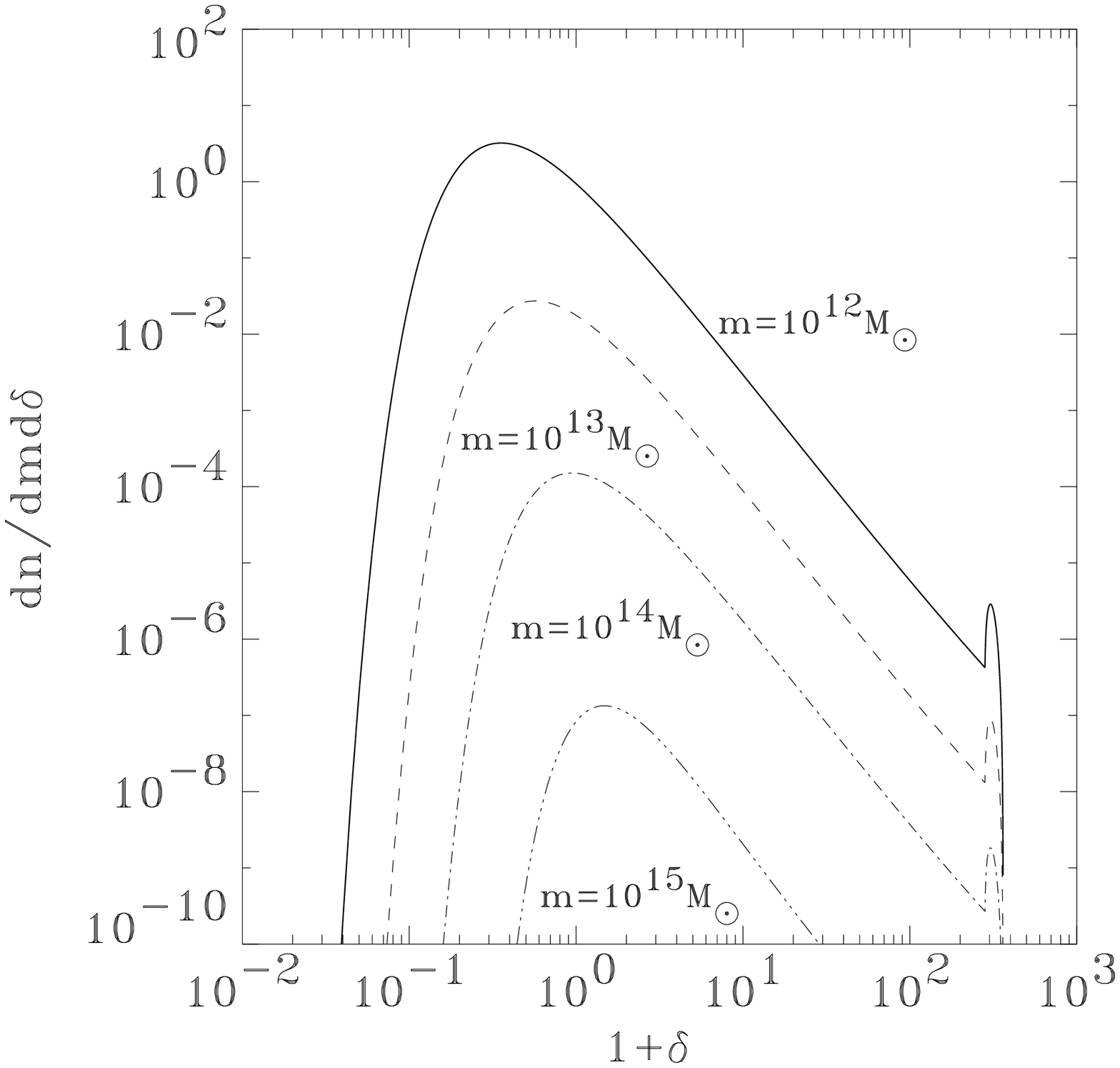}}
\resizebox{2.9in}{!}{
\includegraphics{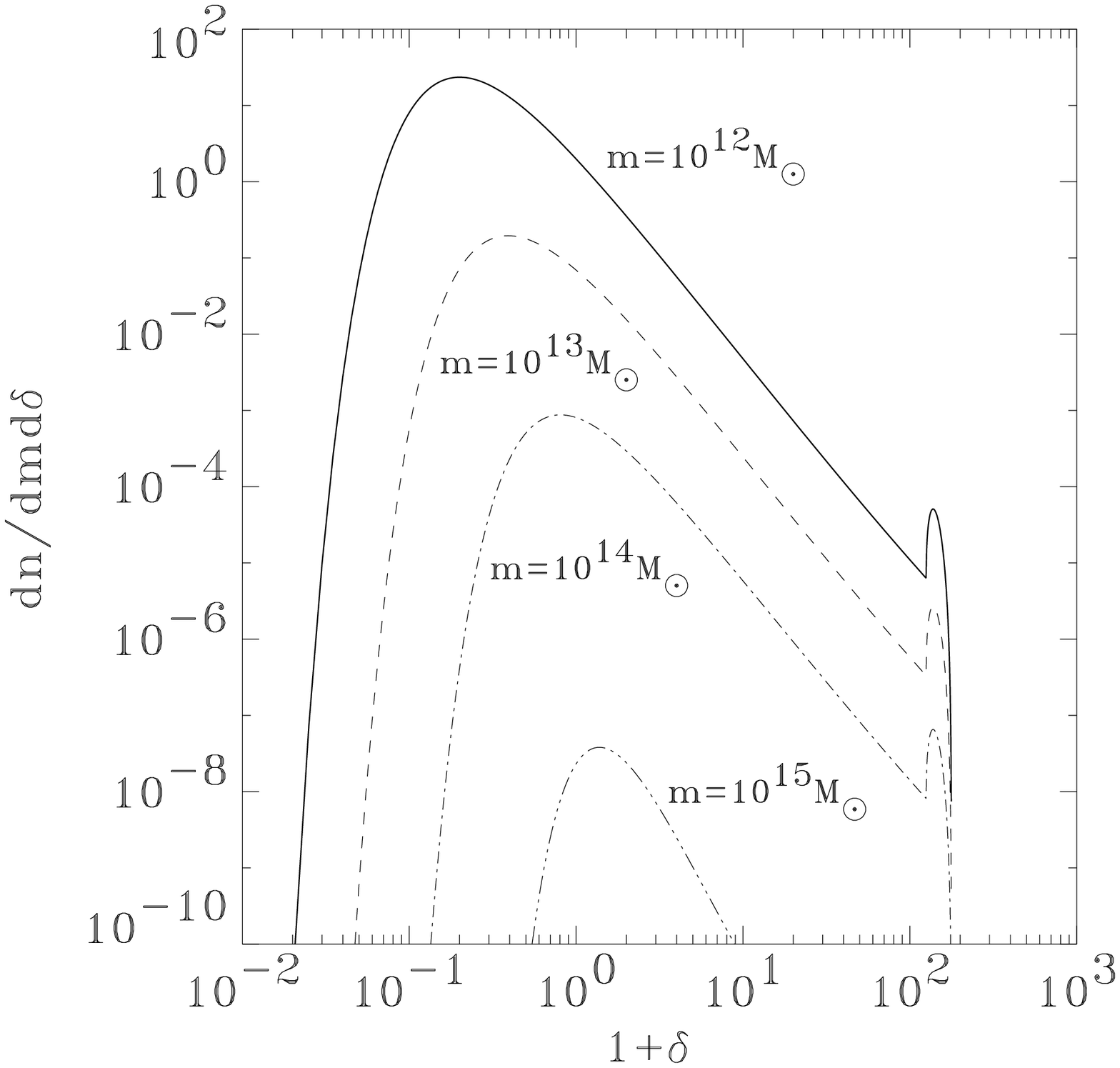}}
\caption{\label{fig:efofm} Slices of the double distribution function
  at various fixed values of the mass for 
$\Omega_{\rm m}+\Omega_{\rm \Lambda} =
1$ (left panel) and  Einstein-deSitter (right panel) universes. The
  units of the double distribution are number of objects per ${\rm Mpc
  ^3}$ per $10^{15} {\rm M_\odot}$.}
\end{figure*}

\begin{figure*}
\resizebox{2.9in}{!}{
\includegraphics{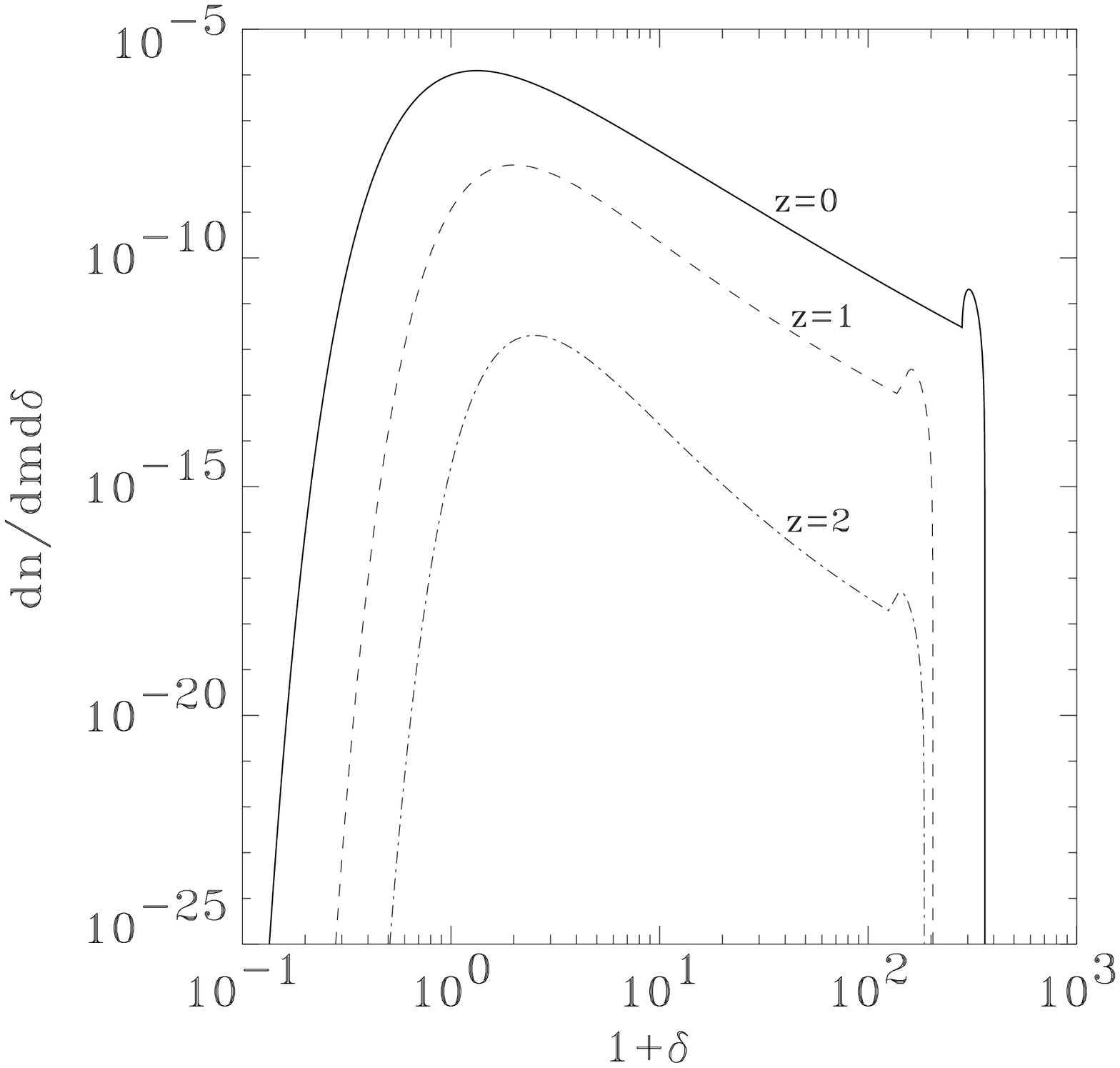}}
\resizebox{2.9in}{!}{
\includegraphics{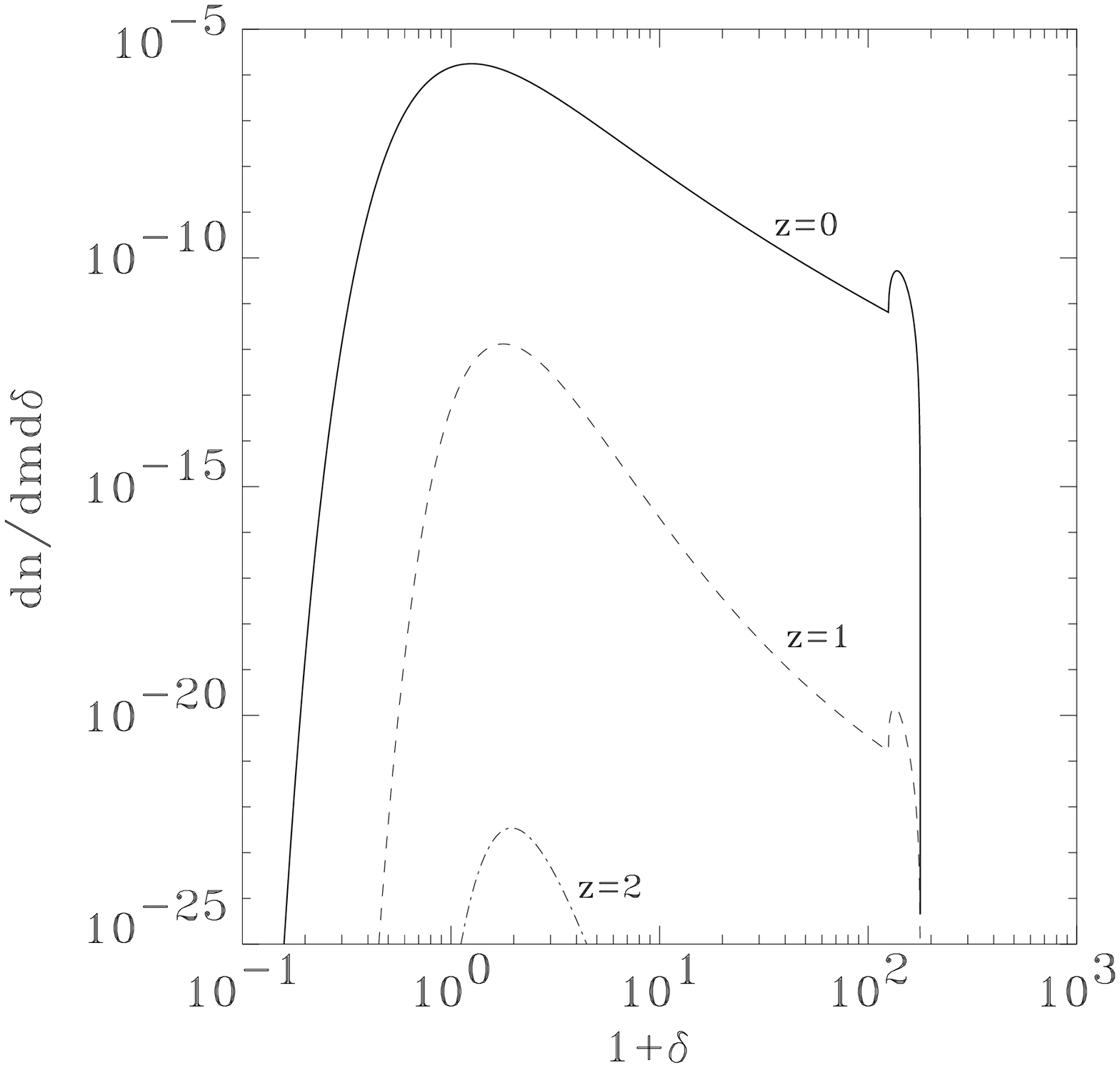}}
\caption{\label{fig:efofz} Slices of the double distribution function
  at $m=5.5\times10^{14}{\rm ,M_\odot}$ and for various values of
  redshift $z$, 
for $\Omega_{\rm m}+\Omega_{\rm \Lambda} =
1$ (left panel) and  Einstein-deSitter (right panel) universes. The
  units of the double distribution are number of objects per ${\rm Mpc
  ^3}$ per $10^{15} {\rm M_\odot}$.}
\end{figure*}

\begin{figure*}
\resizebox{2.9in}{!}{
\includegraphics{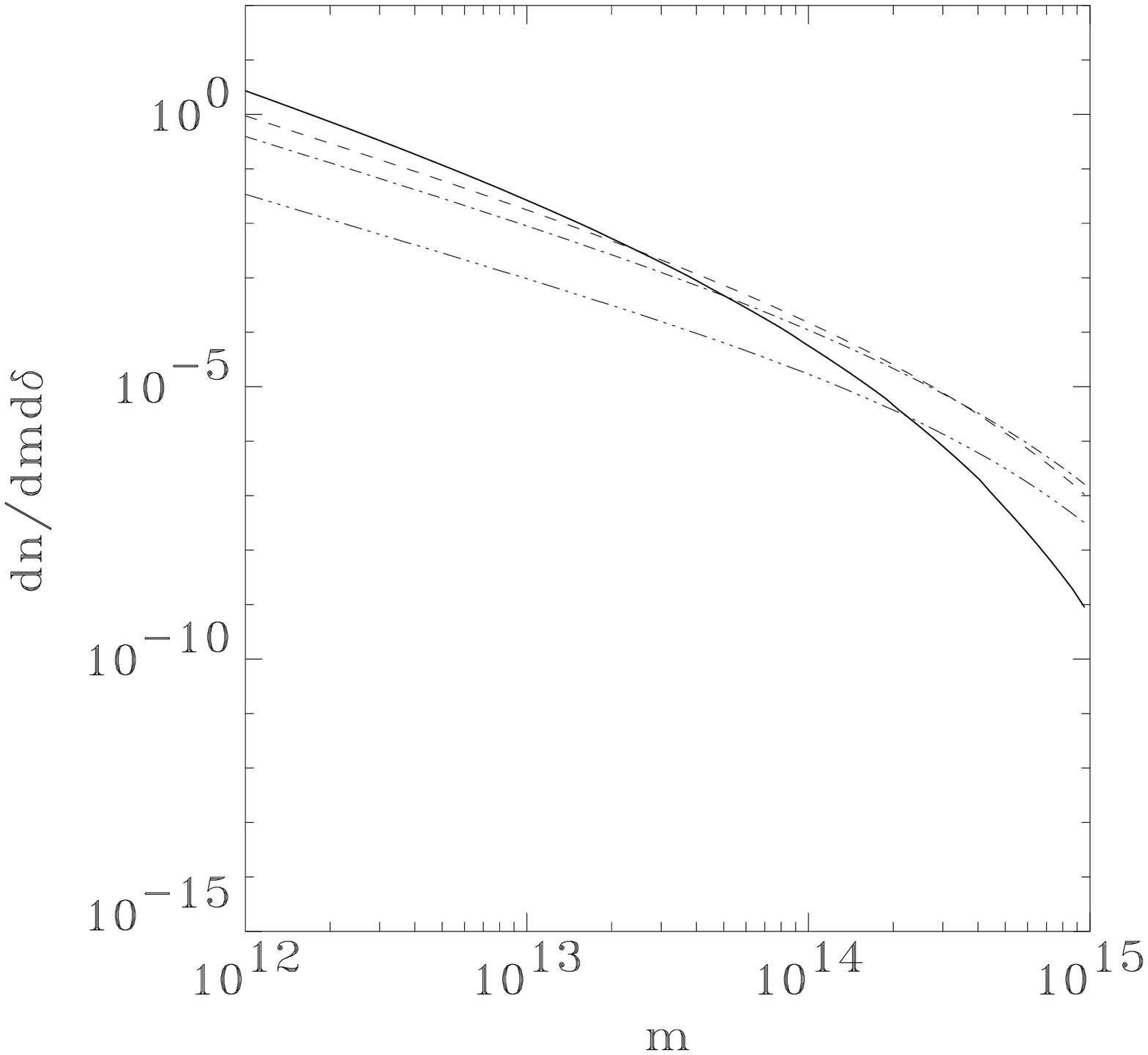}}
\resizebox{2.9in}{!}{
\includegraphics{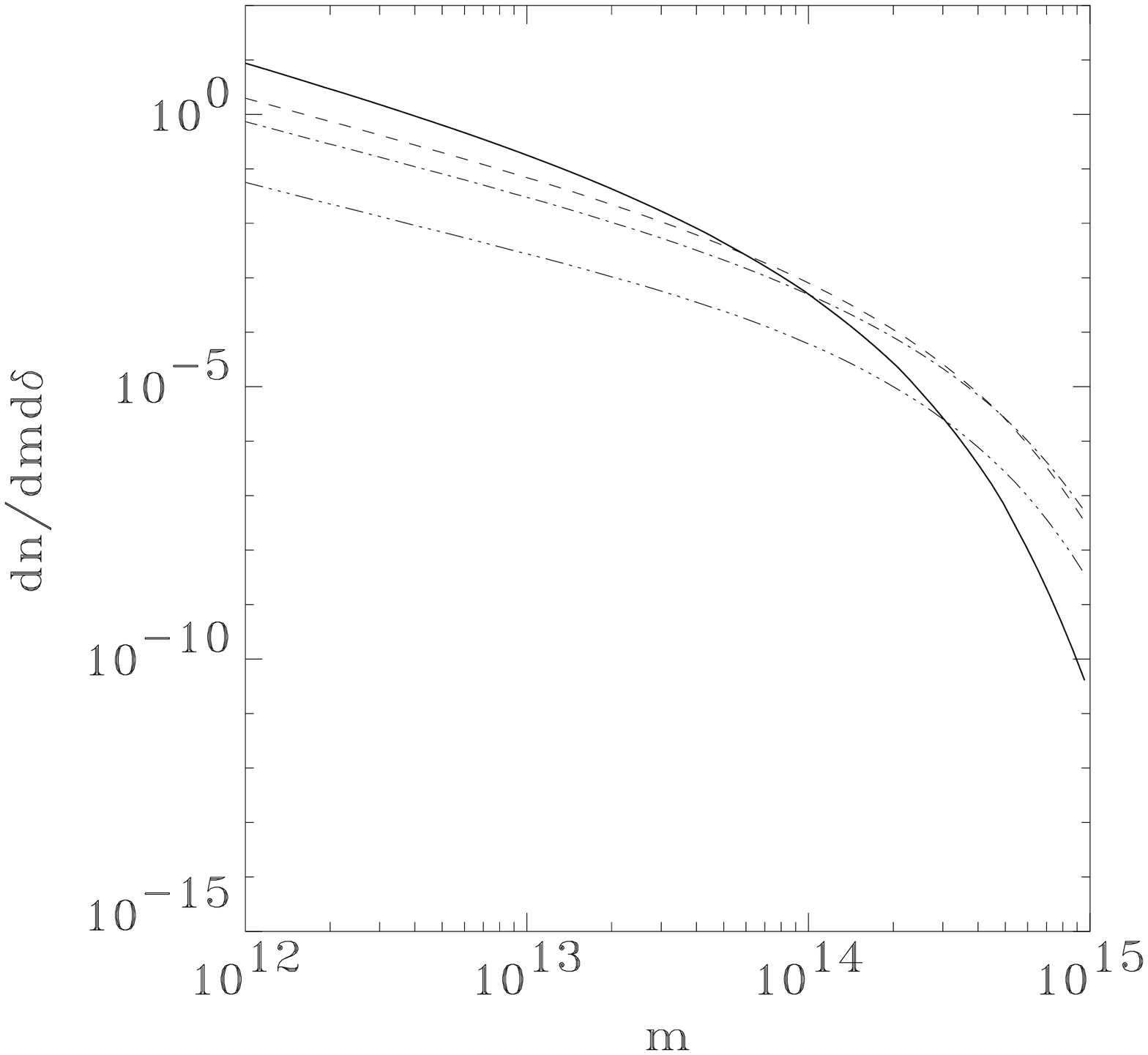}}
\caption{\label{fig:efofd_low} Slices of the double distribution function
  at constant values of $\delta$ 
for $z=0$, $\beta=2$ and for $\Omega_{\rm m}+\Omega_{\rm \Lambda} =
1$ (left panel) and  Einstein-deSitter (right panel) universes. Solid
  line: $\delta = -0.5$; dashed line: $\delta=0$; dot-dashed line:
  $\delta=0.5$; double-dot--dashed line: $\delta=3$. The
  units of the double distribution are number of objects per ${\rm Mpc
  ^3}$ per $10^{15} {\rm M_\odot}$.}
\end{figure*}

\begin{figure*}
\resizebox{2.9in}{!}{
\includegraphics{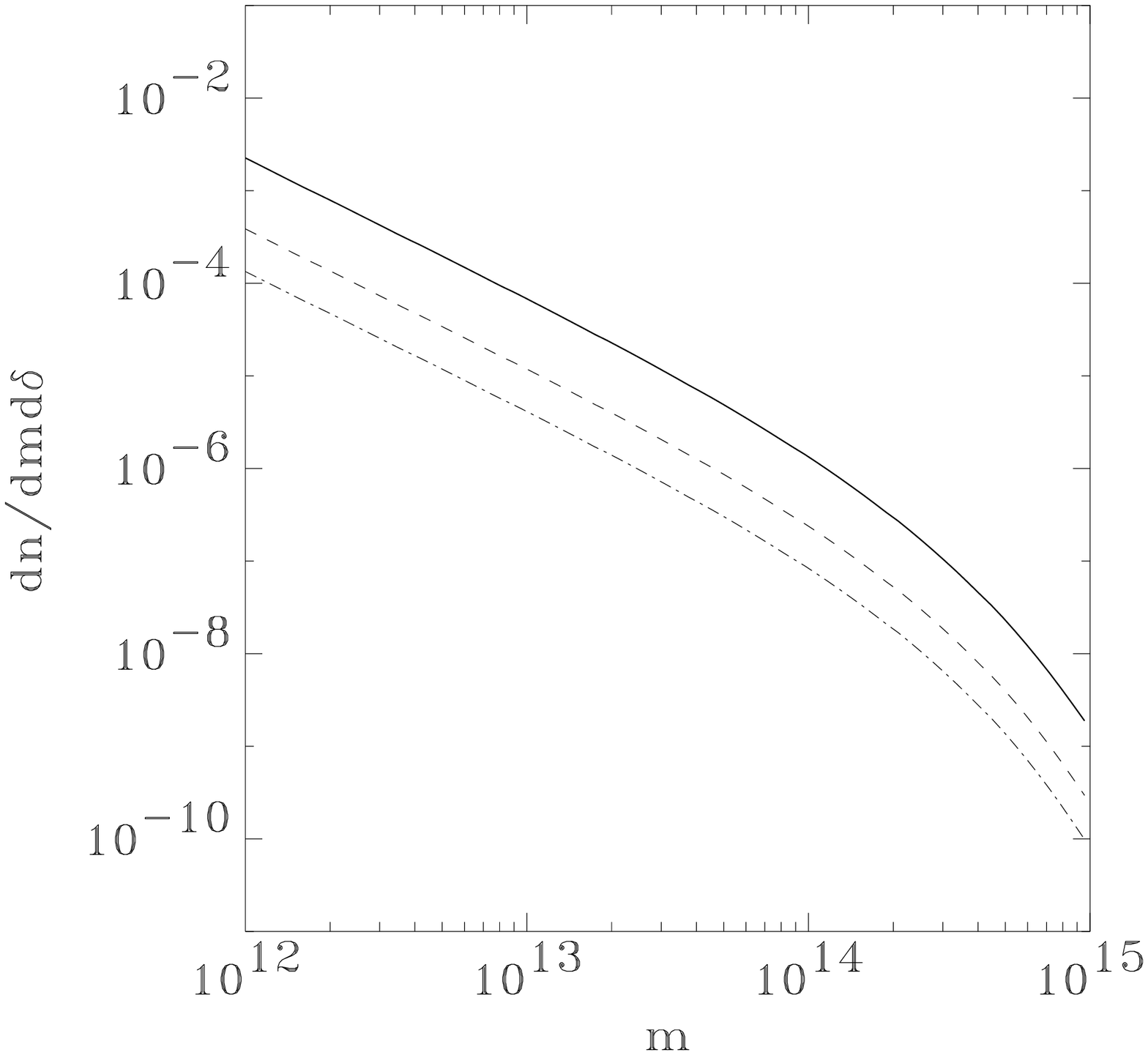}}
\resizebox{2.9in}{!}{
\includegraphics{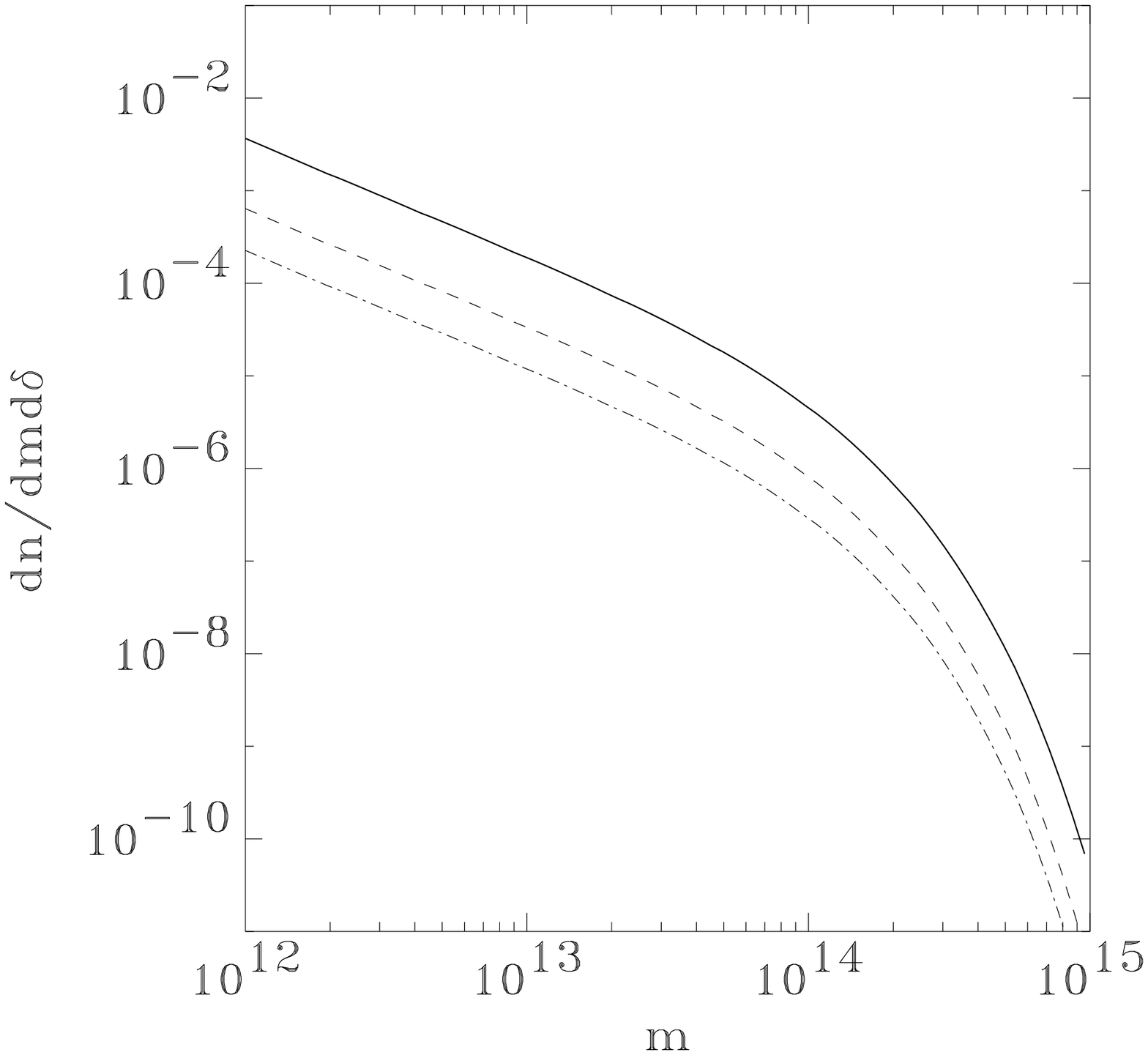}}
\caption{\label{fig:efofd_high} Slices of the double distribution function
  at constant values of $\delta$ 
for $z=0$, $\beta=2$, and for $\Omega_{\rm m}+\Omega_{\rm \Lambda} =
1$ (left panel) and  Einstein-deSitter (right panel) universes. Solid
  line: $\delta = 10$; dashed line: $\delta=20$; dot-dashed line:
  $\delta=30$. The
  units of the double distribution are number of objects per ${\rm Mpc
  ^3}$ per $10^{15} {\rm M_\odot}$.}
\end{figure*}

The features of the double distribution are demonstrated in more quantitative 
detail in Figs.~\ref{fig:efofm}-\ref{fig:efofd_high}. Figures 
\ref{fig:efofm} and \ref{fig:efofz} show slices of the double
distribution at constant values of mass. In Fig.~\ref{fig:efofm},
different curves correspond to different values of the central object
mass. In Fig.~\ref{fig:efofz}, all curves are for an object mass
$m=5.5 \times 10^{15} {\rm \, M_\odot}$, and different curves
correspond to different redshifts. Their most prominent feature is
the pronounced peak at a relatively low value of $|\delta|$,
indicating that for each given pair of $z$ and $m$, there is a
preferred, ``most probable'' value of the local environment density
contrast. As we can see in Fig.\ref{fig:efofm}, the location of
this peak moves to higher values of the density contrast as the mass
of the object increases: small structures are preferentially located 
in relative isolation, while larger structures are more likely to be found in
clustered environments. This result fits well in the picture of
hierarchical structure formation, as smaller structures tend to be
merged into higher-mass objects as time progresses. Lower-mass
objects which are initially part of underdensities are less probable
to undergo mergers, and hence are more likely to survive at late times
than objects which are initially part of overdensities. Conversely,
higher-mass structures are more likely to be parts of overdensities
where they can accumulate mass more easily through mergers with smaller
structures.  

Note, however, that in the hierarchical structure formation picture,
the mass scale where the exponential suppression of collapsed
structures sets in increases with time. Thus, any given mass scale
starts out as being a ``high mass'' at early times and eventually
becomes a ``lower mass'' as it enters the power-law regime of
the Press-Schechter mass function. Hence, according to the argument
we used to explain Fig.~\ref{fig:efofm}, the double distribution for
any given mass scale should peak at increasing $\delta$ values with
increasing redshift. This is because a particular mass scale used to be closer
to the high-mass end of the halo distribution in the past than it is
today. Indeed, this is the trend seen in Fig.~\ref{fig:efofz}. 
As we would expect, the peak of the distribution moves to
higher $\delta$ values with increasing redshift.  
The significantly
more pronounced suppression of this mass scale in high redshifts in
the Einstein-deSitter universe is due to the different power-spectrum
normalization in the two cosmological models. Because of our choice 
in the power-spectrum normalization,
the exponential suppression in the number density of structures sets
in at low masses  in the Einstein-de Sitter case than in the concordance
universe. 
Thus, there is a tendency to 
see more structures of higher mass in our concordance results than in
the Einstein-de Sitter case, despite the intuitive expectation that a
higher $\Omega_{\rm m}$ universe should have more massive structures
at late times due to its ability to continue to form structures even
at the present epoch. This would indeed have been the case if 
the power-spectrum had been normalized in the same way. 

\begin{figure*}
\resizebox{2.9in}{!}{
\includegraphics{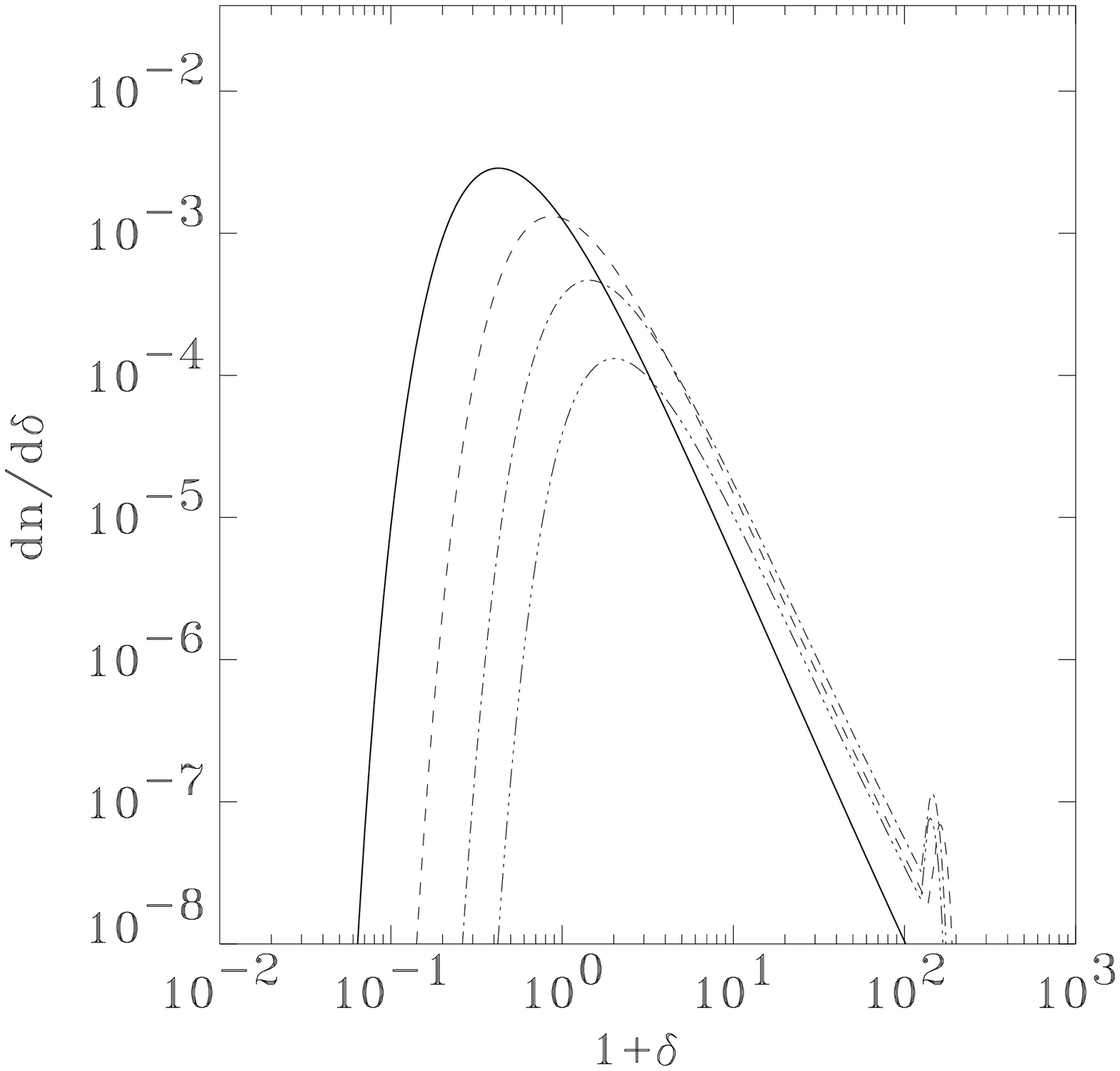}}
\resizebox{2.9in}{!}{
\includegraphics{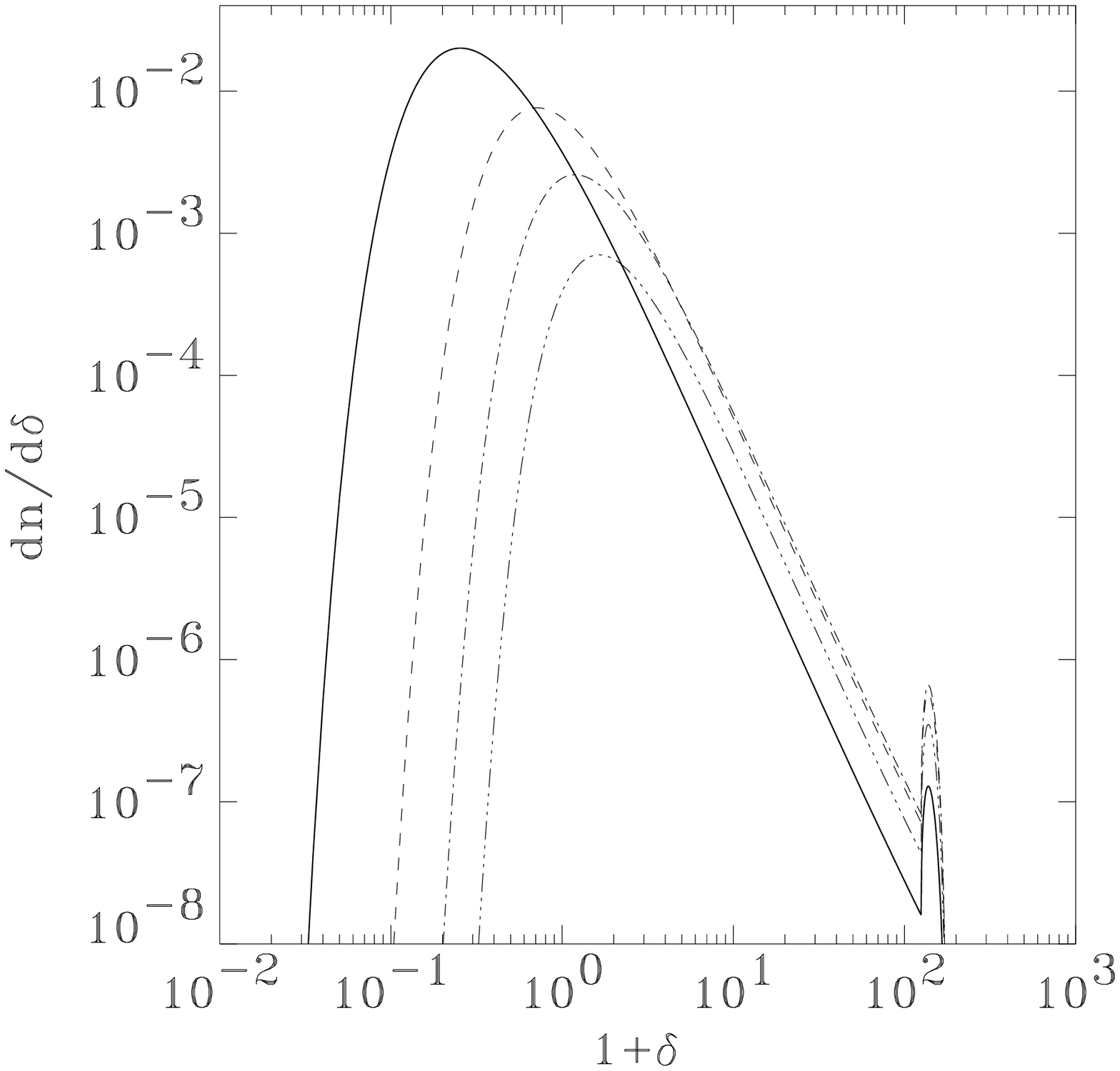}}
\caption{\label{fig:nz} Distribution of structures 
with respect to local density contrast, 
  $dn/d\delta (> 10^{12} {\, \rm M_\odot})$, 
for $\beta = 2$ and for $\Omega_{\rm m}+\Omega_{\rm \Lambda} =
1$ (left panel) and  Einstein-deSitter (right panel) universes. 
  Solid line: $z=0$; dashed line: $z=1$; dot-dashed line: $z=2$;
 double-dot--dashed line: $z=3$. The
  units of  $dn/d\delta$ are number of objects per ${\rm Mpc
  ^3}$.}
\end{figure*}

That halos of a given mass are more strongly clustered with increasing
redshift was also found by \cite{mw00}, who used $\Delta_8 (m)$ (the rms
overdensity in the number of haloes more massive than some mass scale
after smoothing with a spherical top-hat filter of comoving radius $8
h^{-1}$ Mpc) as a measure for halo clustering. A tendency of higher
mass objects to be found in overdense regions was discussed by
\cite{mw96} and \cite{st02}, who interpreted it by viewing halos today
as progenitors of future larger-scale structures viewed at ``high'' or
``low'' redshift.

In addition to the main peak at low $|\delta|$, an additional, much
lower and sharper peak can be seen right before the critical
overdensity cutoff. This peak is the result of the change of the
functional form of the conversion relation between linearly
extrapolated and true density contrast close to virialization, when
application of the spherical collapse model would lead $\delta$ to
diverge. The particular shape of the peak is an artifact of the
recipe we adopted for dealing with the virialization regime, and
carries no physical meaning (the shape of the peak is the shape of the
high-$\delta$ end of $d\ed/d\delta$). 
However, since the boundary conditions we use for $\ed(\delta)$ and
its derivative {\em are} physical, we do expect to
have some form of local maximum at the high-$\delta$ end of the
double distribution. Still, as discussed in the previous section, the
effect of the details or even the existence of this local maximum on
the physical quantities of interest is negligible. 

The high-$\delta$ cutoff occurs at higher values of $\delta$ in the
concordance universe than in the Einstein-deSitter universe. This is a
result of the different density contrast achieved at virialization by
structures in the two different cosmologies. In the Einstein-deSitter
case this density contrast is always $18\pi^2$, while in the
concordance universe it is always higher and increases with time.
At high redshifts, before the effect of $\Lambda$ becomes
significant, $\delta_{\rm c}$ is very close to $18\pi^2$ in the concordance
universe as well, as can be seen in Fig.~\ref{fig:efofz}.

\begin{figure*}
\resizebox{2.9in}{!}{
\includegraphics{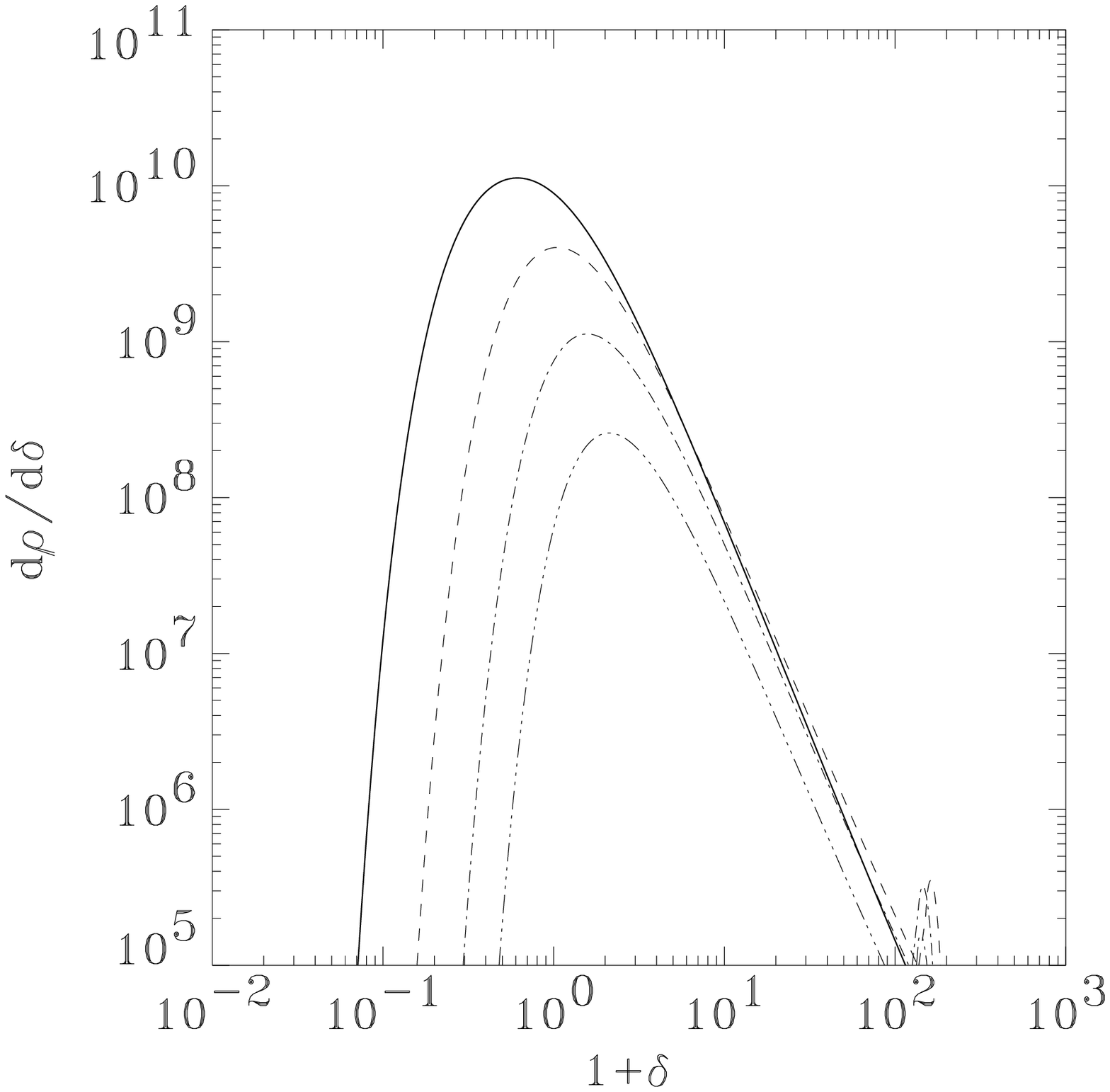}}
\resizebox{2.9in}{!}{
\includegraphics{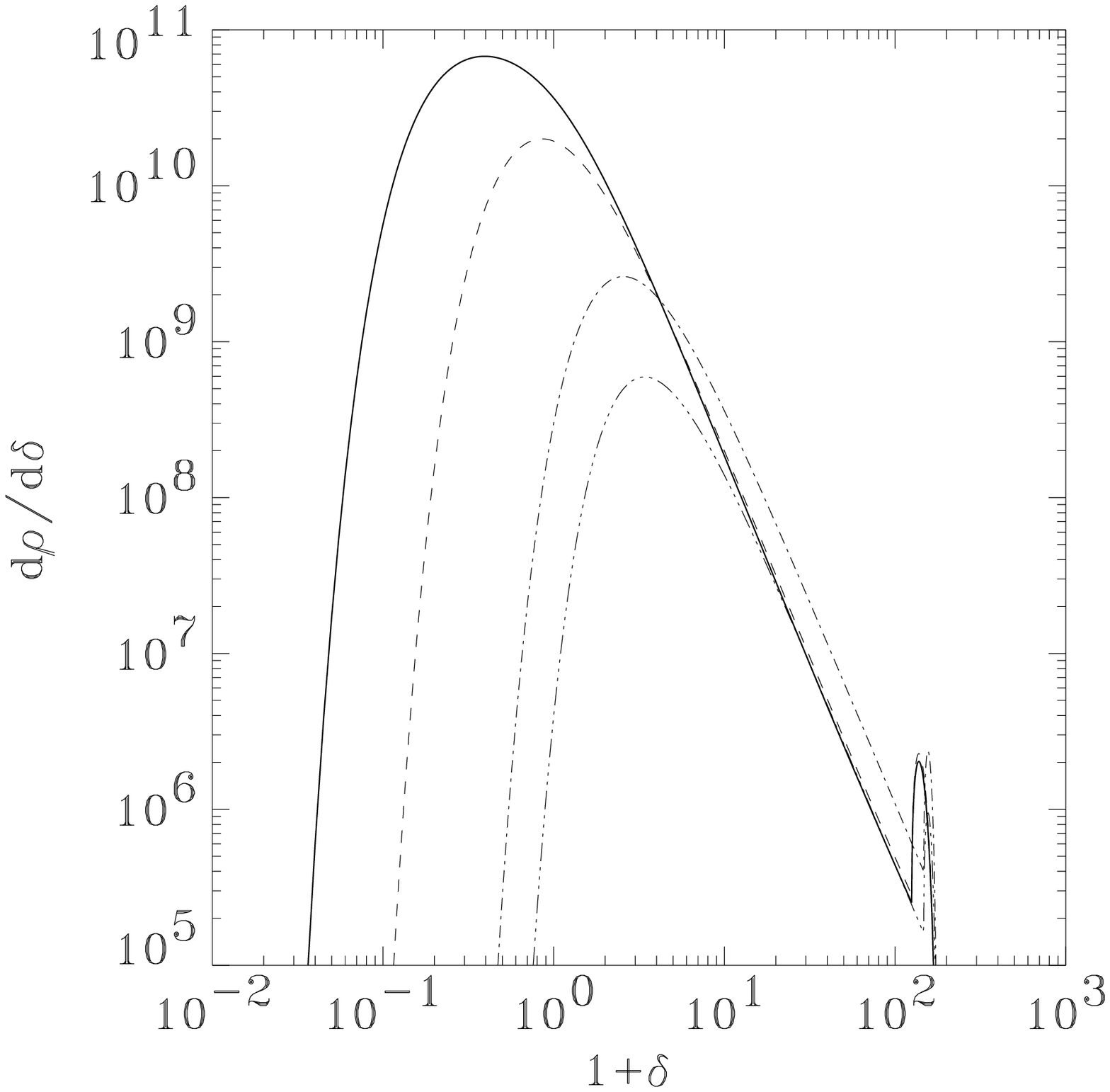}}
\caption{\label{fig:rhoz} Distribution of density of matter inside
  collapsed structures with respect to local density contrast, 
  $d\rho/d\delta (> 10^{12} {\, \rm M_\odot})$, for $\beta =2$ and
for $\Omega_{\rm m}+\Omega_{\rm \Lambda} =1$ 
(left panel) and  Einstein-deSitter (right panel) universes.
  Solid line: $z=0$; dashed line: $z=1$; dot-dashed line: $z=2$;
 double-dot--dashed line: $z=3$.
The units of  $d\rho/d\delta$ are ${\rm M_\odot}$ per ${\rm Mpc
  ^3}$.}
\end{figure*}

\begin{figure*}
\resizebox{2.9in}{2.9in}{
\includegraphics{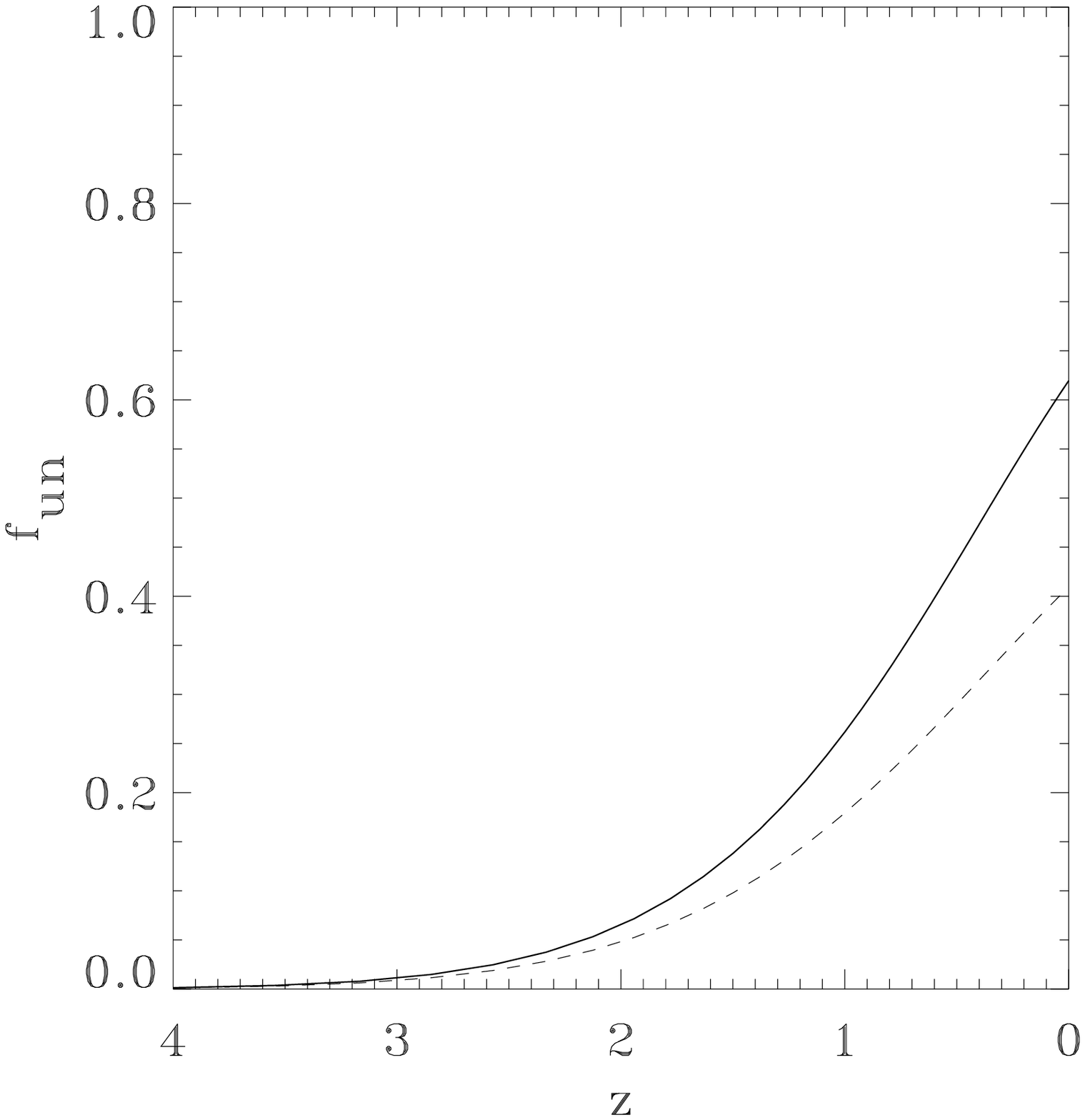}}
\resizebox{2.9in}{2.9in}{
\includegraphics{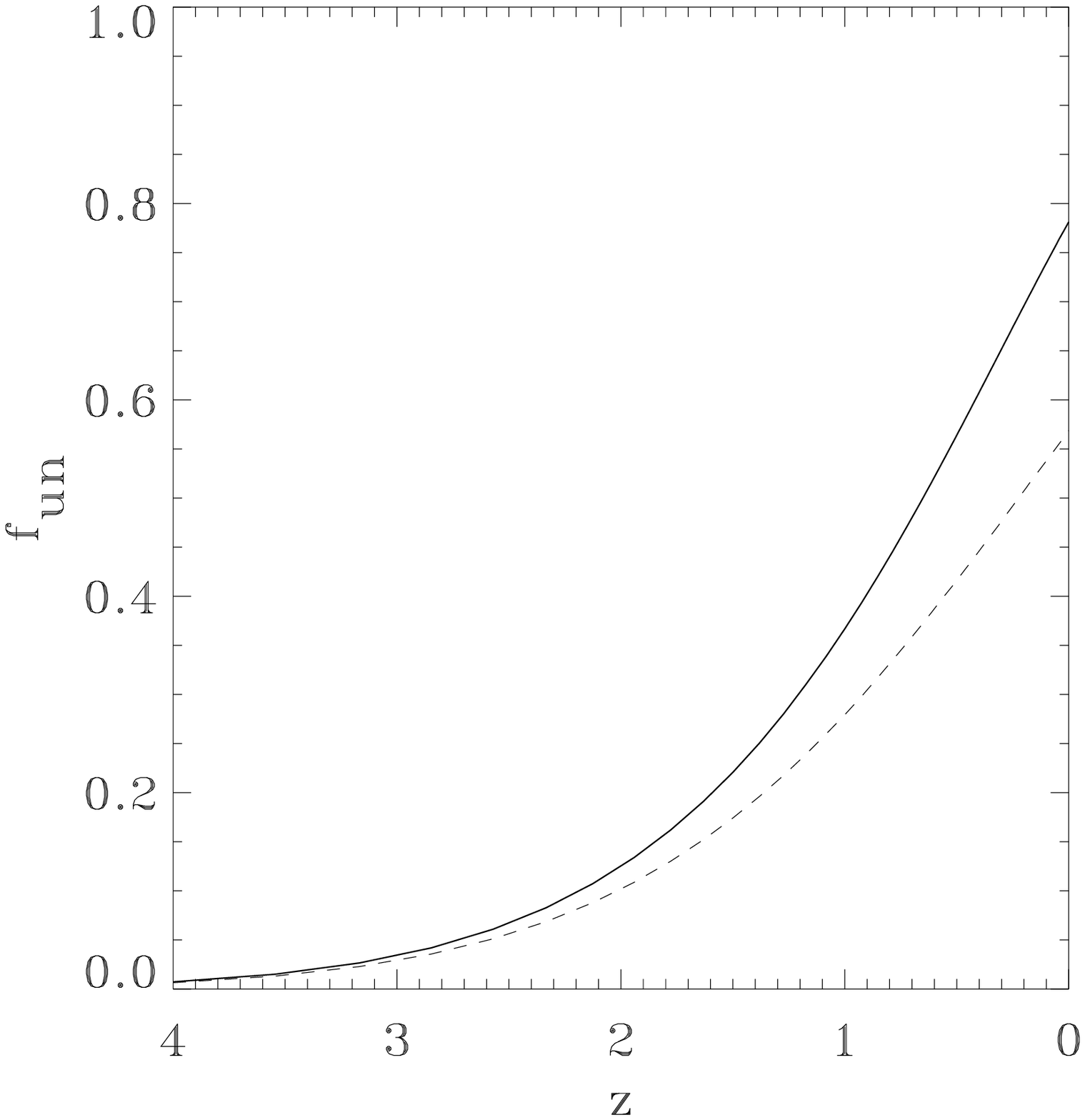}}
\caption{\label{fig:fun} Fraction by number $f_{n,{\rm un}}$
 (solid line) and my mass $f_{\rho,{\rm un}}$ (dashed line)
of objects of mass $>10^{12} {\, \rm M_\odot}$ living in underdense regions, 
as a function of redshift, 
for $\Omega_{\rm m}+\Omega_{\rm \Lambda} =1$ (left panel) and 
Einstein-deSitter (right panel) universes.}
\end{figure*}

Figures \ref{fig:efofd_low} and \ref{fig:efofd_high} show slices of
the double distribution at various fixed values of $\delta$, with $z=0$
and $\beta=2$. Unlike the constant-mass slices, the constant-$\delta$
slices do not exhibit a peak (other than the global maximum imposed by
the minimum-mass cutoff). Hence, there does not exist ``most probable'' 
mass at each given value of $\delta$. Hence, the hierarchical behavior
of the number density of structures seen in the Press-Schechter mass
function (which also exhibits no global maximum but instead 
diverges at low masses) also extends to the constant-$\delta$ slices
of the double distribution. 
Figure \ref{fig:efofd_low} shows slices corresponding
to relatively low values of $|\delta|$ ($\delta = -0.5, 0, 0.5$ and
$3$, close to the distribution peak in $\delta$). 
At the high-mass end of the distribution, the abundance of
objects increases with increasing $\delta$, while in the low mass end
of the distribution the trend is reversed, and the object abundance
increases with decreasing $\delta$. This is in agreement with the
behavior observed in the constant-$m$ slices.

Figure \ref{fig:efofd_high} shows slices corresponding to high values
of $\delta$ ($\delta = 10, 20$ and $30$), farther from the distribution peak.
In this case, the curves do not cross, and an increase of $\delta$
simply results in an overall suppression of object abundance:
structures of all masses are unlikely to be found overly clustered.
This is because the final stages of collapse proceed rather quickly
compared to the time spent around turnaround. The likelihood of a
region observed in its late stages of collapse but before virialization
is then low because the lifetime of this phase is small.

In Fig.~\ref{fig:nz} we plot $dn/d\delta(>10^{12} {\rm M_\odot})$ as
a function of $1+\delta$ for different values of redshift. It is
striking that at $z=0$, the distribution peaks at negative $\delta$
values (around $\delta=-0.6$ in the concordance and $-0.7$ in the
Einstein-deSitter universe), indicating that the most probable
location for a collapsed object of mass $>10^{12} {\, \rm M_\odot}$ is
an {\em underdense} environment. For the specific mass range, this
trend is reversed by $z=1$, when the preferred location of these
objects is close to the universe mean ($\delta =0$). This
time-evolution pattern is independent of cosmology, as it is present
both in the concordance and the Einstein-deSitter universes, 
and appears rather to be a characteristic of the
hierarchical nature of structure formation. Parameters of this
distribution can be calculated using Eqs. (\ref{parnmean}) and
(\ref{parnvar}) which, for the concordance cosmology and $z=0$ give
$\langle \delta\rangle_{\rm n} = 0.43$ and $\sigma_{\rm \delta, n}=4.36$. The
large value of the variance shows that the distribution is
significantly broad. However, the positive value of the mean is an
artifact of the asymmetric boundaries of the distribution and its long
high-$\delta$ tail. This is demonstrated by the
notably different locations of the mean and the median. The value of
the latter is $\delta=-0.22$, therefore more structures in this range
reside inside underdensities.

\begin{figure*}
\resizebox{2.9in}{!}
{
\includegraphics{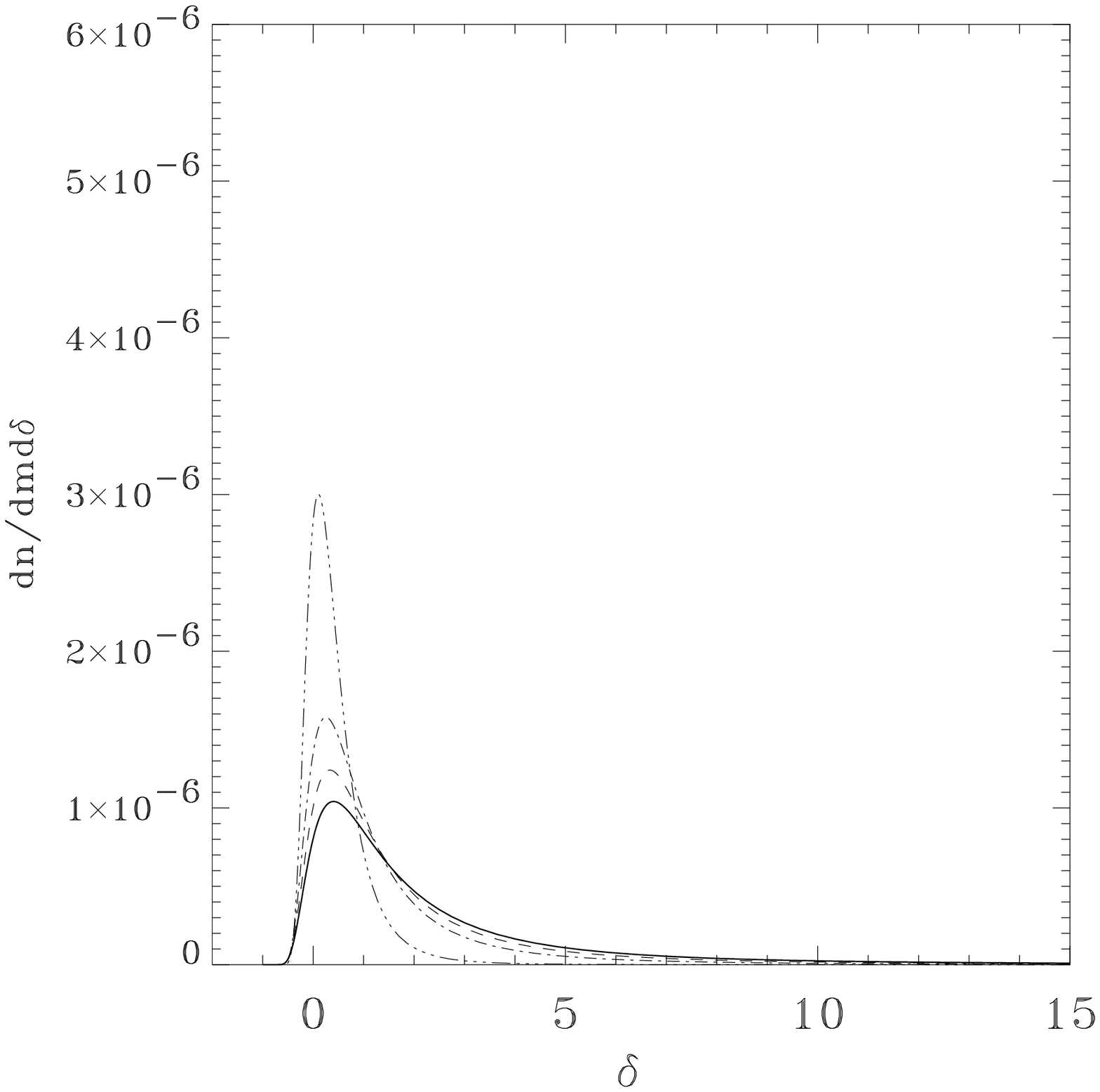}}
\resizebox{2.9in}{!}{
\includegraphics{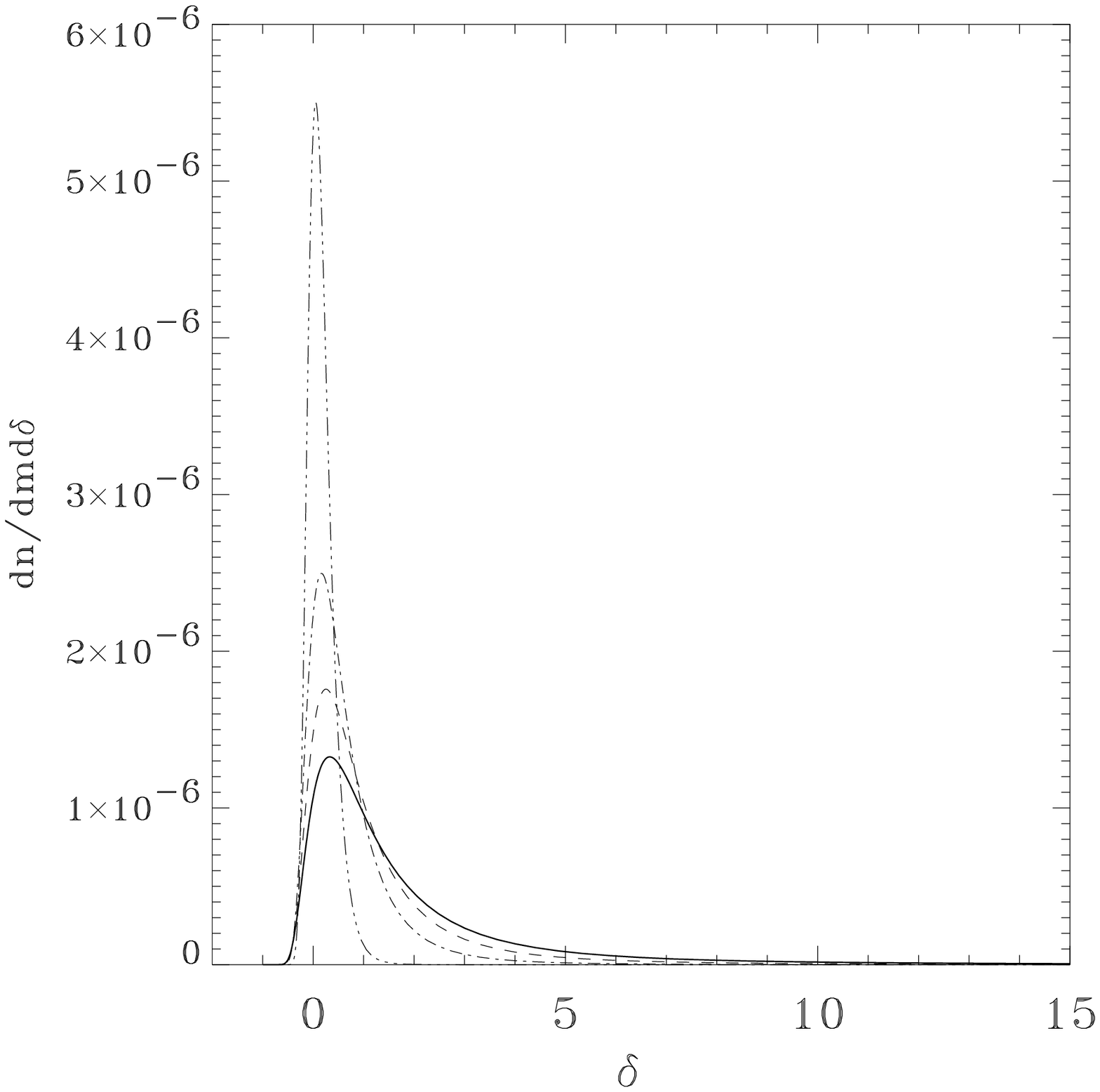}}
\caption{\label{fig:efofbeta} Slices of the double distribution function
  at $m=5.5\times10^{14}{\rm ,M_\odot}$ and for different values of
  the clustering scale parameter $\beta$, 
for $\Omega_{\rm m}+\Omega_{\rm \Lambda} =
1$ (left panel) and  Einstein-deSitter (right panel) universes,
  plotted in linear scale. Solid
  line: $\beta=1.5$; dashed line: $\beta=2$; dot-dashed
  line: $\beta=3$; double-dot--dashed line: $\beta = 10$. The
  units of the double distribution are number of objects per ${\rm Mpc
  ^3}$ per $10^{15} {\rm M_\odot}$.}
\end{figure*}

\begin{figure*}
\resizebox{2.9in}{!}{
\includegraphics{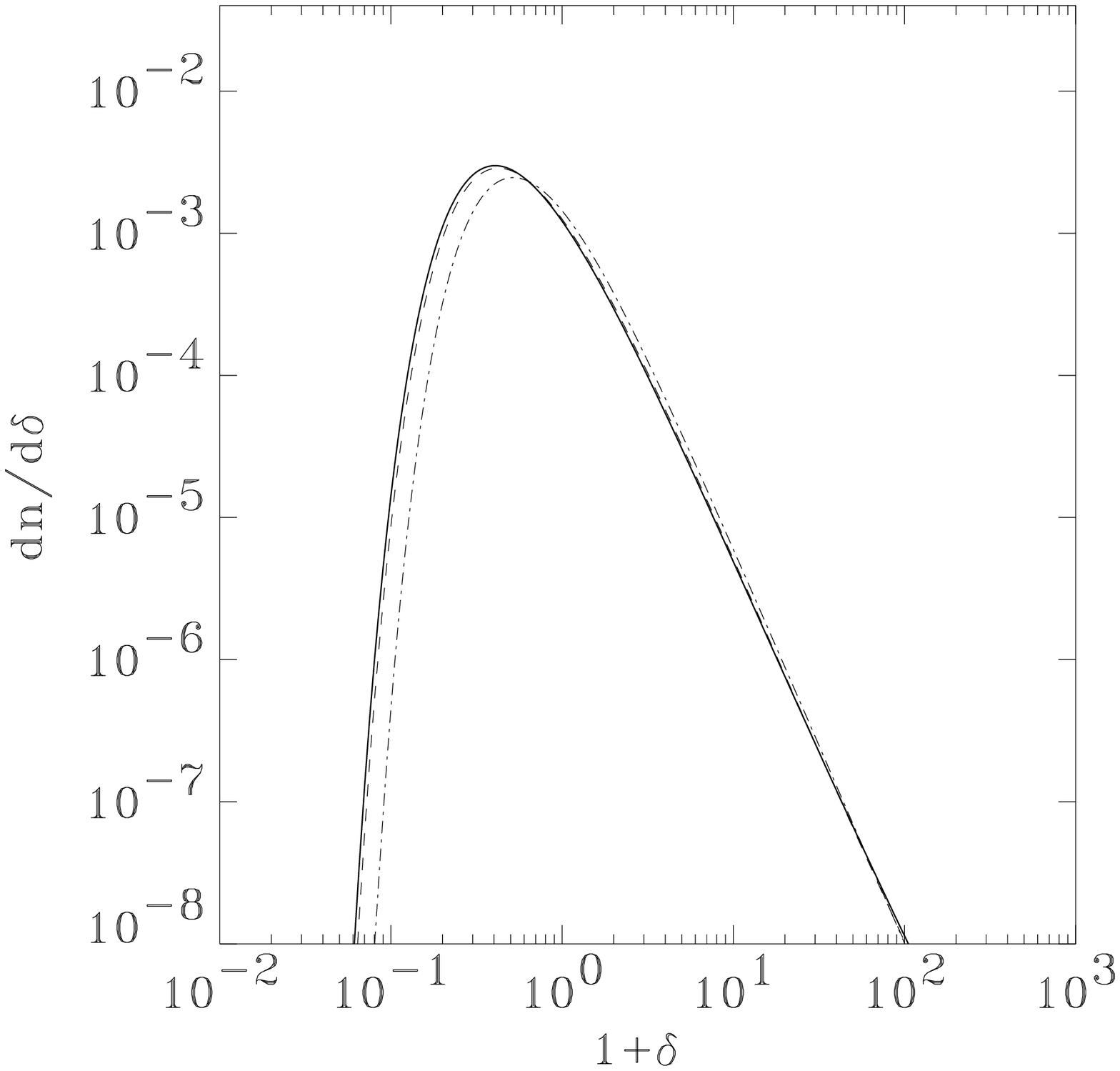}}
\resizebox{2.9in}{!}{
\includegraphics{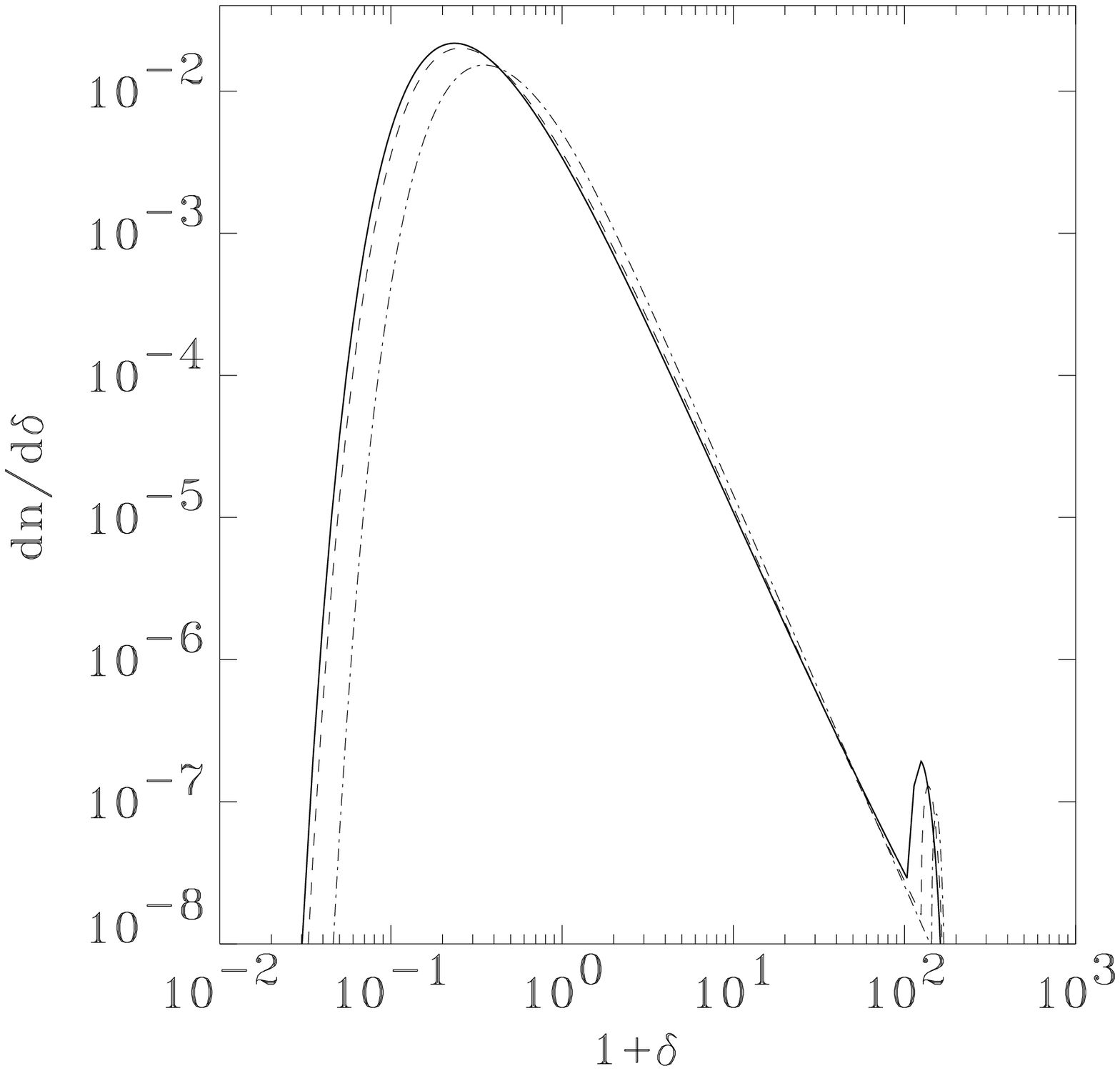}}
\caption{\label{fig:nb} Distribution of structures of mass larger than
  $10^{12} {\, \rm M_\odot}$ with respect to local density contrast, 
  $dn/d\delta$, for $\Omega_{\rm m}+\Omega_{\rm \Lambda} = 
1$ (left panel) and  Einstein-deSitter (right panel) universes,  
at $z=0$, and for
  $\beta=1.5$(solid line), $\beta=2$ (dashed line) and
  $\beta=10$(dot-dashed line).
The  units of  $dn/d\delta$ are number of objects per ${\rm Mpc
  ^3}$.}
\end{figure*}

Figure \ref{fig:rhoz} is the matter-density counterpart of Fig.~\ref{fig:nz}, 
as it shows $d\rho/d\delta(>10^{12} {\rm M_\odot})$
as a function of $1+\delta$ for the same values of redshift as in
Figure \ref{fig:nz}. Again, at the current cosmic epoch, 
the distribution peaks at negative values of $\delta$. Most of the
virialized matter in the universe today appears to reside inside isolated 
objects rather than in clusters (note that decreasing the value of
$m_{\rm min}$ will only enhance this result since the trend towards
isolation is more pronounced for the lower-mass objects).
The trend of the peak with time (towards
larger $\delta$ for higher redshifts) is duplicated here as well.
In particular, note that at present, a significant fraction 
of the mass lies in moderately underdense regions.  Equations (\ref{parrhomean}) and
(\ref{parrhovar}) give for this distribution (in the concordance
cosmology and for $z=0$), $\langle \delta \rangle
_{\rho} = 1.20$ and $\sigma_{\delta,\rho} = 6.23$. The median of this
distribution is at $\delta=0.20$, a positive value.

Finally, Fig.~\ref{fig:fun} shows the evolution with redshift of the 
fractions by number and by mass, $f_{n,{\rm un}}$ and
$f_{\rho,{\rm un}}$, of objects with $m>10^{12} {\rm \, M_\odot}$, 
living inside underdense regions.
At high redshifts, when the mass of such objects 
is well above the exponential suppression cutoff, 
practically none of them are found inside underdensities. 
This trend is reversed as the redshift 
decreases. In the $\Omega_{\rm m}+\Omega_\Lambda=1$ universe, an equal
 number of these structures are 
located inside underdensities by redshift $0.3$ and 
by the current cosmic epoch, about 60\% by number (but only 40\% by
mass)  
of these structures 
are located inside underdensities. 

Figure \ref{fig:efofbeta} demonstrates the effect of changing the
clustering scale parameter on the double distribution. Slices of the
double distribution along $m=5.5\times 10^{14}{\, \rm M_\odot}$ are
plotted (in linear axes)  as a function of $\delta$, and for $\beta =
1.5$ (solid line), $2$ (dashed line), $3$ (dot-dashed line) and $10$
(double-dot--dashed line). The location of the peak appears to be
extremely insensitive to the value of $\beta$ for moderately low
values. It very slowly moves towards $\delta=0$ with increasing
$\beta$, as it should (increasing $\beta$ results in averaging the
overdensity over increasingly large volumes). Note that even as
$\beta$ approaches 1, the peak will {\em not }move towards $\delta_{\rm c}$,
as a result of our correction for the central-object
contamination. This makes our formalism particularly suitable to study
the properties of matter very close but outside a virialized structure
(e.g. the local density of accreted gas). 

The effect of $\beta$ on an integral quantity is shown in Fig.~\ref{fig:nb}, 
which shows  $dn/d\delta(>10^{12} {\rm M_\odot})$
for $\beta = 1.5$ (solid line), $\beta=2$ (dashed line) and $\beta=10$
(dot-dashed line). Again, the results are extremely
insensitive to the value of $\beta$, which gives us confidence about
the robustness of the location of the peak
of our distributions. 

As we have seen in this section, the correlations between the
mass of collapsed structures and their environmental conditions are
not only non-trivial, but also evolve dramatically with cosmic epoch
(for a wide range of masses, the most probable 
$\delta$ changes sign between redshifts
of zero and a few). It would therefore be desirable and of high
interest to use the double distribution to investigate the effect 
of the environment on systems and phenomena the properties of which
depend sensitively on the state of their surrounding space and matter.
In the next chapter, we will use the environmental information encoded
in the double distribution to assess the effect of the environmental
factor on an important aspect of cosmic structure formation: cosmic
accretion shocks.
\chapter{Cosmic Accretion Shocks}
\label{chapter:chap3}

\section{Overview}

\indent

In this chapter, we utilize the double distribution of dark matter halos 
which we derived and studied in the previous chapter, to study the
effect of the environmental factor on the statistical 
properties of cosmic accretion shocks. 
In order to explicitly distinguish between effects of the environment
and effects of the underlying distribution of accretor masses, we
explore two variations of our model. 
In the first, all collapsed cosmic structures are
assumed to live in a similar environment. The properties of the
accretion shock surrounding each such structure is then simply
determined by the mass of the structure and the redshift of
interest. The mass function of collapsed structures is, in this case,
the Press-Schechter mass function. In the context of the second model,
the distribution of collapsed structures with respect to both mass and
environment is described by the double distribution. The properties of
the accretion shock around each structure depends both on the mass of
the structure as well as the environment of the structure (whether the
structure resides inside an underdensity or an overdensity). 


The formalism describing  
the properties of a single accretion shock around a cosmic structure
is presented in section \ref{sshock}. The Mach number of the accretion
shock surrounding each collapsed object is derived from the
temperature jump across the shock. The temperature of the gas behind
the shock is simply taken to be the virial temperature of the
collapsed object. The temperature ahead of the shock depends on the
model used each time. For the Press-Schechter--based model, it is
simply the mean temperature of the intergalactic medium at the epoch
under consideration\footnote{Although we include the 
effect of reionization (the cosmic epochs we 
consider are post-reionization and we
take the temperature of the diffuse intergalactic gas at a density equal to the
cosmic mean to be $\sim 10^4$ K), we do not include filament
preheating in our models. Hence, we 
treat all accretion shocks as external shocks.}. 
For the double-distribution--based model, the
temperature also depends on the density of the environment of the
collapsed structure. The deviation from the mean intergalactic medium
temperature is calculated by assuming that it is only due to adiabatic
heating (cooling) because of the relative local (de)compression with respect
to the cosmic mean density.  
Once the Mach number of the accretion shock has bene determined, the
accretion velocity (in the shock frame) is then calculated from it,
and used to derive properties of the structure such as the accreted
mass current and the kinetic energy crossing the shock surface per
unit time.

Combining the properties of a single shock with an underlying
distribution of accretors (either the Press-Schechter mass function or
the double distribution of cosmic structures), in section
\ref{popshocks}
we derive the statistical properties of the population accretion
shocks. The calculated quantities include 
the distribution of accretor number density,
accretor mass, shock surface, mass current, and kinetic power crossing
the shocks, with respect to accretion shock Mach number. 
In addition, we calculate the cosmic history of the kinetic power
and of the mass current integrated over shock Mach number, as well as
the cumulative kinetic energy and number of baryons which have been
processed through accretion shocks by each redshift. 

Our results are presented in section \ref{results}. We find that the
shock environment alters the physical impact of shocks on the
intergalactic medium, as well as the cosmic history of the shock
population. The most prominent environmental effect is the
development, at low redshifts, of a natural bi-modality in the 
distribution, with respect to shock
Mach number, of the kinetic power crossing accretion
shocks. Concerning the physical impact of shocks on the intergalactic
medium, we find that the
cumulative energy input of accretion shocks by redshift $\sim 3$ is
comparable to the energy required to reionize the universe. In
addition, more than a third of all baryons in the universe have been
shocked in accretion processes by the present cosmic epoch.
Finally, we
comment on the components of the shock populations found in 
cosmological simulations \cite{min_shock,RKJ03} with which our models
are directly comparable.

\section{Properties of a single shock}\label{sshock}

\indent

Throughout this chapter, we assume an adiabatic equation of state, and
we consider all shocks to be non-radiative.  We also assume that any
individual collapsed object as well as its accretion shock are
spherically symmetric.
We will take the accretion shock position
around each structure to coincide with the virial radius of each
structure.

The Mach number of a shock, $\mach$, is defined as the ratio of the
velocity of the accreted material in the shock frame to the adiabatic sound
speed of the accreted material.
The Mach number is related to the temperature jump across the shock through
\cite{ll}
\begin{equation}\label{jumpgen}
\frac{T_2}{T_1}=\frac{\left[2\gamma \mach^2 -\gamma+1\right]
\left[(\gamma-1)\mach^2+2\right]}{(\gamma+1)^2\mach^2}
\end{equation}
where $T_1$ and $T_2$ are the pre-shock and post-shock temperatures
correspondingly, and $\gamma$ is the ratio of specific heats (assuming
that this remains constant across the shock). 
For a $\gamma=5/3$ gas, Eq.~(\ref{jumpgen}) becomes
\begin{equation}\label{jumpspec}
\frac{T_2}{T_1}=\frac{(5\mach^2-1)(\mach^2+3)}{16\mach^2}\,.
\end{equation}
In the limit $\mach \gg 1$ this equation is further simplified, 
\begin{equation}\label{highmachlim}
\mach = \sqrt{\frac{16}{5}\frac{T_2}{T_1}}\,.
\end{equation}
The pre-shock temperature can be written in terms of the adiabatic
sound speed of the pre-shock material $\cs$, 
\begin{equation}
kT_1 = \frac{\mu m_{\rm p}}{\gamma}\cs^2\,,
\end{equation}
where $\mu$ is the mean molecular weight of the accreted gas, and 
$m_{\rm p}$ is the proton mass.  If we also take 
$T_2$ to be the virial temperature of the accreting structure, 
\begin{equation}
 T_2 = T_{\rm vir}
= \frac{\mu m_{\rm p}}{k}\frac{Gm^{2/3}(4\pi f_{\rm c} \rho_{\rm
m,0})^{1/3}
(1+z)}{3^{1/3}5}\,,
\end{equation}
then the ratio $T_2/T_1$
becomes
\begin{eqnarray}\label{jumpvir}
\frac{T_2(m,z)}{T_1}&=&\frac{T_{\rm vir}(m,z)}{T_1}\nonumber \\
&=& 2.7 \times 10^3\Omega_{\rm m}\left(\frac{f_c}{18\pi^2}\right)^{1/3}
(1+z)\nonumber \\
&&\times \left(\frac{m}{m_8}\right)^{2/3}
\left(\frac{15 {\rm \,\, km \,s^{-1}}}{\cs}\right)^2\,,\nonumber \\
\end{eqnarray}
where $m$ is the object mass, $f_c = {\rho_{\rm vir}/\rho_{\rm m}}$
 is the compression factor for a virialized object (which
may vary with virialization redshift, depending on the cosmological
model), $z$ is the virialization redshift, 
$h$ is the dimensionless Hubble parameter, and $m_8 = 5.96 \times
10^{14} h^{-1} \Omega _{\rm m} \msol$ is the mass included in a sphere
of radius $r_8=8 h^{-1} {\, \rm Mpc}$ assuming the mean matter density
inside the sphere to be equal
to the cosmic mean.

In Appendix \ref{apa} we compare this result with the Bertschinger
similarity solution for an $\Omega_{\rm m}=1$ universe \cite{bert85},
 and we find it to be in excellent
agreement in the high-$\mach$ regime, where the Bertschinger solution
is applicable.

The surface area of a spherical shock around a structure of
mass $m$ is
\begin{equation}\label{surf}
S_1(m) = 4 \pi r_{\rm v}(m,z)^2
\end{equation}
where $r_{\rm v}$ is the virial radius of the structure, 
\begin{eqnarray} \label{rvir}
r_{\rm v}&=&
 1.4 h^{-1} {\rm \, Mpc }\times \nonumber \\
&&
\left(\frac{m}{m_8}\right)^{1/3}
\left(\frac{f_{\rm c}}{18\pi^2}\right)^{-1/3}
(1+z)^{-1}\,.
\nonumber \\
\end{eqnarray}

The mass current, defined as the rate at which mass crosses the
surface of a single accretion shock
around a structure of mass $m$ at an epoch $z$, is
\begin{eqnarray}\label{j1eq}
J_1 = \frac{dm}{dt} &=&  
4\pi r_{\rm v}^2(m) v_1\rho_{\rm b}(z) (1+\delta_{\rm s})\nonumber \\
&=& 4\pi r_{\rm v}^2(m)\Omega_{\rm b}\rho_{\rm c}(z) (1+\delta_{\rm s})
\mach c_{\rm s,1}
\end{eqnarray}
where $\rho_{\rm b}$ is the cosmic baryon density at the epoch of
interest and $(1+\delta_{\rm s})$ is the density enhancement 
(with respect to the cosmic mean) just outside the shock.

Finally, the kinetic power crossing
  the accretion shock around a single structure of mass $m$ is
\begin{equation}\label{p1eq}
P_1 = \frac{dE}{dt} = \frac{1}{2}\frac{dm}{dt}v_{1}^2 = 
 \frac{1}{2}J_1v_{1}^2 \,.
\end{equation}
Hence, 
\begin{equation}
P_1 = 2 \pi r_{\rm v}(m)^2 \Omega_{\rm b}\rho_{\rm c}(z)(1+\delta_{\rm
s}) \mach^3 c_{s1}^3 \,.
\end{equation}

\section{Properties of the population of Cosmic Accretion 
Shocks}\label{popshocks}

\indent

The quantities we will use to describe the statistical properties of the
population of cosmic accretion shocks are: 
\begin{itemize}

\item The ``number distribution'' of shocks with respect to Mach
      number. This is defined as the distribution of the comoving
      number density of accreting structures per
  logarithmic Mach number interval of their respective accretion
  shocks, 
\begin{equation}
\frac{dn}{d\ln \mach} = \mach \frac{dn}{d\mach}\,,
\end{equation}
with units number of structures per comoving 
${\rm Mpc^3}$.

\item The ``surface distribution'' with respect to Mach number.
This is defined as the shock surface area per
  logarithmic Mach number interval per comoving volume under consideration, 
\begin{equation}
\frac{dS}{d\ln \mach} = \mach \frac{dS}{d\mach}
\end{equation}
with units of
${\rm Mpc}^{-1}$ (since it represents a ratio of shock surface over space
volume).

\item The ``accretor mass'' distribution of shocks with respect to
      Mach number. This is defined as the distribution of the comoving 
      mass density in accreting structures per logarithmic Mach number
      interval of their respective accretion shocks, 
\begin{equation}
\frac{d\rho}{d\ln \mach} = \mach \frac{d\rho}{d\mach}\,,
\end{equation}
with units of $\msol$ per comoving ${\rm Mpc^3}$. 

\item The ``mass current distribution'' with respect to Mach number. 
This is defined as the  comoving mass current density crossing shock 
surfaces of
  logarithmic Mach number between $\ln \mach$ and $\ln \mach+d\ln
  \mach$,
\begin{equation}
\frac{dJ}{d\ln\mach} = \mach\frac{dJ}{d\mach}
\end{equation}
with units of $\msol {\rm \, yr^{-1}\, Mpc^{-3}}$.

\item The ``integrated mass current'', $J$, which is the comoving mass current
      density crossing shock surfaces of any Mach number at a given
      cosmic epoch, with units of $\msol {\rm \, yr^{-1}\, Mpc^{-3}}$.

\item the ``cumulative processed mass'', $\int _{t_{\rm i}}^t J dt$,
      which is the total mass density processed by shocks of any Mach
      number since some initial epoch $t_i$, expressed as a
      non-dimensional shocked baryon fraction. 

\item The ``kinetic power distribution'' with respect to Mach
      number. This is defined as the comoving kinetic power density crossing 
shock surfaces of
logarithmic Mach number between $\ln \mach$ and $\ln\mach+d\ln\mach$, 
\begin{equation}
\frac{dP}{d\ln \mach} =\mach \frac{dP}{d\mach}
\end{equation}
with units ${\rm erg \, s^{-1} \,}$ 
${\rm Mpc^{-3}}$.

\item The ``integrated kinetic power'', $P$, which is the comoving kinetic
      power density crossing shock surfaces of any Mach number at a
      given cosmic epoch, with units of ${\rm erg \, s^{-1} \,}$ 
${\rm Mpc^{-3}}$.

\item The ``cumulative processed kinetic energy'', $\int _{t_{\rm
      i}}^{t} P dt$, which is the total kinetic energy density 
      processed by shocks
      of any Mach number since some initial cosmic epoch $t_{\rm i}$,
      with units of ${\rm eV}$ per baryon in the universe. 

\end{itemize}

In order to explicitly identify the environmental effects on the
statistical properties of accretion shocks, we will use two different 
models to calculate these quantities. The
first will assume that all structures are accreting gas of a
single temperature, and that the population of accreting
objects is well described by the Press-Schechter mass function. In
this case, there exists a one-to-one correspondence between accretor 
mass and Mach number of the associated accretion shock. 

The second model assumes that the accreted gas has a distribution of
densities and hence temperatures (where adiabatic heating and cooling
are assumed to calculate the relation between local density and
temperature). The distribution of accretors with respect to both mass
and local over-(or under-)density is described by the double
distribution of cosmic structures. In this case, there is a
distribution of possible Mach numbers for the 
accretion shock around a structure of a given mass, depending on the
local overdensity of the accreted material.

Features of our results exclusive to the second model will then be an
effect of the environment in which accreting structures reside.

\subsection{Accreted Material of a Single Temperature}

\indent

If we assume that all collapsed 
objects accrete baryons of a single temperature, then at
a given redshift, all objects with an accretion shock of a given Mach number 
will have the same mass,
\begin{eqnarray}\label{mofmach1}
m = m(\mach, z) &=&4.2\times10^9 h^{-1} \, {\rm M_\odot} \times \nonumber \\
&&  \left(\frac{18\pi^2}{\Omega_{\rm m}
f_{\rm c}}\right)^{1/2}
\left(\frac{c_{\rm s}}{15 {\,\, \rm km \, s^{-1}}}\right)^3\ \times 
\nonumber \\ 
&&
\left[\frac{(5\mach^2-1)(\mach^2+3)}{16\mach^2}\right]^{3/2} 
(1+z)^{-3/2}\,.\nonumber \\
\end{eqnarray}
We will also assume that the mass distribution of collapsed objects
can be described by the Press-Schechter mass function \cite{ps74, lc93}
\begin{equation}
\frac{dn}{dm}(m,z) = \sqrt{\frac{2}{\pi}}\frac{\rho_{m,0}}{m^2}
\left|\frac{d\ln \sigma}{d\ln m }\right|
\exp\left\{-\frac{\left[\tilde{\delta}_c(z)\right]^2}{2\left[\sigma(m)\right]^2
}\right\} 
\end{equation}
where $\tilde{\delta}_c(z)$ is the linearly extrapolated overdensity
of an object which collapses at redshift $z$, $\sigma(m)$ is the
square root of the variance of the
linearly extrapolated field smoothed at a mass scale $m$, and $\rho_{m,0}$ is
the cosmic mean matter density at the present time.

In this case, the statistical quantities describing the population of accretion
shocks can be derived in a straight forward way.
The number distribution of shocks with respect to Mach number is 
\begin{equation}\label{ps1}
\frac{dn(\mach,z)}{d\ln \mach} =
\mach 
\frac{dn}{dm}\left[m(\mach,z),z\right]\frac{\partial m}{\partial
\mach}
(\mach,z)
\end{equation}
where $m=m(\mach,z)$ is given by Eq.~(\ref{mofmach1}).

Similarly, the surface distribution with respect to Mach number is 
\begin{equation}
\frac{dS}{d\ln \mach} = \mach \frac{dn}{dm}4\pi r_{\rm
  v}^2\left .\frac{\partial m}{\partial \mach}\right|_z
\end{equation}
while the accretor mass distribution is
\begin{equation}
\frac{d\rho}{d\ln\mach} = \mach \, m\frac{dn}{dm}
\left .\frac{\partial m}{\partial \mach}\right|_z\,.
\end{equation}
The mass current distribution in this model becomes 
\begin{equation}
\frac{dJ}{d\ln\mach}(\mach,z) = \mach 
\frac{dn}{dm}\left. \frac{\partial m}{\partial \mach}\right|_z
J_1\,,
\end{equation}
where $J_1(\mach,z)$ is given by Eq.~(\ref{j1eq}) with $\delta_s=3.13$
(the overdensity factor in the Bertschinger solution (see Appendix \ref{apa})
just outside the
shock), while the integrated mass current is
\begin{equation}
J(z) = 
\int_{\mach=1}^\infty \mach 
\frac{dn}{dm}\left. \frac{\partial m}{\partial \mach}\right|_z
J_1
d\mach \,,
\end{equation}
and the cumulative processed mass is 
\begin{equation}
\int _{t_{\rm i}} ^{t_0} J dt =
\int _{t_{\rm i}} ^{t_0}
\int_{\mach=1}^\infty \mach 
\frac{dn}{dm}\left. \frac{\partial m}{\partial \mach}\right|_z
J_1
d\mach dt \,,
\end{equation}
which, in a concordance cosmology becomes 
\begin{eqnarray}
\int _{t_{\rm i}} ^{t_0} J dt = \nonumber \\
\frac{1}{H_0}
\int_{0}^{z_{\rm i}} \int_{\mach=1}^\infty
\frac{dn}{dm}\left. \frac{\partial m}{\partial \mach}\right|_z
\frac{J_1 dz \, d\mach  }{(1+z)\sqrt{\Omega_\Lambda + \Omega_{\rm
m}(1+z)^3}}
\,,\nonumber \\
\end{eqnarray}
where $H_0$ is the Hubble parameter and 
$z_{\rm i}$ is the redshift corresponding to time $t_{\rm i}$. We use
$z_{\rm i}=10$.

Finally, the kinetic power distribution is
\begin{equation}\label{pslast}
\frac{dP}{d\ln \mach}(\mach,z) = 
\mach 
\frac{dn}{dm}\left. \frac{\partial m}{\partial \mach}\right|_z
P_1
\end{equation}
where $P_1(\mach,z)$ is given by Eq.~(\ref{p1eq}), while the
integrated kinetic power is
\begin{equation}
P(z) = 
\int_{\mach=1}^\infty \mach 
\frac{dn}{dm}\left. \frac{\partial m}{\partial \mach}\right|_z
P_1
d\mach \,,
\end{equation}
and the cumulative processed kinetic energy is
\begin{eqnarray}
\int _{t_{\rm i}} ^{t_0} P dt = \nonumber \\
\frac{1}{H_0}
\int_{0}^{z_{\rm i}} \int_{\mach=1}^\infty
\frac{dn}{dm}\left. \frac{\partial m}{\partial \mach}\right|_z
\frac{P_1 dz \, d\mach  }{(1+z)\sqrt{\Omega_\Lambda + \Omega_{\rm
m}(1+z)^3}}
\,.\nonumber \\
\end{eqnarray}

\subsection{Accreted Material of Varying Temperature}

\indent

In the second variation of our model, we wish to relax the assumption that the
temperature of the accreted material is the same for all structures. 
Assuming that adiabatic heating or cooling is 
the only process that causes deviations of the
temperature of the gas outside collapsed structures
from its mean value, we can
relate the local sound speed, $c_{\rm s}$, to the local overdensity or
underdensity, $\delta$, where 
\begin{equation}
\delta = \frac{\rho_{\rm local} - \rho_{\rm m}}{\rho_{\rm m}}\,.
\end{equation}
Since for adiabatic heating and cooling $c_{\rm s}^2\propto \rho_{\rm
  local}^{\gamma-1}$, we get
\begin{equation}\label{thecs}
c_{\rm s}= c_{\rm s,avg} \left(\delta+1\right)^{(\gamma-1)/2}\,, 
\end{equation}
where $c_{\rm s,avg}$ is the ``cosmic average'' sound speed 
(the sound speed of the intergalactic
medium at a density equal to the cosmic mean at
 the epoch of interest). This formalism can accommodate 
the case where a process (such as reionization)
heats the universe almost homogeneously, therefore
increasing the average temperature (and consequently the cosmic 
average sound speed
$c_{s, \rm avg}$).

In order to make further progress and be able to calculate measures of
the statistical properties of the population shocks in this
approximation, we need an analytical model for the {\em environment}
of collapsed structures. For this purpose, we will use the double
distribution (DD) of collapsed structures with respect to mass and local
overdensity which we derived and studied in the previous chapter. 
In the context of the double distribution, we defined the ``local
environment'' of a collapsed structure through the {\em clustering
  scale parameter}, $\beta$. The clustering scale parameter is a free
parameter in the DD model, and is 
defined so that  the ``environment'' of an object of mass $m$ be a
surrounding region in space which encompasses mass $\beta m$. 
The double distribution was found to be given by Eq.~(\ref{dd}).

In Eq.~(\ref{dd}), $\ed_\ell$ is the local linearly extrapolated
overdensity (or underdensity), which is related to the true 
(calculated from the
spherical evolution model) overdensity of the environment sphere {\em
  including the local object}, $\delta_{\ell}$ through the exact relations
given in chapter 2, 
or through the useful approximation represented by 
Eq.~(\ref{apconv}), which is accurate at
a better than $2\%$ level throughout its domain for all cosmologies of
interest.

Because in the problem of cosmic accretion shocks we are interested in 
the properties (density and sound speed) of the material right outside
the shock surface, we will adopt a small value for the clustering
scale parameter, $\beta=1.1$. Our results, however, are not sensitive
to the exact value of $\beta$ since, as we saw in chapter 2, the
properties of the double distribution (when
calculated as a function of $\delta$ rather than $\delta_\ell$) 
depend very mildly on $\beta$ for small values of $\beta$.

\subsubsection{Distribution of Sound Speeds}

\indent

From the double distribution and Eq.~(\ref{thecs}), 
we can immediately calculate
the number density of collapsed objects of mass
$m>m_{\rm min}$ embedded 
in a medium of local sound speed between $c_{\rm s}$ and 
$c_{\rm s}+d c_{\rm s}$, 
\begin{equation}\label{ncs}
\frac{dn}{dc_{\rm s}}dc_{\rm s} = 
dc_{\rm s} \frac{d\delta}
{d c_{\rm s}} \int_{m=m_{\rm min}}
^{\infty} \frac{dn}{dmd\delta} dm\,,
\end{equation}
Similarly, the density of matter in collapsed objects of mass $>m_{\rm
min}$ embedded in a medium of local sound speed between $c_s$ and $c_s
+d c_s$ is
\begin{equation}\label{rcs}
\frac{d\rho}{d c_{\rm s}}d c_{\rm s} = 
d c_s \frac{d \delta}
{d c_{\rm s}} \int_{m=m_{\rm min}}^{\infty} m 
\frac{dn}{dmd\delta} dm\,.
\end{equation}
In Eqs.~(\ref{ncs}) and (\ref{rcs}), $d\delta/dc_{\rm s}$ can be found
from Eq.~(\ref{thecs}),
\begin{equation}
\frac{d\delta}{dc_{\rm s}} = \frac{2}{\gamma-1}\frac{1}{c_{\rm s,
    avg}}\left(\frac{c_{\rm s}}{c_{\rm
    s,avg}}\right)^\frac{3-\gamma}{\gamma-1}\,. 
\end{equation}
Hence, if we use $\dcs$ to denote the sound speed in units of the
cosmic average sound speed, 
\begin{equation}
\dcs \equiv \frac{c_{\rm s}}{c_{\rm s,avg}}\,,
\end{equation}
Eq.~(\ref{ncs}) becomes
\begin{equation}
\frac{dn}{d\dcs} = \frac{2}{\gamma-1}
    \dcs^\frac{3-\gamma}{\gamma-1}
\int_{m=m_{\rm min}}
^{\infty} \frac{dn}{dmd\delta} dm\,,
\end{equation}
and Eq.~(\ref{rcs}) becomes
\begin{equation}
\frac{d\rho}{d\dcs} = \frac{2}{\gamma-1}
    \dcs^\frac{3-\gamma}{\gamma-1}
\int_{m=m_{\rm min}}
^{\infty} m \frac{dn}{dmd\delta} dm\,.
\end{equation}

\subsubsection{Shock Properties}

\indent

In this second variation of our model, the mass of a collapsed object with an
associated accretion shock of Mach $\mach$ is also dependent on
the local overdensity $\delta$, 

\begin{eqnarray}\label{massofmach1}
m = m(\mach,\delta,z)&=&4.2 \times 10^9 h^{-1} \, {\rm M_\odot}  
 \left(\frac{18\pi^2}{f_{\rm c}\Omega_{\rm m}}\right)^{1/2} 
\times \nonumber \\
&& \left(\frac{c_{\rm s,avg}}{15 {\,\, \rm km \, s^{-1}}}\right)^3
\left(\delta+1\right)^{3(\gamma-1)/2}\times \nonumber \\
&&\left[\frac{(5\mach^2-1)(\mach^2+3)}{16\mach^2}\right]^{3/2}
(1+z)^{-3/2}\,.\nonumber \\
\end{eqnarray}
Note that, when $\delta=0$, this equation is reduced to
Eq.~(\ref{mofmach1}), as it should.

Hence, the number distribution of shocks with respect to Mach number 
is
\begin{equation}
\frac{dn}{d\ln \mach} = \mach \int_{\delta=-1}^{\delta_c}
d\delta\frac{dn}{dm d\delta}\left[m(\mach, \delta, z),\delta, z \right]
\left.\frac{\partial m}
{\partial \mach}\right|_{\delta,z}
\end{equation}
where $m(\mach, \delta, z)$ is given by Eq.~(\ref{massofmach1}).

Similarly, the surface distribution with respect to Mach number is
\begin{equation}
\frac{dS}{d\ln \mach} = \mach \int_{\delta=-1}^{\delta_c}
d\delta 4\pi r_{\rm v}^2 \frac{dn}{dm d\delta}\left.\frac{\partial m}{\partial
  \mach}\right|_{\delta,z}\,,
\end{equation}
while the accretor mass distribution is
\begin{equation}
\frac{d\rho}{d\ln \mach} = \mach \int_{\delta=-1}^{\delta_c}
d\delta \, m\frac{dn}{dm d\delta}\left.\frac{\partial m}{\partial
  \mach}\right|_{\delta,z}\,.
\end{equation}

The mass current distribution is 
\begin{equation}
\frac{dJ}{d\ln\mach}=\mach \int_{\delta=-1}^{\delta_c} d\delta
J_1 \frac{dn}{dm d\delta}\left.\frac{\partial m}{\partial
  \mach}\right|_{\delta,z}\,,
\end{equation}
where $J_1(\mach, \delta, z)$ given by Eq.~(\ref{j1eq}). In this case,
$\delta_s = \delta$, the local overdensity outside the shock as given
by the DD. The integrated mass current is then 
\begin{equation}
J(z) = \int _{\mach=1}^\infty \int_{\delta=-1}^{\delta_c} d\mach \,
d\delta \, 
J_1 \frac{dn}{dm d\delta}\left.\frac{\partial m}{\partial
  \mach}\right|_{\delta,z}\,,
\end{equation}
and the cumulative processed mass is 
\begin{eqnarray}
\int_{t_{\rm i}}^{t_0} J dt = \nonumber \\ \int _{z_{\rm 1}}^0
\int _{\mach=1}^\infty \int_{-1}^{\delta_c} 
\frac{ d\mach \, d\delta\, dz \,
J_1 \frac{dn}{dm d\delta}\left.\frac{\partial m}{\partial
  \mach}\right|_{\delta,z}
}{H_0(1+z)\sqrt{\Omega_\Lambda+\Omega_{\rm m}(1+z)^3}}
\,. \nonumber \\
\end{eqnarray}

Finally, the kinetic power distribution is
\begin{equation}
\frac{dP}{d\ln \mach} = \mach \int_{\delta=-1}^{\delta_c} d\delta
P_1\frac{dn}{dm d\delta}\left.\frac{\partial m}{\partial
  \mach}\right|_{\delta,z}\,,
\end{equation}
with $P_1(\mach, \delta, z)$ given by Eq.~(\ref{p1eq}), while the
integrated kinetic power is
\begin{equation}
P(z) = \int _{\mach=1}^\infty \int_{\delta=-1}^{\delta_c} d\mach \,
d\delta \, 
P_1 \frac{dn}{dm d\delta}\left.\frac{\partial m}{\partial
  \mach}\right|_{\delta,z}\,,
\end{equation}
and the cumulative processed kinetic energy is
\begin{eqnarray}
\int_{t_{\rm i}}^{t_0} P dt = \nonumber \\ \int _{z_{\rm 1}}^0
\int _{\mach=1}^\infty \int_{-1}^{\delta_c} 
\frac{ d\mach \, d\delta\, dz \,
P_1 \frac{dn}{dm d\delta}\left.\frac{\partial m}{\partial
  \mach}\right|_{\delta,z}
}{H_0(1+z)\sqrt{\Omega_\Lambda+\Omega_{\rm m}(1+z)^3}}
\,. \nonumber \\
\end{eqnarray}

\section{Results}\label{results}

\subsection{Distribution of Environmental Sound Speeds}

\indent

\begin{figure*}
\resizebox{3in}{!}{
\includegraphics{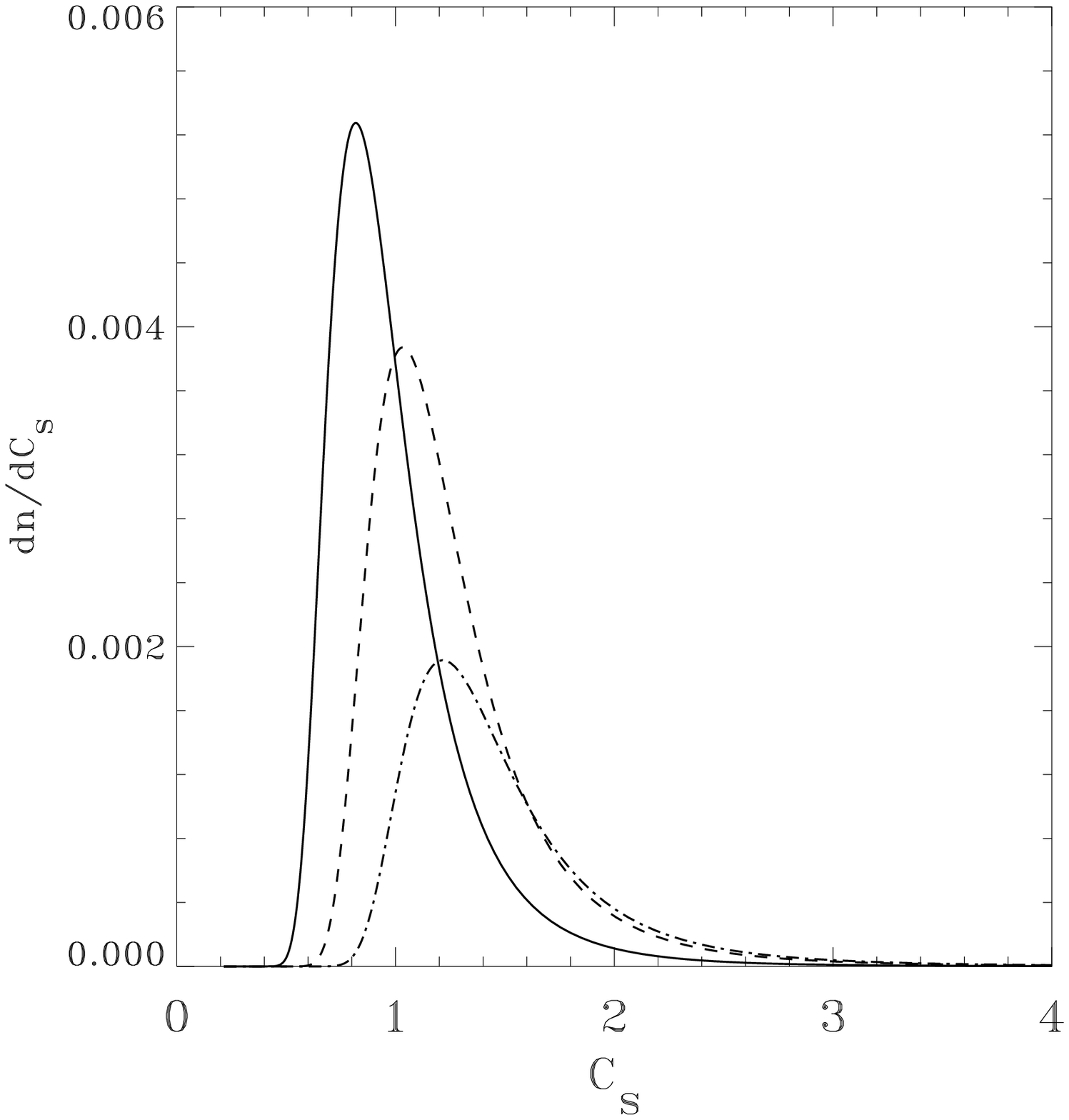}}
\resizebox{3in}{!}{
\includegraphics{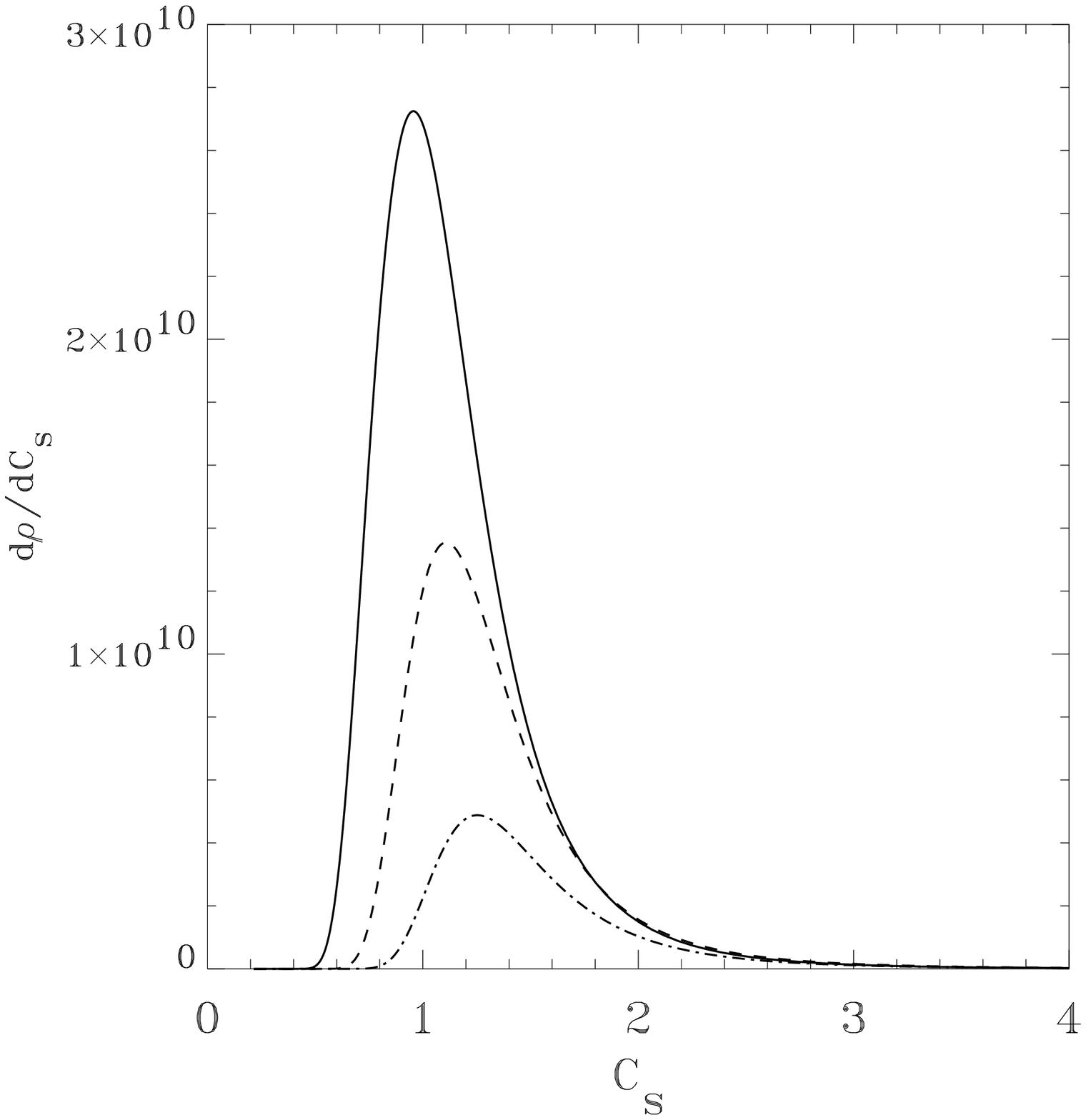}}
\caption{\label{fig:csdists} Distribution of number of objects (left panel, in 
units of objects per comoving Mpc$^3$) and mass density (right panel, in units of 
${\rm M_\odot \, Mpc^{-3}}$) per $C_{\rm s}$ interval, for objects with mass 
$>10^{12} {\rm \, M_\odot}$, in a WMAP concordance universe. 
Solid line: $z=0$; dashed line: $z=1$; dot-dashed line: $z=2$.}
\end{figure*}

In Fig.~\ref{fig:csdists} we plot the distribution of sound speeds of
the material in which collapsed objects of mass greater than $10^{12}
{\, \rm M_\odot}$ are embedded, for different redshifts. 
The left panel shows the comoving
number density of objects per interval of the dimensionless sound
speed $\dcs$, $dn/d\dcs$, and the right panel shows the mass-density
counterpart $d\rho/d\dcs$. All curves correspond to a concordance {\it Wilkinson
  Microwave Anisotropy Probe} (WMAP) universe ($\sigma_8=0.84$,
$h=0.71$, $\Omega_{\rm m}=0.27$ and $\Omega_{\rm b}=0.04$, \cite{Sper03}).
As expected from the results of chapter 2, which show that the most
probable density contrast of the material surrounding collapsed structures
decreases with time, the most probable sound speed of the material
surrounding collapsed structures of mass $>10^{12} {\, \rm M_\odot}$
increases with redshift. The overall suppression of the curves with
increasing redshift is due to the fact that fewer structures more
massive than a fixed cutoff have had enough time to collapse at large
redshifts. At the present epoch, the most probable sound speed for the
masses under consideration is only a fraction of the cosmic  average
sound speed, due to the fact that structures in this range of masses
are more likely to be found inside underdensities. 
Note however that the picture presented here is incomplete, as shock heating
in filaments will further modify the sound speed distribution,
favoring higher values of the sound speed.

\subsection {Properties of Accretion Shocks: Effects of the Local Environment}

\indent

In this section we again assume a WMAP concordance universe. We
focus on post-reionization redshifts, hence we assume a
cosmic average sound speed of
$15 {\rm \, km/s}$, corresponding to a temperature of $\sim 10^4$K,
for a fully ionized plasma with $\mu=0.59$ ($25\%$ He by mass). 

\begin{figure*}
\resizebox{1.95in}{!}{
\includegraphics{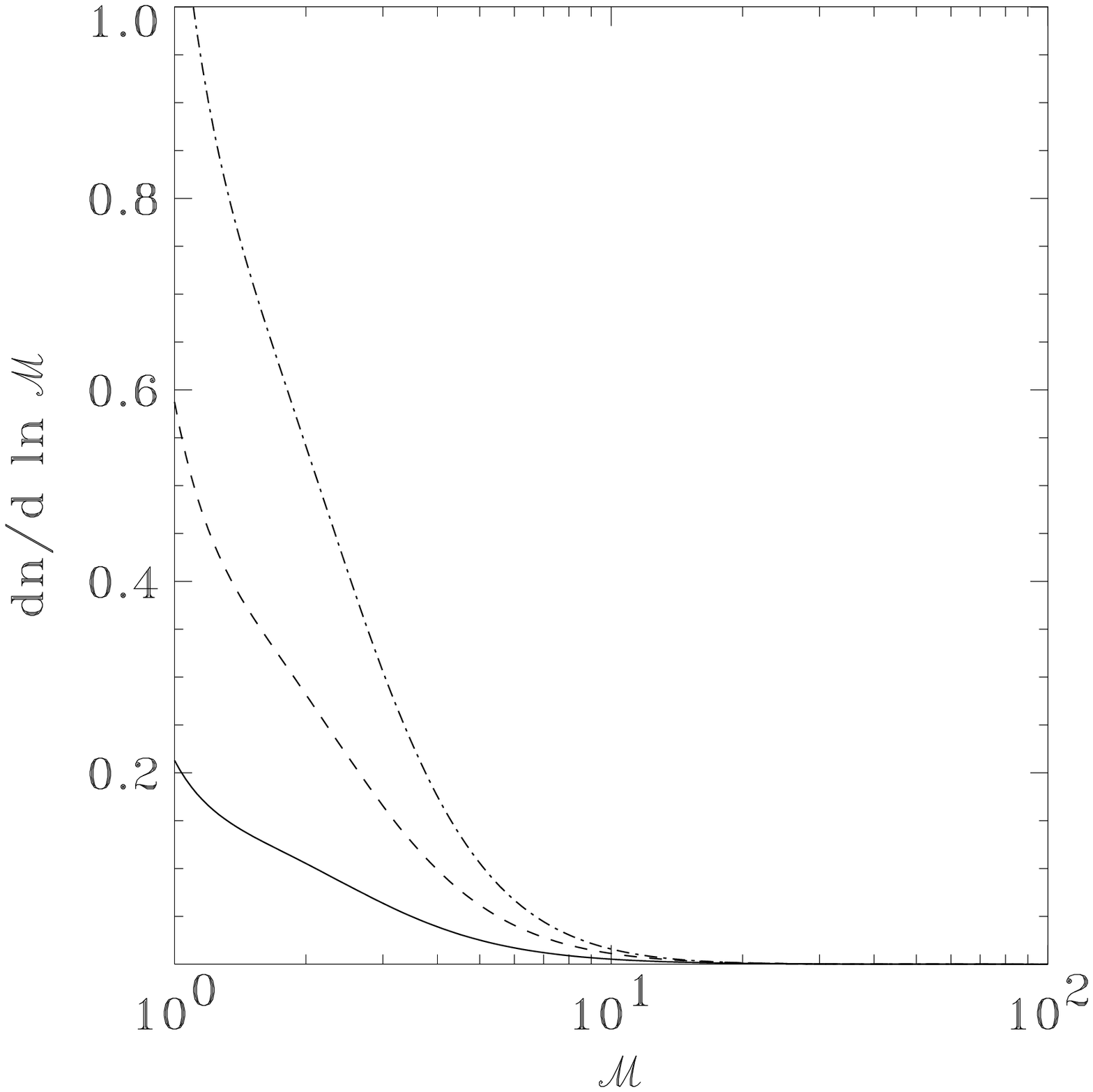}}
\resizebox{1.95in}{!}{
\includegraphics{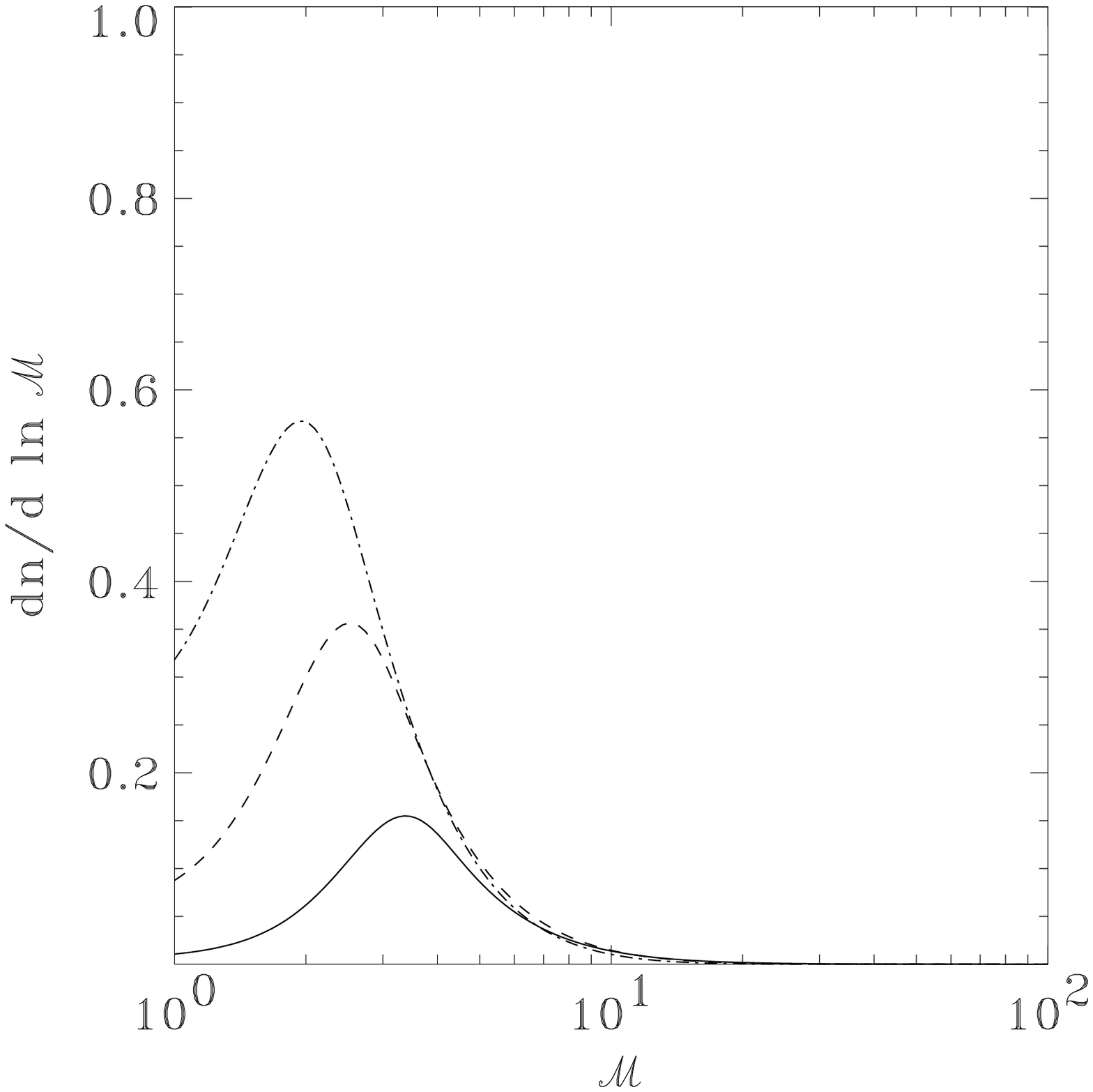}}
\resizebox{1.95in}{!}{
\includegraphics{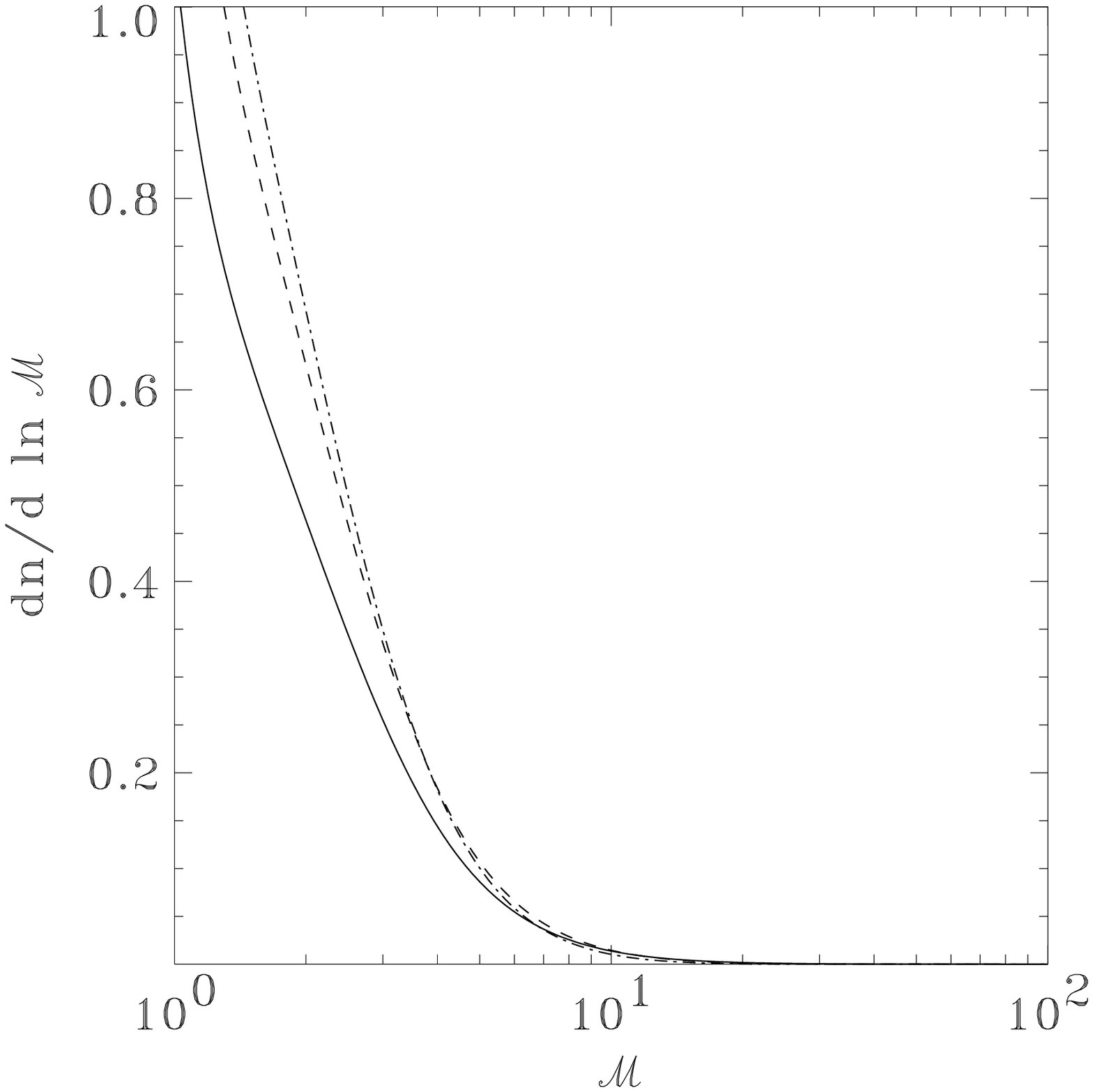}}
\caption{\label{fig:machdists} Number distribution of shocks 
per logarithmic Mach number interval for the
Press-Schechter--based model (left panel) and the double-distribution--based
models (middle and left panels). The middle panel models implement a 
halo mass cutoff identical to that of the Press-Schechter case, while 
the right panel model assumes no mass cutoff. 
The units of the vertical axes are number of
objects per comoving Mpc$^3$.
Solid line: $z=0$; dashed line: $z=1$; dot-dashed line: $z=2$.}
\end{figure*}

In Fig.~\ref{fig:machdists} we plot the number distribution of shocks with
respect to Mach number,
$dn/d\ln \mach$, for the Press-Schechter--based (hereafter PS) model 
(left panel) and two implementations of the double-distribution--based
(hereafter DD) model (middle and right panel). 

The distribution in the PS model monotonically
increases for decreasing $\mach$ since as we have discussed in this
case there is a one-to-one correspondence between mass and Mach
number, hence the number distribution of shocks simply follows the
Press-Schechter mass function. The requirement $\mach \ge 1$ imposes
then a cutoff in the mass of the dark mater halos that can act as a
host to an accretion shocks. This mass cutoff is the mass of the
object whose virial temperature exactly equals the temperature of the
diffuse gas, and it is $m_{\rm min} = 8.1\times 10^9 \msol$ at $z=0$, 
 $m_{\rm min} = 3.8\times 10^9 \msol$ at $z=1$ and  $m_{\rm min} =
 2.2\times 10^9 \msol$ at $z=2$. The decrease of $m_{\rm min}$ with
 increasing redshift is due to the fact that the virial temperature of
 an object of a fixed mass is larger if the object virializes at high
 redshift than if it virialized today. In an Einstein-deSitter Universe
 this is a simple result of the higher cosmic matter density in the
 past (the compression factor of a virialized object is constant in
 time, so since the cosmic density is larger in the past, the virial
 density which is the compression factor times the cosmic density is
 also larger in the past, resulting to a higher virial
 temperature). In the concordance universe, there are two competing
 factors mediating this effect. The cosmic matter density is still
 larger in the past, however the compression factor is now itself
 redshift-dependent, and increases with decreasing redshift. However,
 this increase of the compression factor is not steep enough to counteract
 the decrease in cosmic density, and still the virial density
 decreases with decreasing virialization redshift. 

The middle panel of Fig.~\ref{fig:machdists} shows a variation
of our DD model which includes only objects of
a mass larger than the mass cutoff of the PS model in the
corresponding redshift. The mass cutoff, in combination with the
double-distribution--imposed distribution of pre-shock sound speeds for each
accretor mass, results in a $\mach$ number distribution which peaks at
$\mach \lesssim 4 $, a position which is defined by the combination of
the virial temperature at the mass cutoff (which is the most populated
available mass bin) and the sound speed
corresponding to the most probable environment (as given
by the double distribution) at that particular mass. 
If the additional constraint $\mach\ge 1$ did not exist, then the area
below the curves corresponding to the same redshift in the left and 
middle panels would be the same, as it would represent simply the
total number of objects with mass larger than the mass
cutoff. However, some of these objects are ``lost'' in the double
distribution model because they are embedded in material of
temperature equal or larger than their virial temperature and hence
they cannot harbor shocks. This effect is more pronounced at higher
redshifts, while at $z=0$ where very few objects are ``lost'' from the
distribution as described above, the areas below the two solid lines in
the left and middle panels are almost equal. 

The right panel of Fig.~\ref{fig:machdists} shows the
DD model where the only restriction in the
participating objects is the requirement that $\mach \ge 1$. In this
case, the distribution does not turn over, and it is in fact at an
overall higher level that the corresponding PS
model. This is because the DD model tends to move
smaller objects to higher $\mach$ bins, because in the double-distribution
picture smaller objects preferentially reside inside underdense
regions. Therefore, smaller objects
 accrete cooler gas and harbor shocks of
higher $\mach$ than in the Press-Schechter case. When no mass cutoff
is imposed, smaller objects (that are excluded in the Press-Schechter
picture because their virial temperatures are below the cosmic mean
temperature) ``leak'' inside the $\mach$ distribution, because in the
double-distribution picture they can lie inside an underdense region
(void) and accrete material cooler than the cosmic mean and cooler
than their own confined gas. As a result, a larger total number of
objects participate in the double distribution and the overall level
of the curves is higher than in the PS model.

Note that the effect described above is moderated by the fact that
some objects also ``leak'' out of the distribution, because they lie
in overdense regions where the diffuse gas has a temperature larger
than their virial temperature (it is this effect that caused the
``loss of signal'' that is observed in the middle-panel curves).

\begin{figure}
\resizebox{5.5in}{!}{
\includegraphics{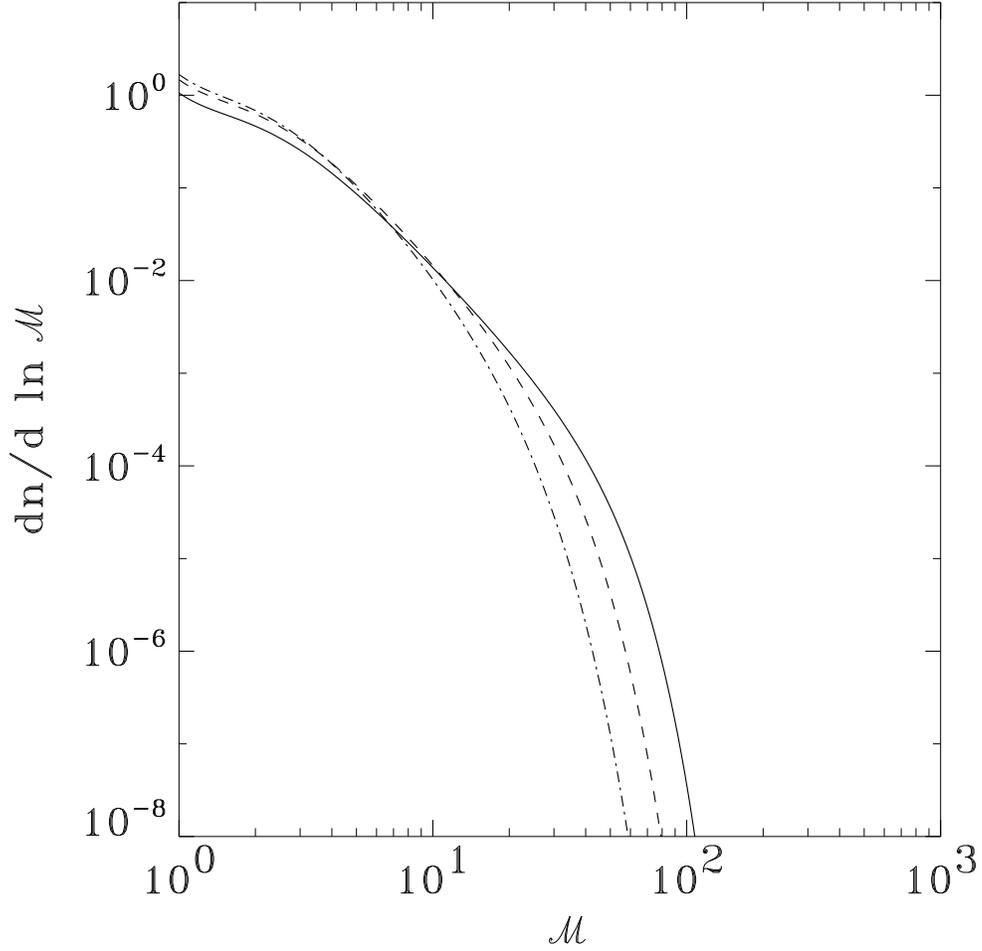}}
\caption{\label{fig:machloglog} Distribution of number density of objects
per logarithmic Mach number interval for the double-distribution--based
model without a mass cutoff, plotted in logarithmic scale.
The units of the vertical axes are number of
objects per comoving Mpc$^3$.
Solid line: $z=0$; dashed line: $z=1$; dot-dashed line: $z=2$.}
\end{figure}

In all panels of Fig.~\ref{fig:machdists} the amplitude of the
distribution increases with increasing redshift. This is because the
number distribution plotted here is dominated by the low-mass objects,
which are merged into larger halos as time progresses. In 
Fig.~\ref{fig:machloglog} we plot the double distribution without cutoff
model (right panel of Fig.~\ref{fig:machdists}) using a logarithmic
scale for the vertical axis, so that the behavior of the large-mass
(and consequently high-$\mach$) objects can be better
illustrated. At these high $\mach$s, the behavior of the distribution
with redshift is reversed, as more high-mass objects are formed as
time progresses. 

\begin{figure*}
\resizebox{1.95in}{!}{
\includegraphics{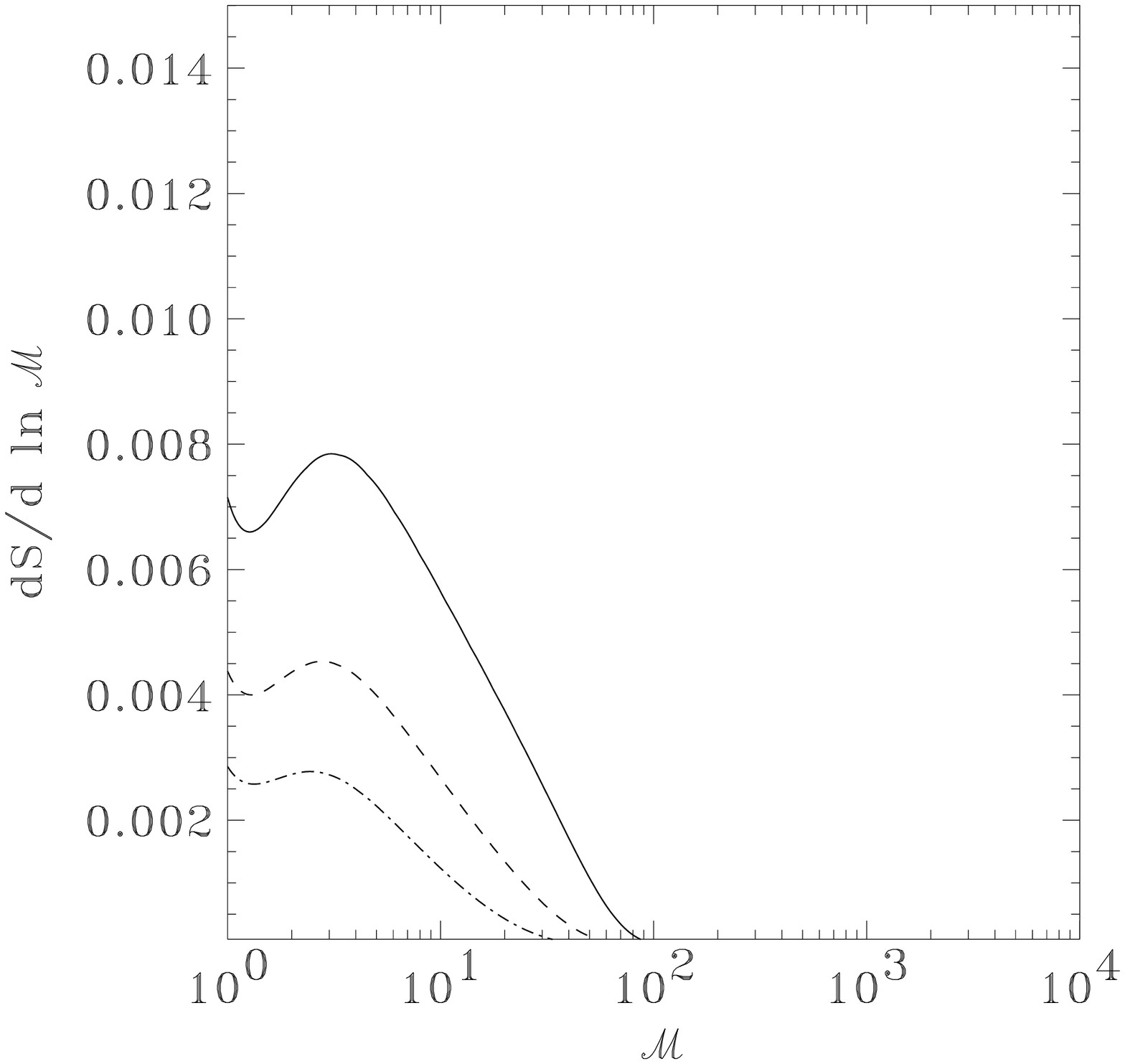}}
\resizebox{1.95in}{!}{
\includegraphics{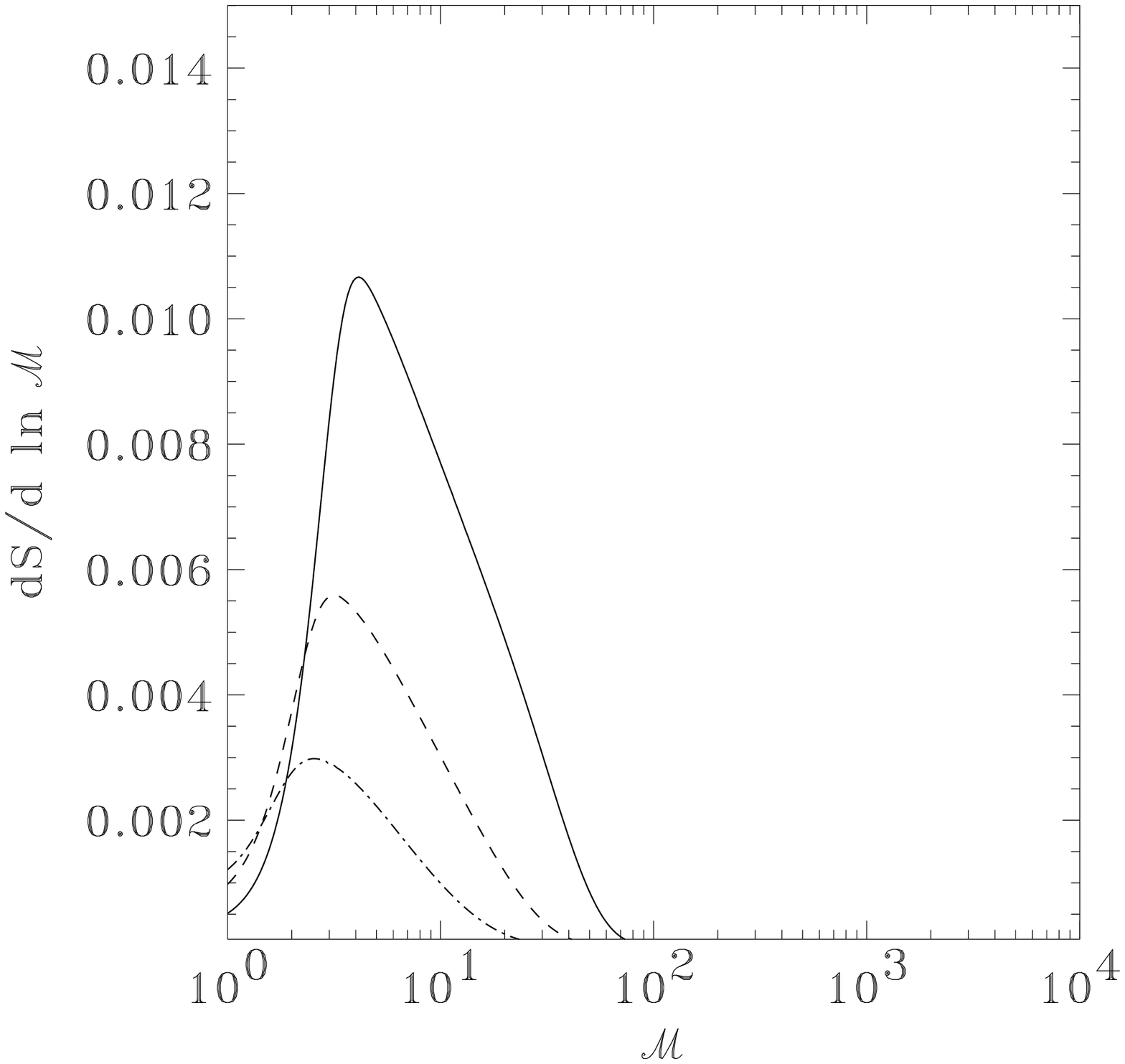}}
\resizebox{1.95in}{!}{
\includegraphics{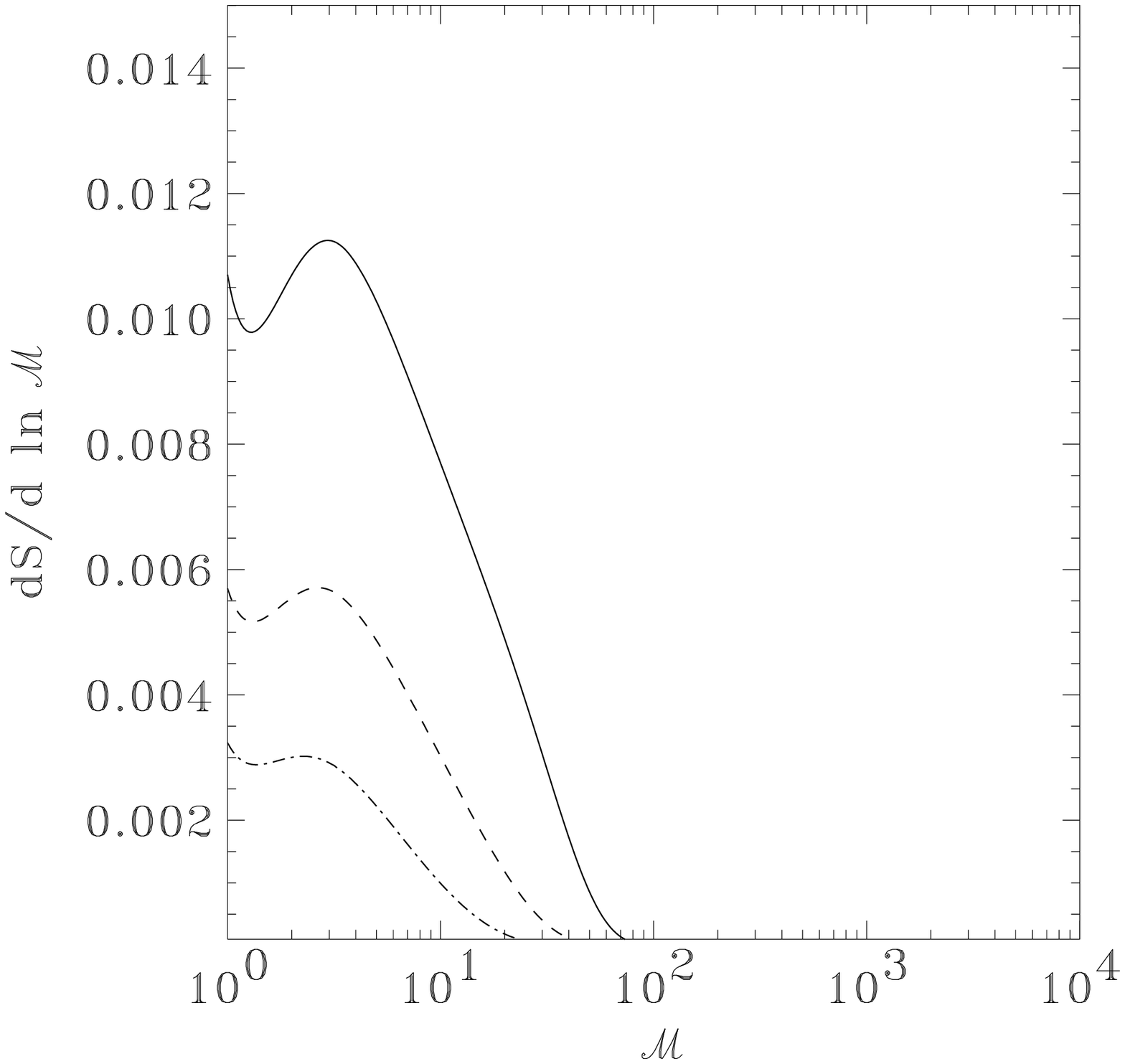}}
\caption{\label{fig:surfdists} Distribution of spatial density of
shock surface  
per logarithmic Mach number interval ($dS/s\ln\mach$)for the
Press-Schechter--based (left panel) and the double-distribution--based
(right panel) models. The units of the vertical axes are comoving Mpc$^{-1}$.
Solid line: $z=0$; dashed line: $z=1$; dot-dashed line: $z=2$.}
\end{figure*}

Figure \ref{fig:surfdists} shows the distribution of spatial density of
shock surface  per logarithmic Mach number interval $dS/d\ln\mach$,
with $(dS/d\ln\mach)d\ln \mach$ being
the total surface of shocks in a certain comoving volume $V$ with 
logarithmic Mach number between $\ln \mach$ and $\ln \mach + d\ln
\mach$ divided by that volume. 

This quantity is frequently used to
characterize the statistical properties of shocks in cosmological
simulations as it can be calculated without need for identification of
collapsed structures and assignment of shock-hosting
gridpoints to specific structures. The left panel shows this
distribution for the PS model while the middle and
right panels
correspond to the DD models, with (middle panel) and
without (right panel) mass cutoffs. The behavior of this quantity is
similar to the number distribution with respect to $\mach$ in that it
is dominated by the low-mass structures, and therefore the existence
of a low-mass cutoff significantly affects its qualitative
behavior.

Although the number and surface distributions of shocks 
with respect to Mach number are dominated by the low-mass objects and
their corresponding low--Mach-number shocks, the accretor mass, the
mass accretion rate and the energetics of shocks are dominated by
high-mass objects. 
This can be immediately verified by simple analytic arguments. We
consider for simplicity the Press-Schechter--based variation of our
model. For masses high enough that the primordial density fluctuation 
power spectrum can be regarded as a power law but low enough that the 
exponential mass cutoff is not affecting the results,  
$dn/dm \propto m^{-2}$. In addition, $\mach \propto m^{1/3}$ (in the
high-$\mach$ limit), 
while $r_{\rm v} \propto m^{1/3}$.
Hence, $J_1 \propto m$ and $P_1 \propto m^{5/3}$. Equations
(\ref{ps1})-(\ref{pslast}) then give for the low-mass dependence of
the various statistical properties of the shock population, 
\begin{eqnarray}
\frac{dn}{d\ln \mach} &\propto &m^{-1} \nonumber \\
\frac{dS}{d\ln \mach} &\propto &m^{-1/3} \nonumber \\
\frac{d\rho}{d\ln\mach}& \propto& m^0 \nonumber \\
\frac{dJ}{d\ln \mach} &\propto &m^0 \nonumber \\
\frac{dP}{d\ln \mach} &\propto& m^{2/3}\,
\end{eqnarray}
accounting for the difference in the low-mass behavior of different
distributions, in particular the divergent behavior of the number
distribution and the convergent nature of the mass, mass current and
kinetic power distributions. 
Note that the low-mass suppression is somewhat stronger
than what is predicted by the simple arguments above, due to the deviation
of $\mach(m)$ from $m^{1/3}$ for $\mach \rightarrow 1$.

\begin{figure*}
\resizebox{1.95in}{!}{
\includegraphics{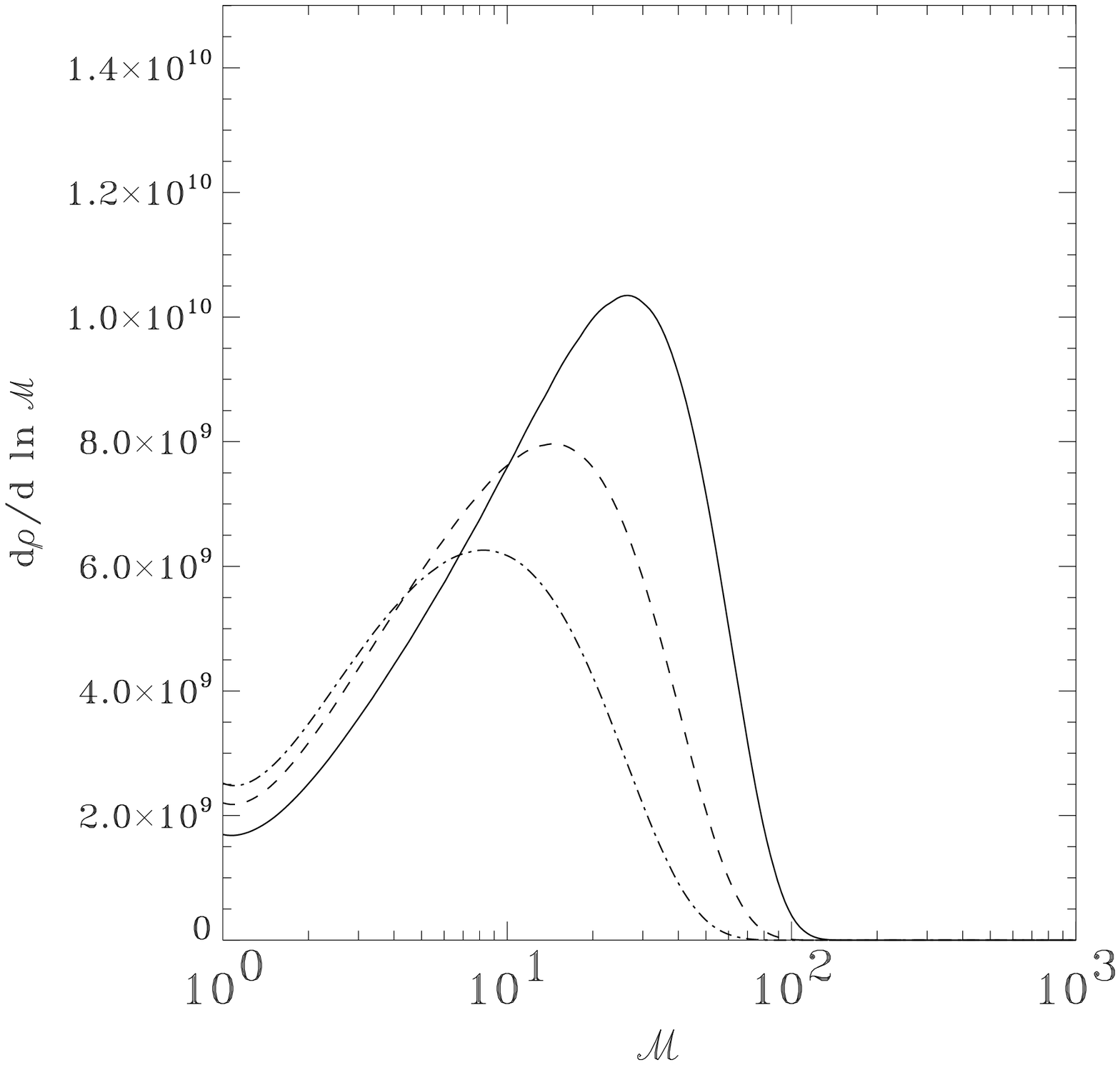}}
\resizebox{1.95in}{!}{
\includegraphics{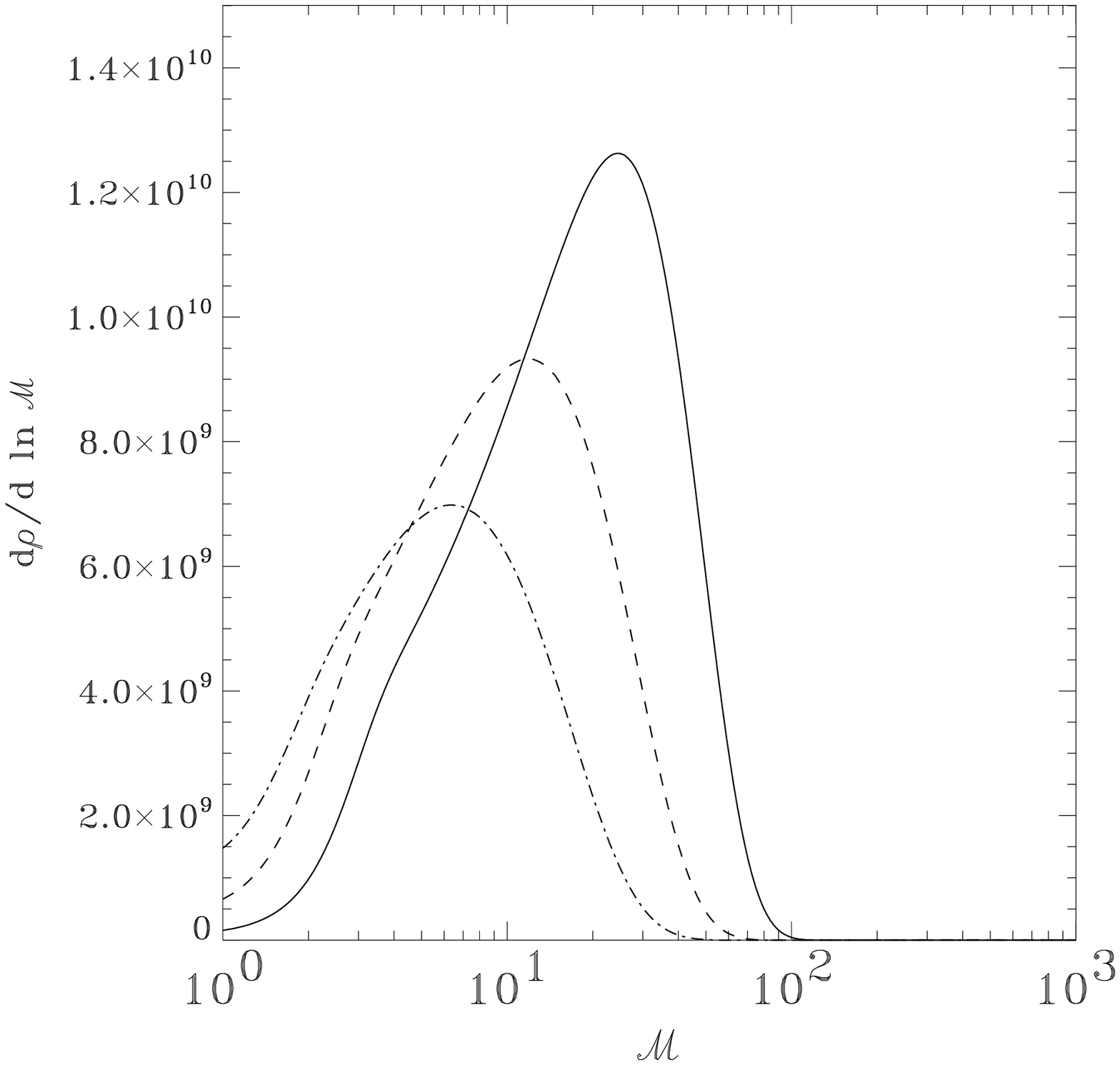}}
\resizebox{1.95in}{!}{
\includegraphics{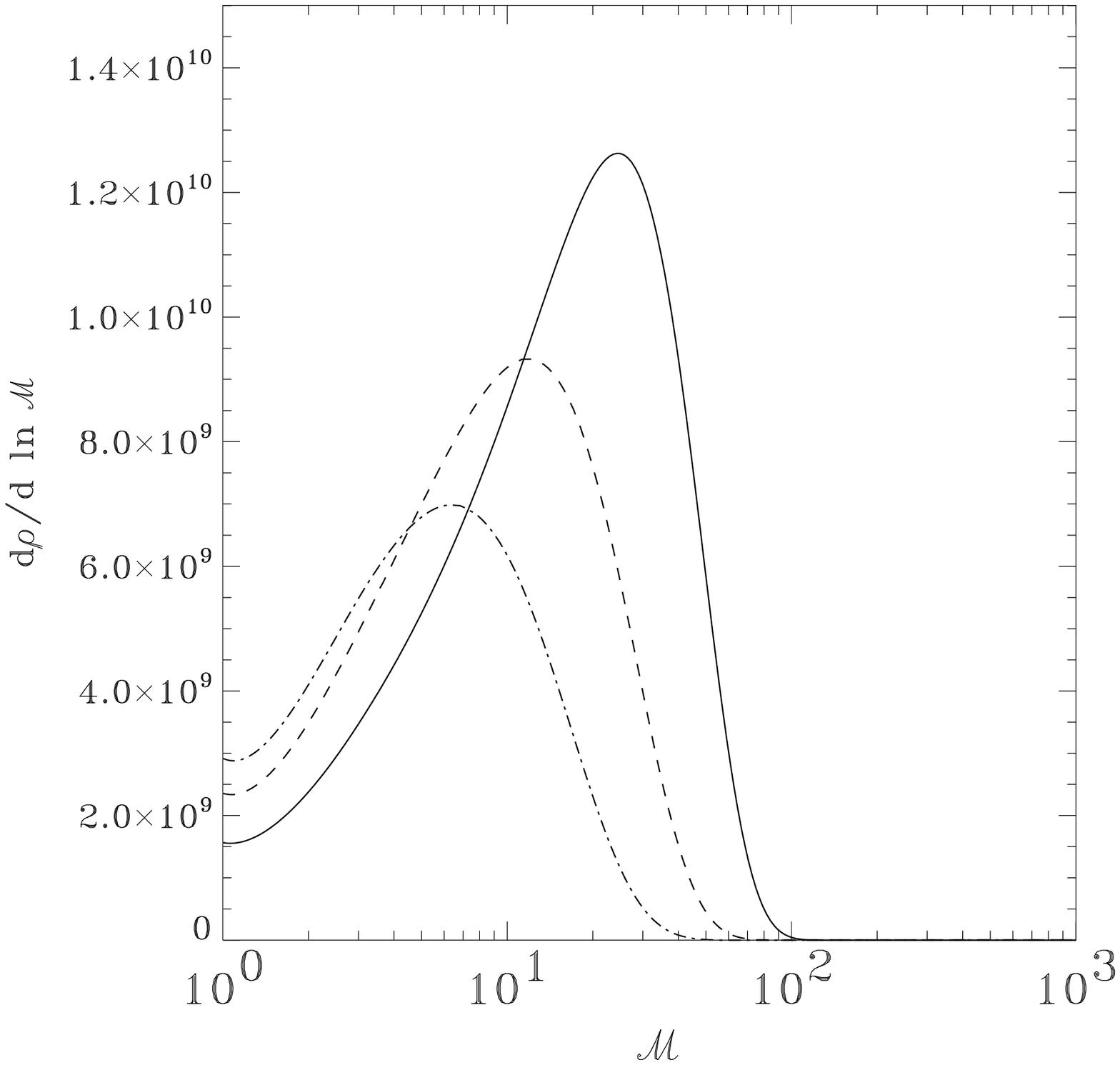}}
\caption{\label{fig:machmass} Accretor mass distribution (mass density
in accretors 
per logarithmic Mach number interval of associated accretion shock, 
$d\rho/d\ln\mach$)for the
Press-Schechter--based (left panel) and the double-distribution--based
models, with a Press-Schechter like mass cutoff (middle panel) and
without an explicit mass cutoff (right panel). 
The units of the vertical axes are ${\rm \msol
 \, comoving \, Mpc^{-3}}$.
Solid line: $z=0$; dashed line: $z=1$; dot-dashed line: $z=2$.}
\end{figure*}

In Fig.~\ref{fig:machmass} we plot the accretor mass
distribution with respect to Mach number, i.e. the mass in objects with
accretion shocks of logarithmic Mach number between  $\log \mach$ and
$\log \mach+d\log \mach$ versus $\mach$. All models exhibit similar
behavior, with the high-$\mach$ objects dominating the
distribution. In this case, the presence or not of a mass cutoff in
the DD models does not have a significant effect, as
it only affects the low-$\mach$ end of the distribution. The location
of the peak is at similar values of $\mach$ in all models, and is an
effect of the high-mass exponential cutoff in collapsed accreting
structures. The effect of a varying environmental overdensity and
sound speed in this case is to increase the height as well as
steepness of decline toward both higher and lower Mach numbers of
the distribution peak.
This is because the double distribution tends to move 
lower-mass objects (which
generally reside in underdensities) towards higher $\mach$s, and
higher-mass objects (which reside in overdensities) towards lower
$\mach$s, causing a greater concentration of accretor mass close to
the peak.

\begin{figure*}
\resizebox{1.95in}{!}{
\includegraphics{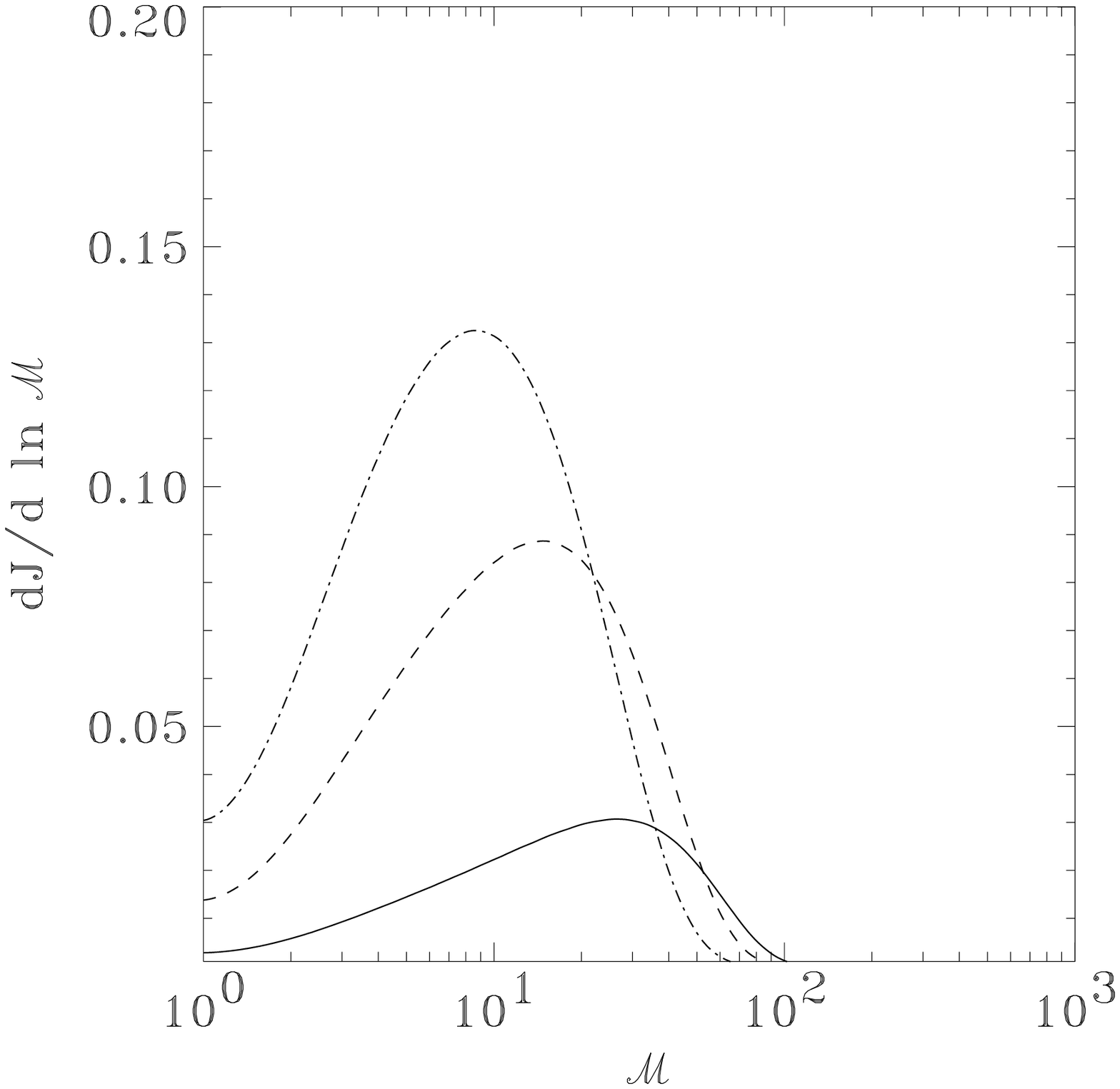}}
\resizebox{1.95in}{!}{
\includegraphics{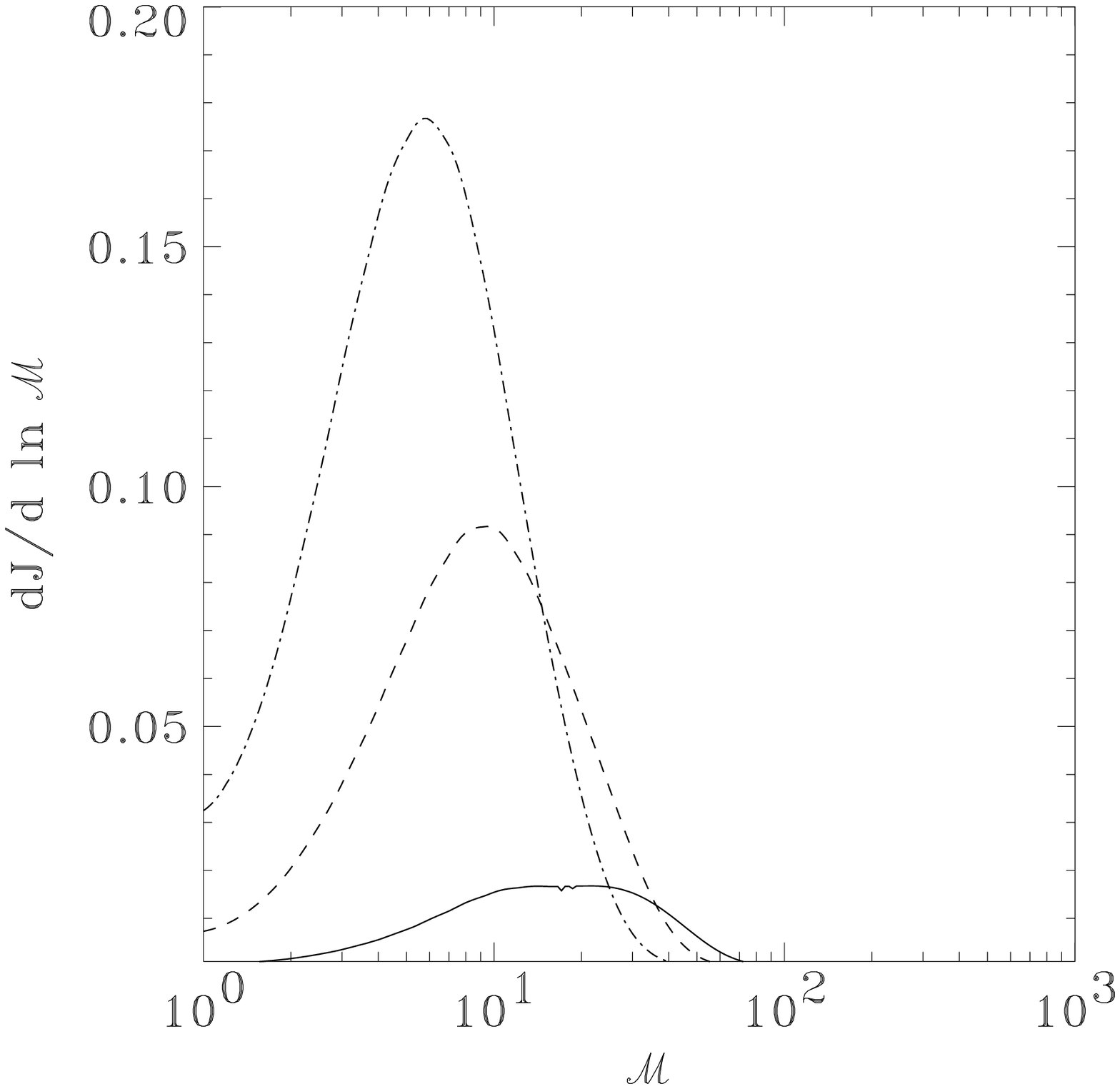}}
\resizebox{1.95in}{!}{
\includegraphics{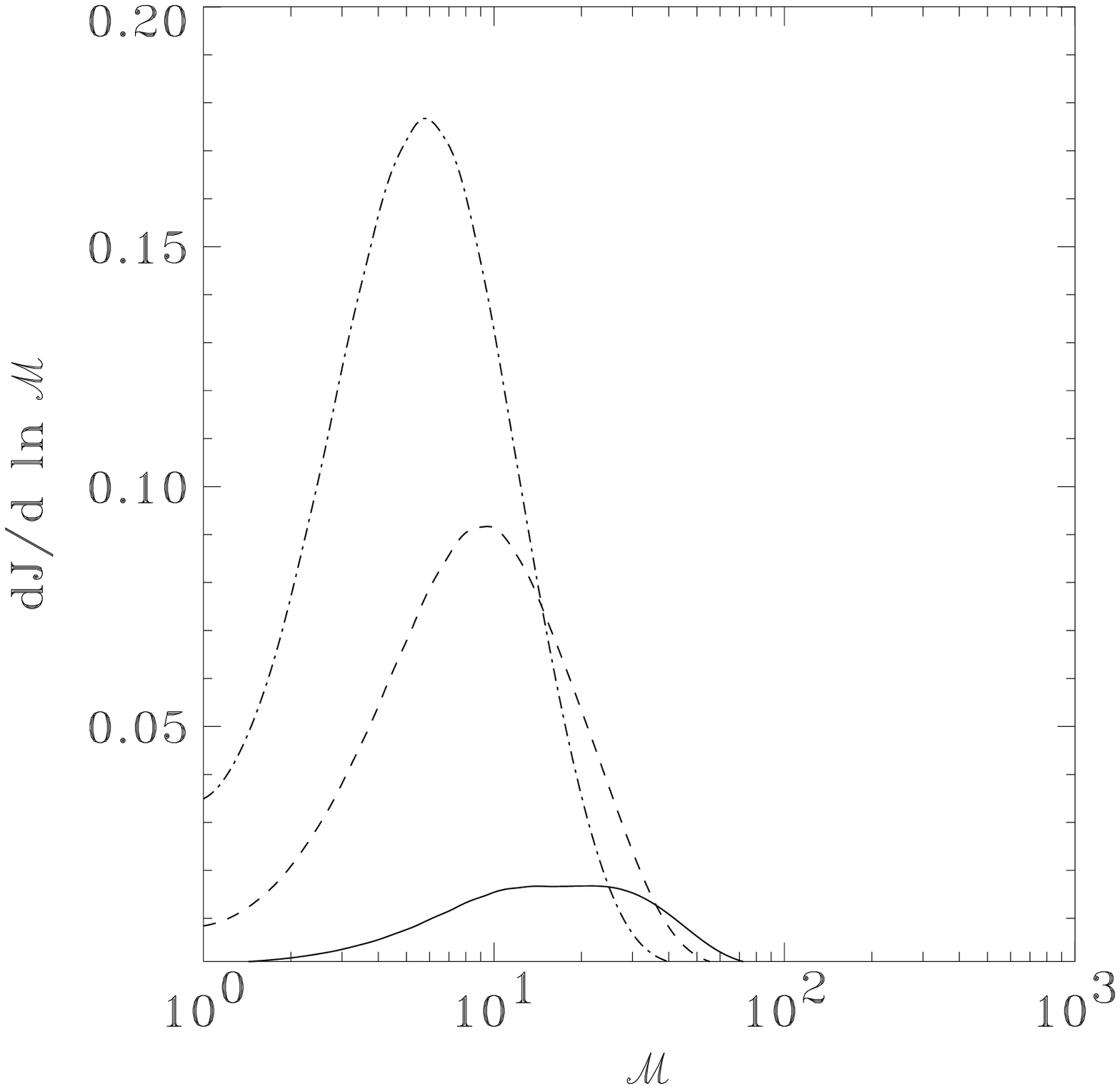}}
\caption{\label{fig:mcdists} Mass current distribution (spatial density of
 mass current
per logarithmic Mach number interval ($dJ/d\ln\mach$)for the
Press-Schechter--based (left panel) and the double-distribution--based
(right panel) models. The units of the vertical axes are ${\rm \msol
 \, comoving \, Mpc^{-3} \, yr}$.
Solid line: $z=0$; dashed line: $z=1$; dot-dashed line: $z=2$.}
\end{figure*}

In Fig.~\ref{fig:mcdists} we plot the mass current distribution, 
$dJ/d\ln \mach$. Again,
the left panel shows the results for the
PS model, the middle panel represents the
DD model with a Press-Schechter mass cutoff and the
right panel is the DD model without a mass cutoff and
with the only restriction being $\mach \ge 1$. 

The environmental effects are
more pronounced in this case, while the existence or not of a mass
cutoff plays no significant role in the properties of the DD
model, again due to the overwhelming dominance of high-mass
objects in the total mass processing rate through accretion
shocks: the existence or not of a low-mass cutoff cannot affect the
mass current distribution because the low-$\mach$ objects process only a
very small fraction of the total accreted mass. 

The effect of taking into account the environmental overdensity 
distribution using the DD in this case is two-fold: On the one hand, 
it spreads out every mass bin to a larger $\mach$ range as in the
previous cases. On the other hand, it
also adjusts the local density of the material just outside
the object, which in turn affects the local value of the mass
current. 

The location of the mass current distribution peak and its evolution
with redshift do not change appreciably between the two models. The
location of the peak represents the high-mass exponential cutoff in the
Press-Schechter mass function (the largest mass scale which has
collapsed by a certain epoch), modulated by the nonlinear mass-$\mach$ 
relation. 
However, the DD model declines more sharply towards lower $\mach$. 
This is because in the
double-distribution picture, low-mass, low-$\mach$ objects reside inside
underdensities and hence they can process less mass than in the
Press-Schechter case. 

Finally, the suppression of the distribution amplitude with decreasing
redshift in all models is a result of the reduction of the mean cosmic
density of the accreted material. The result is more pronounced in the
DD models since, in addition to the reduction of the
mean density, there is also a shift of the most
probable environmental density for structures of a given mass towards
regions increasingly underdense with respect to the cosmic mean as
time increases. 

\begin{figure*}
\resizebox{1.95in}{!}{
\includegraphics{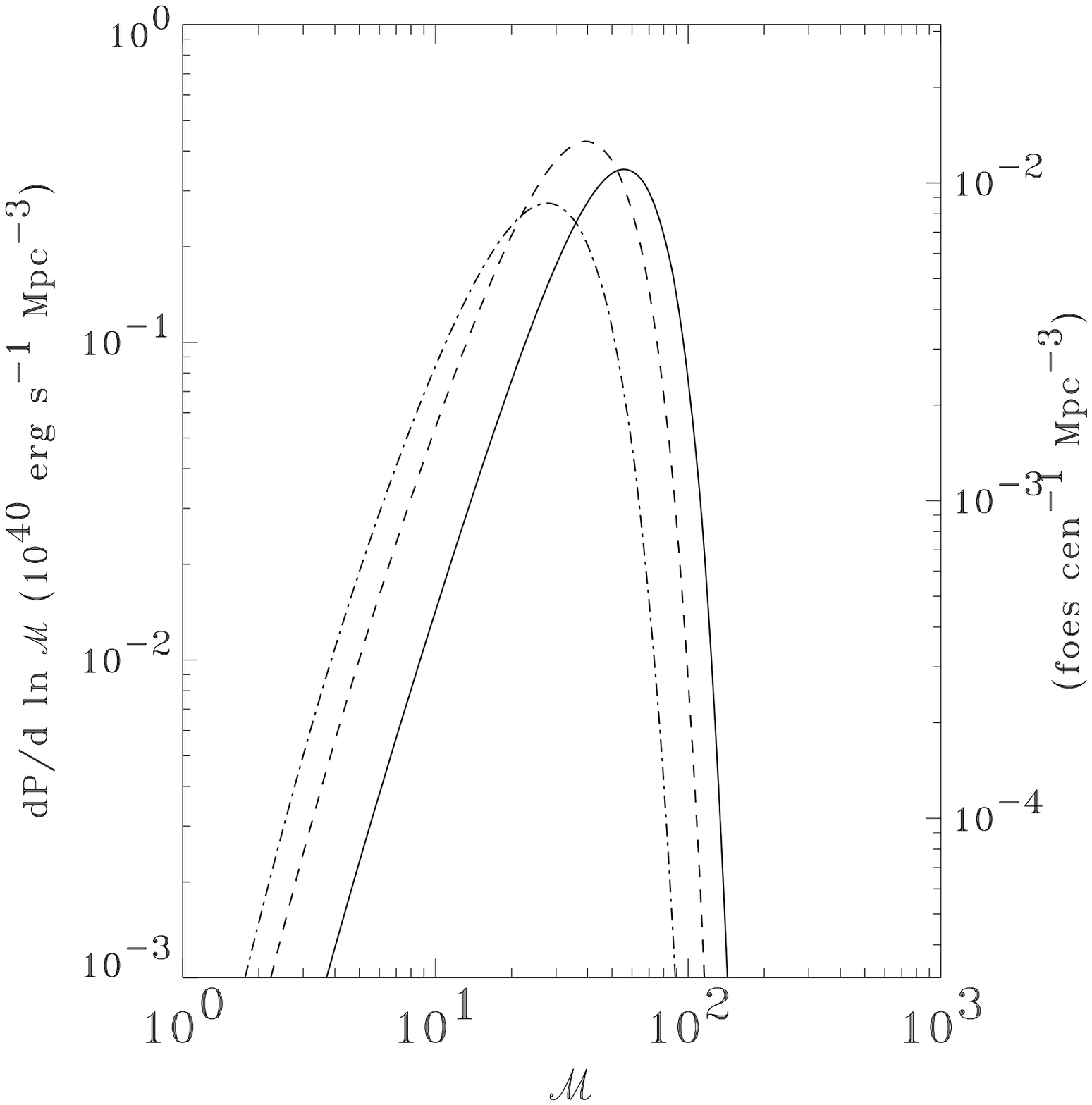}}
\resizebox{1.95in}{!}{
\includegraphics{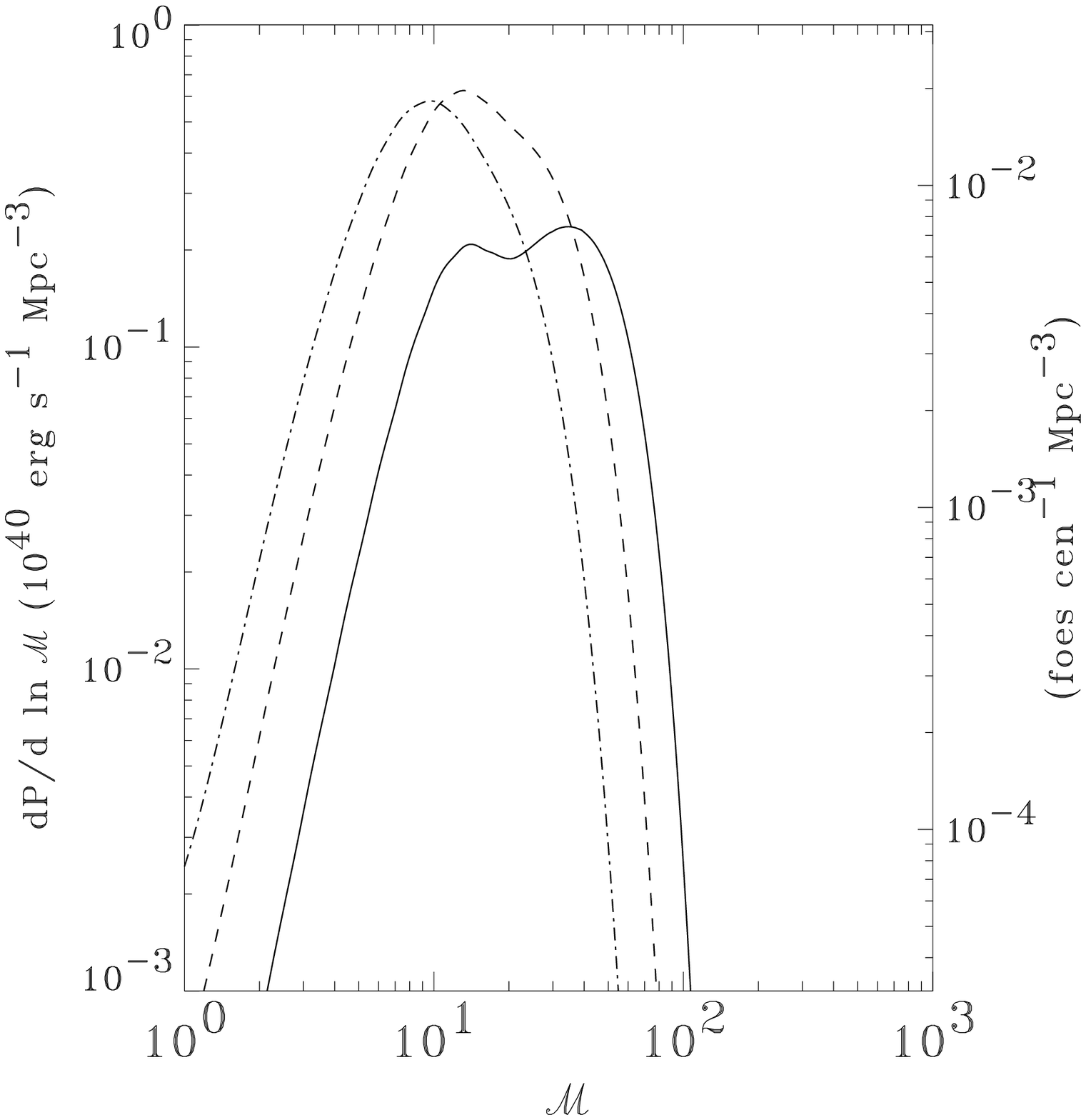}}
\resizebox{1.95in}{!}{
\includegraphics{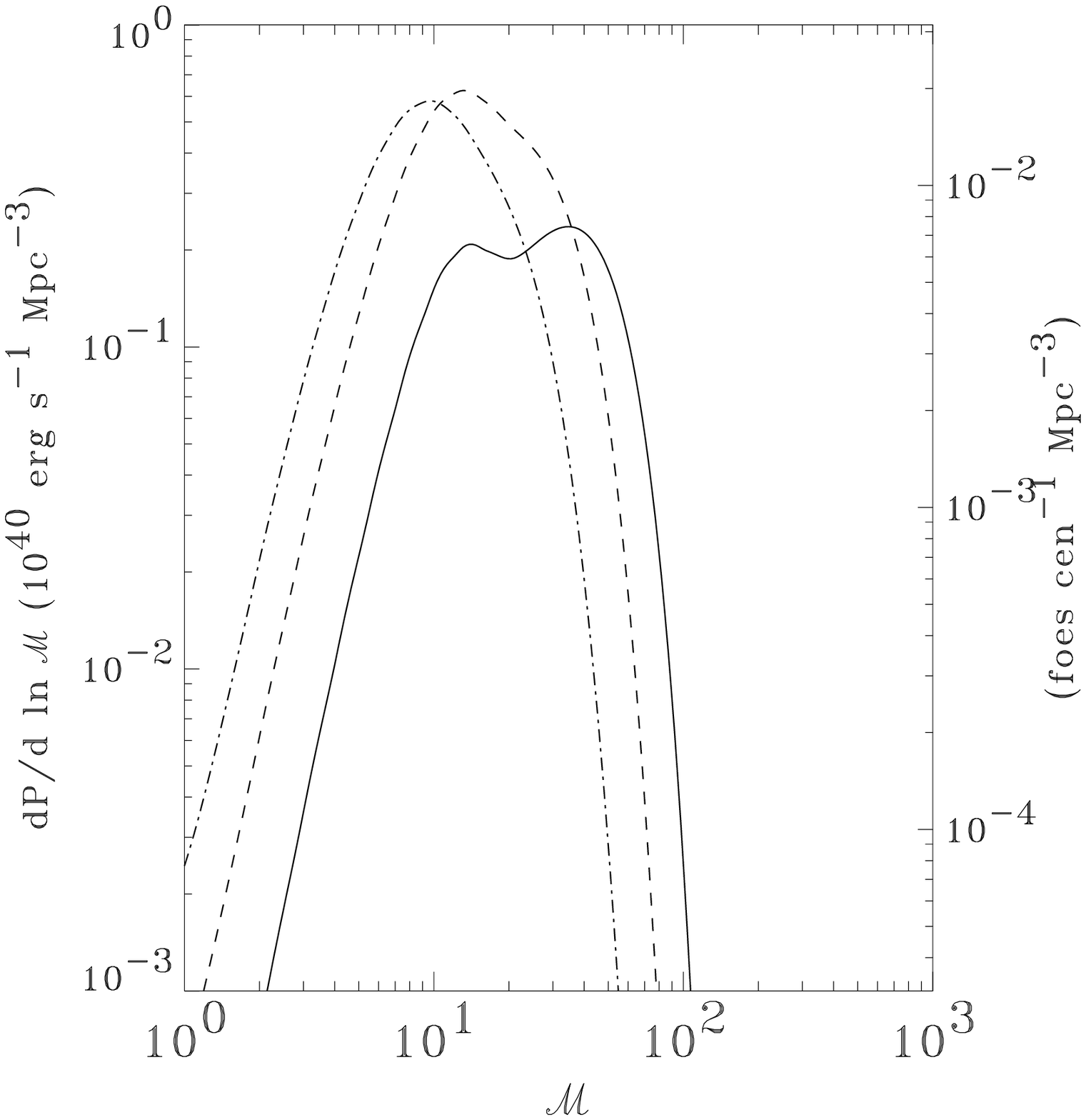}}
\caption{\label{fig:kindists} 
Kinetic power distribution (spatial density of
  kinetic power processed by accretion shocks
per logarithmic Mach number interval, $dP/d\ln\mach$) for the
Press-Schechter--based (left panel) and the double-distribution--based
(right panel) models. The units of the vertical axes are ${10^{40}\rm ergs \,\,
 s ^{-1}}$ ${\rm \, comoving \, Mpc^{-3}}$ (left axis) and $10^{51}$ ergs per 
century per 
comoving ${\rm Mpc^3}$ (right axis).
Solid line: $z=0$; dashed line: $z=1$; dot-dashed line: $z=2$.}
\end{figure*}

In Fig.~\ref{fig:kindists} we plot the kinetic power distribution with
respect to Mach number, for the
PS model, and the two variations of the 
DD model. As in the case of the mass current
distribution, the kinetic power distribution is also dominated by
objects of high mass and Mach number, and the presence of a
lower-mass cutoff does not have a significant effect on the
DD models.

\begin{figure*}
\resizebox{1.95in}{!}{
\includegraphics{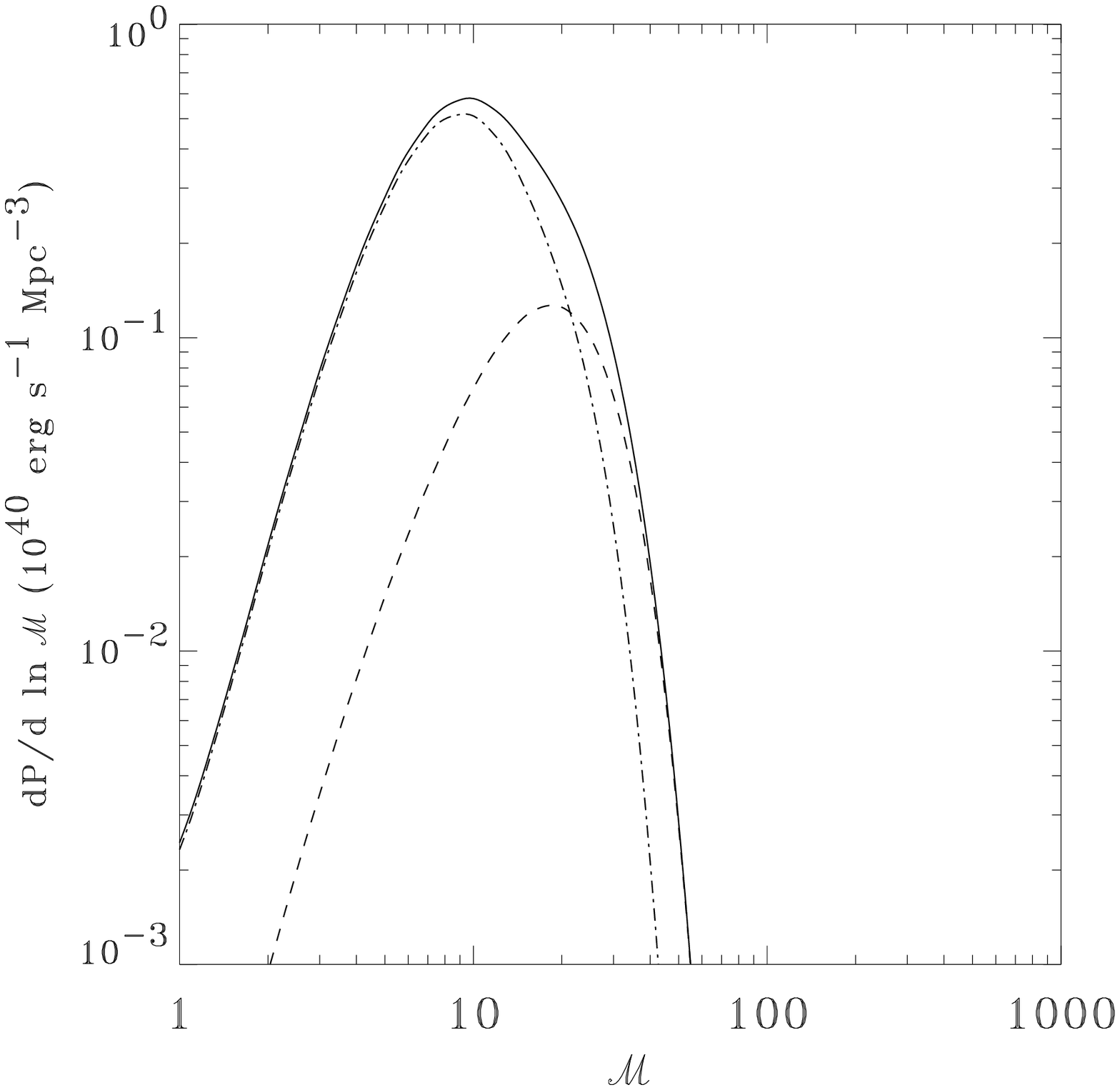}}
\resizebox{1.95in}{!}{
\includegraphics{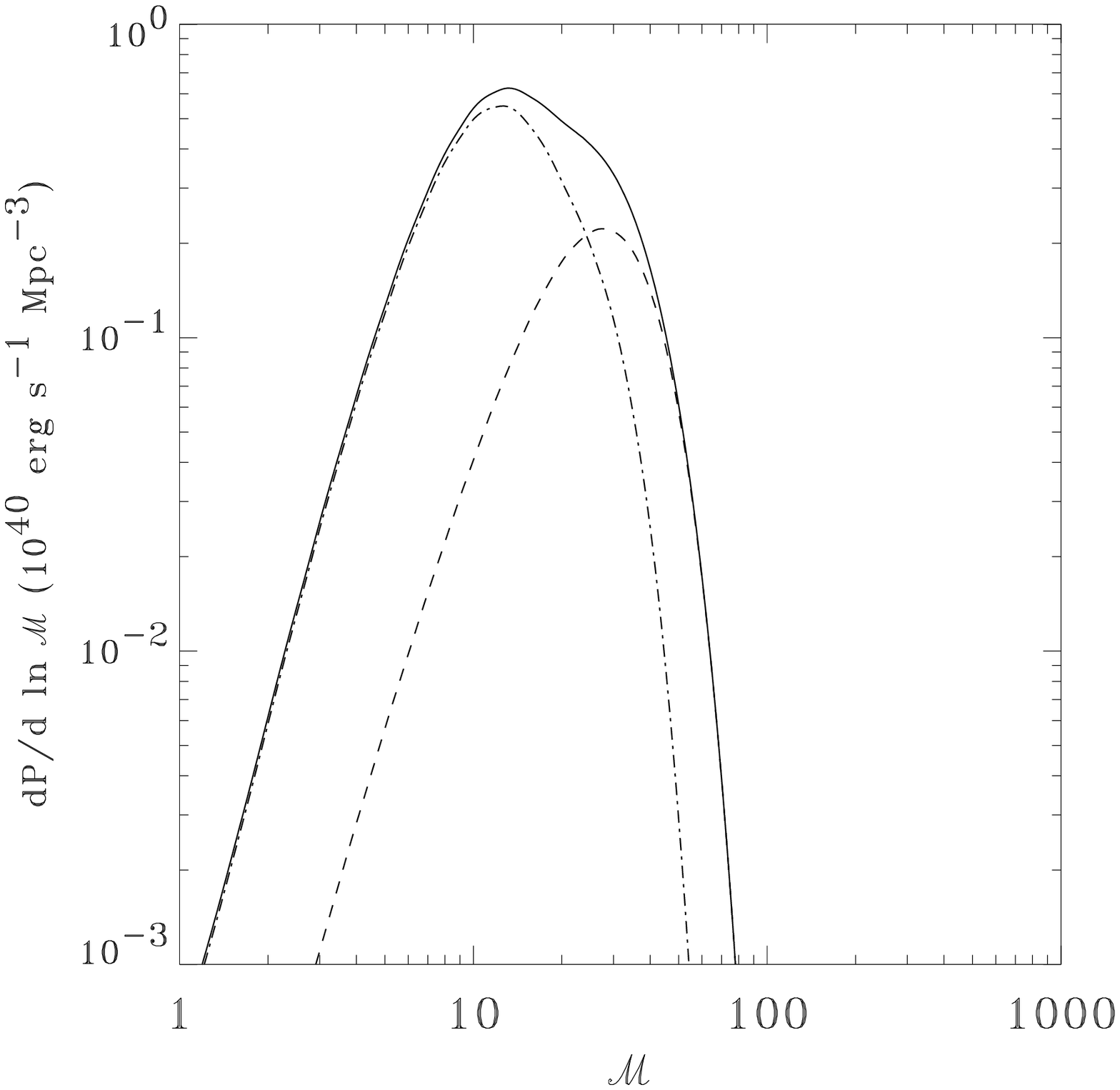}}
\resizebox{1.95in}{!}{
\includegraphics{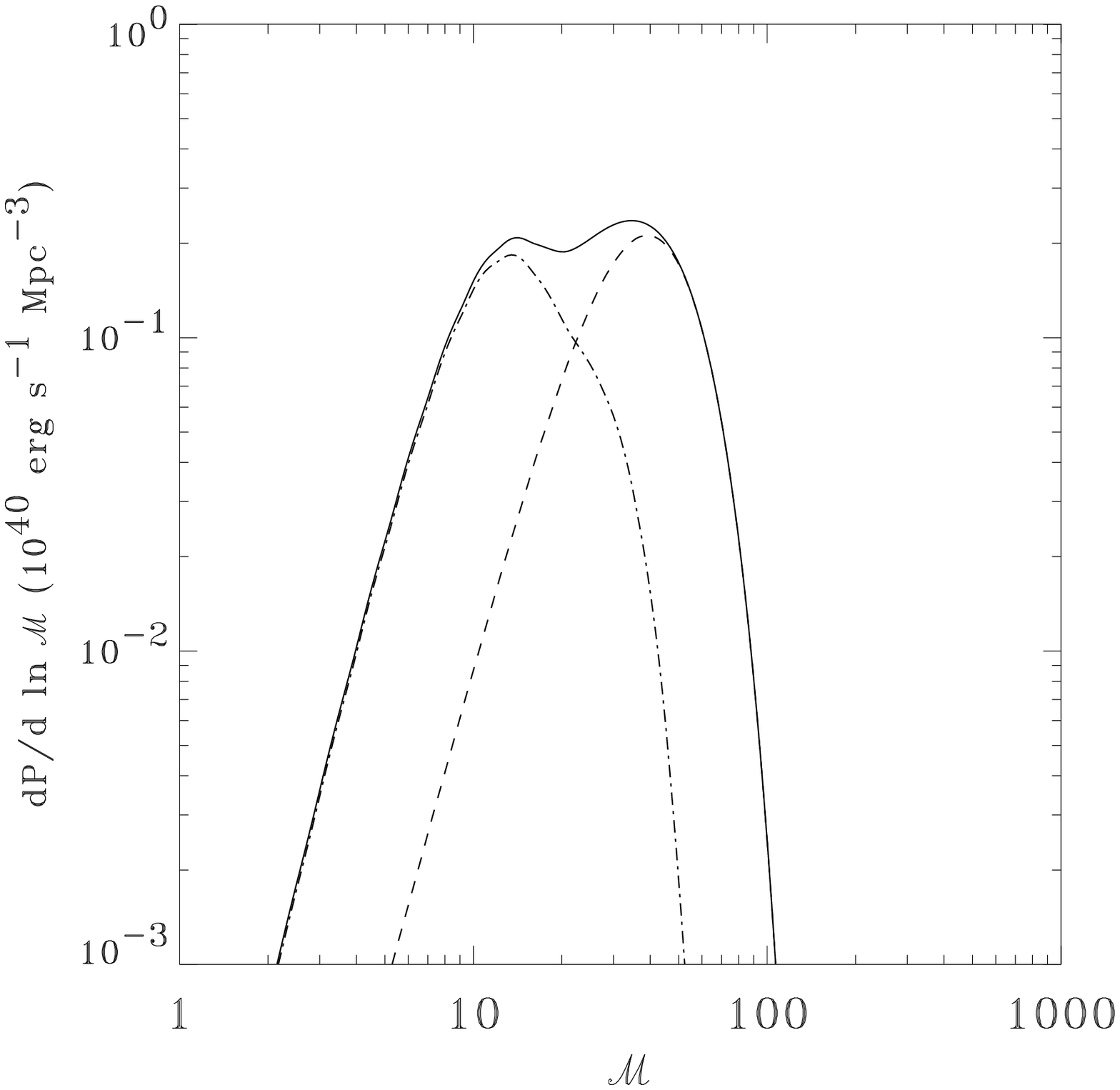}}
\caption{\label{fig:expkind} 
Distribution of spatial density of
  kinetic power 
per logarithmic Mach number interval ($dP/s\ln\mach$)
the double-distribution--based model without a mass cutoff. Left panel:
$z=2$; middle panel: $z=1$; right panel: $z=0$. 
Solid line: overall distribution; dot-dashed line: contribution from
structures with environmental overdensities between $\delta_{\rm ta}$
and $\delta_{\rm v}$ (between turnaround and virialization
overdensities); dashed line: contribution from structures with
environmental overdensities $\delta < \delta_{\rm ta}$.}
\end{figure*}

A most striking environmental effect in this case is that, as
redshift decreases, a  second peak separates out 
in the DD model, which is absent in the
PS models. The presence of this second, high-$\mach$ 
peak is due to the gradual shift of 
increasingly massive structures towards underdense environments. 
To better demonstrate this effect, we plot in Fig.~\ref{fig:expkind}
the components of the distribution produced by structures in
significantly overdense environments 
(dot-dashed line, $\delta > \delta_{\rm ta}$, where
$\delta_{\rm ta}$ is the overdensity of a perturbation turning around
at the specific epoch), and structures in environments of $\delta <
\delta_{\rm ta}$ (dashed line), for the DD model
without a mass cutoff and for three different redshifts. The
contribution from the low-density environments is distinct in $\mach$
space and increasing with decreasing redshift. 

\begin{figure*}
\resizebox{2.9in}{!}{
\includegraphics{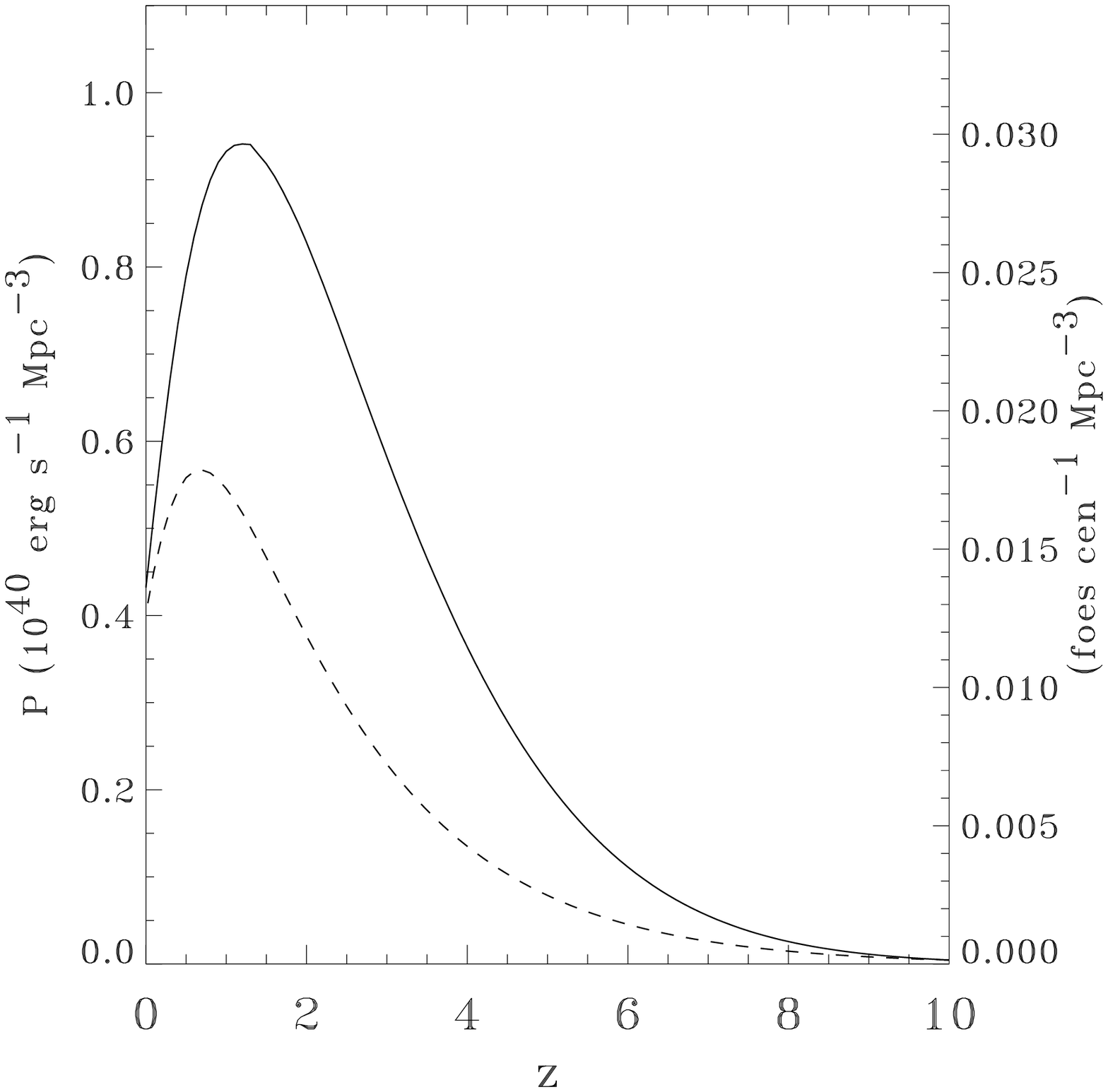}}
\resizebox{2.9in}{!}{
\includegraphics{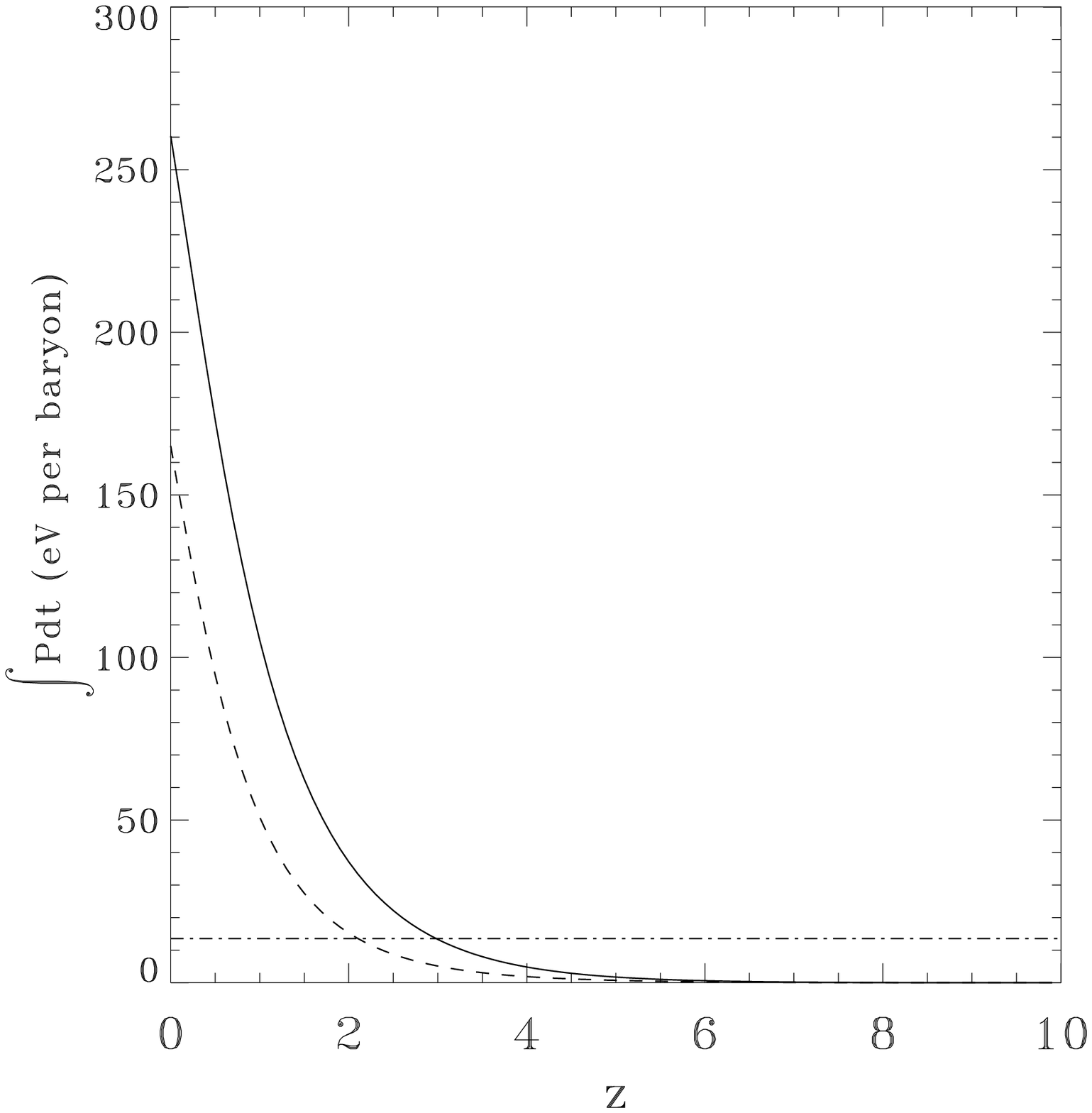}}
\caption{\label{fig:intp} 
Integrated kinetic power over shocks of any Mach number, $P$, for the
Double-Distribution (solid line) and the Press-Schechter (dashed line)
models. Left panel: redshift history of P for $z<10$. Right panel:
$\int P dt$ in units of eV per baryon. The horizontal line in the
right-panel plot corresponds to 13.6 eV per baryon.}
\end{figure*}

Figure \ref{fig:intp} shows the evolution of the
integrated kinetic power over shocks of any Mach number, $P$, for
redshifts between $10$ and $0$. The solid line corresponds to the
DD model and the dashed line to the Press-Schechter
models. In the left panel, $P(z)$ is plotted as a function of $z$, and
is seen to peak near $z\sim 1$ for both models, when the increase with
time of the number and mass of collapsed structures is balanced by the
decreasing mean density of the accreted material due to cosmic
expansion. The effect is more pronounced and the peak occurs at a
slightly higher redshift in the DD model where
environmental effects are accounted for, since the decrease in the
cosmic mean density is accompanied by a decrease in the most probable
overdensity {\em with respect to cosmic mean}. The overall level of
the curve in the DD model is also a factor of $\sim 2$ higher compared
to the PS model at redshifts $\gtrsim 1$, because environmental effects
further enhance the contribution of the larger structures which are
already favored due to their size and accelerating potential. However,
at lower redshifts, when in the DD model the contribution of the smaller
structures becomes comparable to that of the larger structures, the
integrated kinetic power predicted by the two models tends to
converge. 

The right panel
shows the redshift history of $\int P dt$ (the cumulative processed
kinetic energy) in units of eV per baryon {\em in the universe} (as
opposed to per shocked baryon). Again, the solid line is the
DD model while the dashed line is the PS
model.  The horizontal line in this plot corresponds to 13.6 eV per baryon.
From the location of the intersection of the horizontal line with the
$\int P dt$ curve, we can conclude that the energy processed by
accretion shocks {\em alone} by redshift $z\sim 3$ ($\sim 2$ in the PS
model) is of order of magnitude comparable to the energy required to
reionize the universe even in absence of other sources of energy
\footnote{Note that this order-of-magnitude argument is meant
to give a feeling about the amount of energy
processed by shocks as compared to other energy inputs in the IGM. If
one wanted to consider shocks as an actual reionization mechanism, a
detailed modelling of the reionization process would be required. }.

\begin{figure*}
\resizebox{2.9in}{!}{
\includegraphics{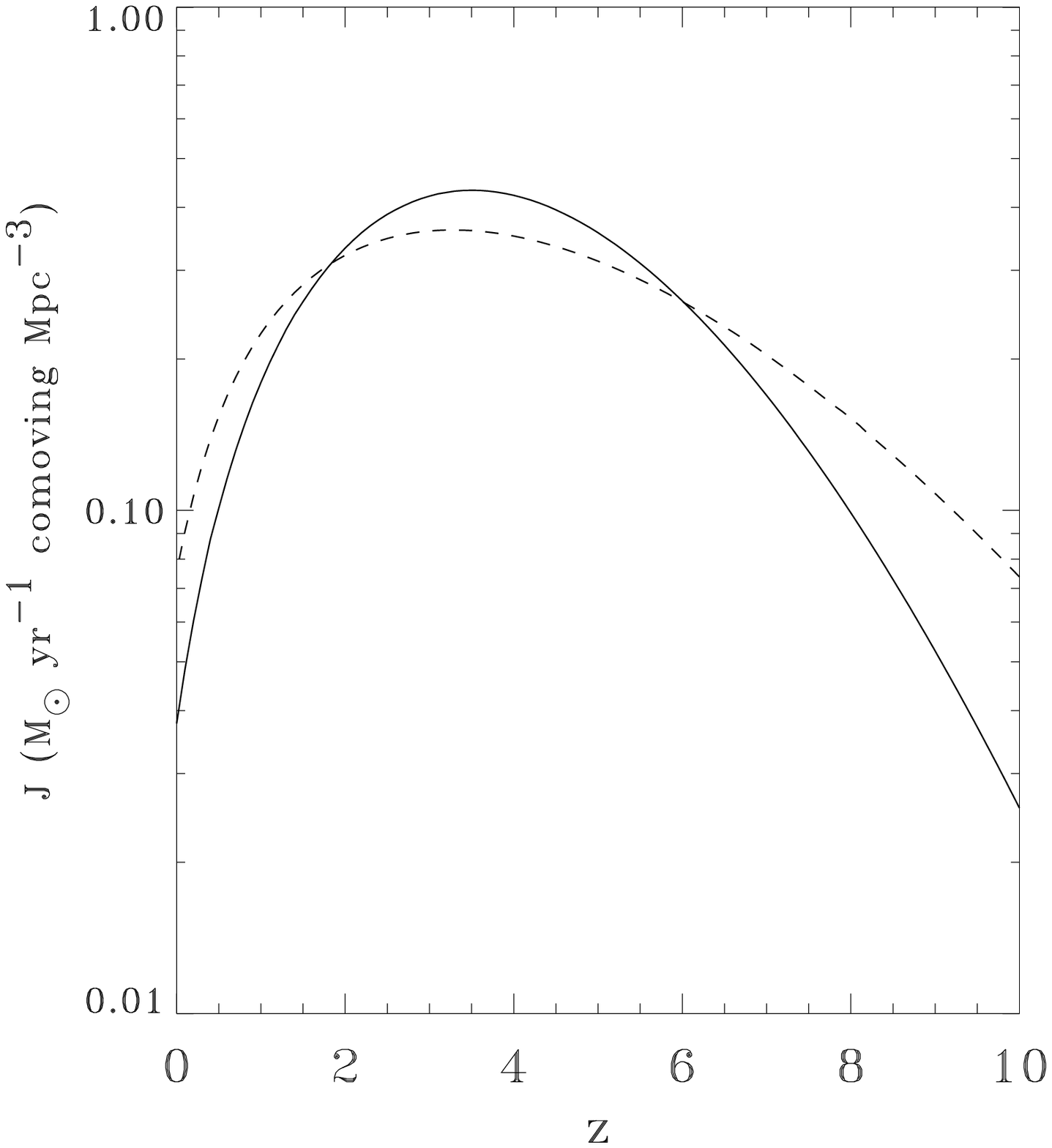}}
\resizebox{2.9in}{!}{
\includegraphics{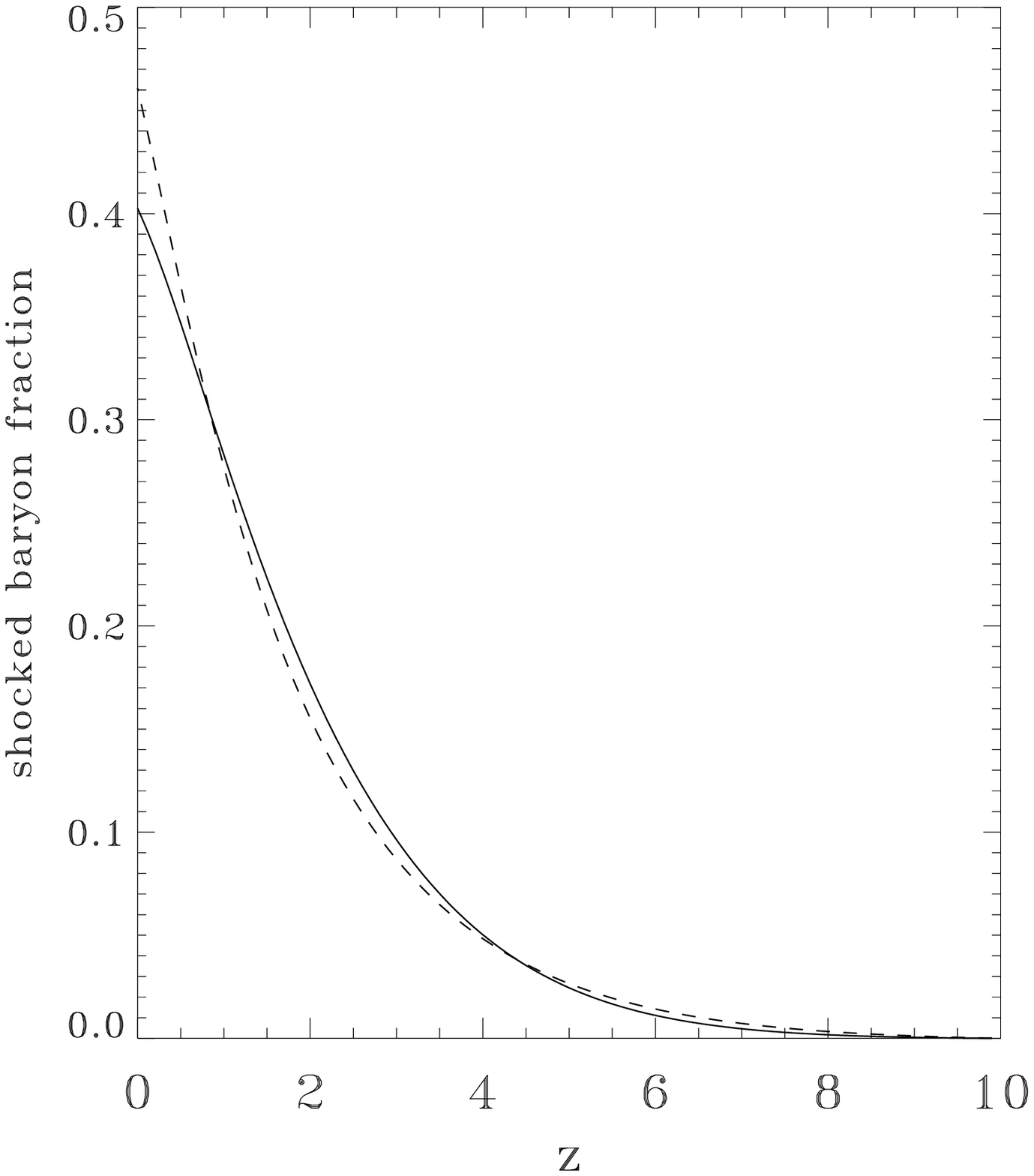}}
\caption{\label{fig:intj} 
Integrated mass current over shocks of any Mach number, $J$, for the
Double-Distribution (solid line) and the Press-Schechter (dashed line)
models. Left panel: redshift history of J for $z<10$. Right panel:
$\int J dt$ expressed as the fraction of baryons in the universe which
have been processed by accretion shocks.}
\end{figure*}

Finally, in Fig.~\ref{fig:intj} we plot the integrated mass current
over shocks of any Mach number, $J$, for the
DD (solid line) and the PS (dashed line)
models. The left panel shows the redshift history of $J$ for $z<10$,
while the right panel shows the redshift history of
$\int J dt$, the cumulative shocked gas mass, 
 expressed as the fraction of baryons in the universe which 
have been processed by accretion shocks. By the current cosmic epoch,
a fraction between $40-50\%$ of baryons have already been processed by
accretion shocks, while $\sim 10\%$ of the baryons have already been
processed by shocks by $z\sim 3$. 

\subsection{Comparison with Cosmological Simulations}

\indent

Caution should be exercized when comparing these calculations to
results of cosmological simulations studying the properties of cosmic
shocks, to ensure that corresponding quantities are being
compared. One should keep in mind that the shocks studied in this chapter
(a) are accretion shocks only, while effects of merger and filament
shocks and their contribution to shock statistics and energetics have
not been included and (b) are all considered to be external shocks,
accreting material which has never been shocked before.

In \cite{RKJ03}, the cosmological simulation results are presented by
classifying shocks as external or internal, depending on the temperature
of the accreted material. In both the kinetic and surface
distributions with respect to Mach number, used by the authors to
describe the statistical properties of cosmic shocks, external shocks
are dominated by filament shocks, while internal shocks are dominated
by merger shocks. The population of accretion shocks is presumably
divided between these two categories, depending on whether structures
accrete void or filament material. Accretion shocks are expected to lie
in the high Mach number end of both internal and external shock
distributions. Hence, no direct comparison between our model and
these results is possible without the inclusion of a detailed model 
of filament-shocked material and the way it modifies the properties of
the accretion shock population. However, properly adjusting our model
to the \cite{RKJ03} simulation parameters, we find that if we consider
the entire population of accretion shocks to be either external (as in
the results of the previous section) or internal with mean density and
temperature values for the filament gas taken from the \cite{RKJ03}
results, our curves lie, as expected,  {\em below} their overall
curves for external or  
internal shocks, respectively, for all relevant quantities, although
they constitute an appreciable fraction of them.

In \cite{min_shock}, the simulated shocks are instead divided in
accretion and merger shocks, the accretion shocks being assigned to
specific structures. The principal quantity used to statistically
describe the shock population itself is the number distribution of
shocks with respect to Mach number. However, a direct comparison is
not appropriate in this case either, because part of the simulated
population of accretion shocks resides inside filaments, which
introduces a bi-modality in the number distribution, especially
pronounced in the $\Omega_{\rm m}=1$ simulation. In addition, the
properties of the shocks even around structures which accrete
principally pristine material are modified by lateral accretion of
hotter gas from the filaments. Finally, the combination of selection
criteria for the collapsed structures inhibits the identification of
a sharp mass cutoff to use in our analytical models. 

Overall, although the models we have used here can reveal the
effect of a certain class of environmental factors (local overdensity
and associated change in temperature) on the statistics of cosmic
accretion shocks, it becomes clear that for a detailed comparison with
simulations and observations, a treatment of the second important
environmental factor, i.e. the filamentary structure of the universe,
is needed.  

A result of this work with direct relevance to the interpretation
of cosmological simulations studying the properties of
cosmic shocks is the effect of the existence of an explicit mass
cutoff to the properties of different shock distributions, since mass
cutoffs are always present in simulations as a result of finite mass
resolution. We have seen that the number distribution as well as
surface distribution of shocks with respect to Mach number are
dominated by the contribution of low-mass objects and hence they are
appreciably affected by the presence of a mass cutoff. The rest of our
shock distributions however, such as the mass current distribution and
the kinetic power distribution are dominated by high-mass objects and
are unaffected by the presence of a mass cutoff. This is consistent
with the findings of \cite{RKJ03}, who performed a series of
simulations with increasing mass and spatial resolution, and found
convergence for their results on kinetic power distribution but not
for the surface distribution. However, the physical impact of
shocks to their environment is better represented by the latter class
of distributions, which are unaffected by the value of the mass
cutoff, and hence the lack of convergence in the number-dominated
distributions need not decrease confidence in the relevance and
robustness of the physical output of such cosmological simulations.

\chapter{Discussion}
\label{chapter:chap6}

\indent

In this thesis, we have derived and explored 
a new tool for the analytical study of hierarchical
structure formation: a double distribution of the number density of
cosmic collapsed structures with respect to mass and local
overdensity. We have done so by introducing a clustering
scale parameter $\beta > 1$, which 
we use to associate with each collapsed object of mass
$m$ a larger environment of mass
$\beta m$.  The scale parameter $\beta$ can be expressed as a function of 
the number of virial radii included in the local environment of each structure.
We found that for reasonable
values $\beta \sim 2$,
the shape of the distribution does not depend sensitively on this
parameter. Integration over linearly extrapolated overdensity returns
the original
Press-Schechter mass function, independently of the value of $\beta$.

We have presented the double distribution in terms of the
true, physical, nonlinear density contrast $\delta$.
However, in calculating the distribution it is 
useful to identify regions using instead
the overdensity obtained via linear analysis,
$\ed$,
extrapolated to the present epoch.
A useful fitting function was given for the
$\delta-\ed$ conversion.

The double distribution
is useful because it allows us to have an explicit analytical
if approximate description of the environment in which collapsed objects
of all masses reside. 
Using the tools we have developed, it can be readily calculated
for any flat cosmology,
and evaluated at any epoch.
Consequently, it offers new insight into the
growth of structure as well as the present
distribution of collapsed objects.

We have evaluated the double distribution and some
of its integral moments for
both a concordance cosmology and an Einstein-de Sitter universe.
Some key results
are that at any redshift, the double distribution
is dominated by a peak
which shifts in mass but is always at
a relatively low value of $|\delta|$.
For each mass,
there is a most probable $\delta$, which
increases with structure mass.
Moreover, at the present epoch in the 
concordance universe, 
the most probable environment
is a modest {\em under}density,
for all objects below about $10^{14} M_\odot$;
thus, underdensities are preferentially
populated by low-mass objects.
Finally, the fraction
of mass in underdensities
increases with time, and in the
concordance cosmology the present
underdense mass fraction in objects of 
$M>10^{12} {\,\rm M_\odot}$ is about 40\%.
These trends can be understood in terms
of hierarchical clustering
in which overdense regions are the site of
vigorous merging that
clears out low-mass objects,
which then find their last refuge in voids.

In addition, we have utilized the double distribution of cosmic
structures 
to investigate analytically the effect of environmental
factors on the properties of cosmic accretion shocks around collapsed
structures. For this purpose, we have explored two different models for 
the cosmic shock population. The first used the Press-Schechter mass
function to describe the underlying population of collapsed, accreting
objects. All such objects were assumed to accrete material of the same
density and temperature. This was our ``control'' model, which did not
include any environmental effects. The second model used the
double distribution of collapsed structures \cite{pfd} to describe the
distribution of accreting objects with respect to both their mass and
local environment overdensity or underdensity. The overall mass
distribution of objects is the same as in the first model, as the
double distribution integrates to the Press-Schechter mass function. 

We found that the number and surface distributions of shocks with
respect to Mach number are dominated by
the contribution from low-mass objects and hence peak at low Mach
numbers for both models. The contrary is true for distributions
describing the physical impact of accretion shocks on their
environment, such as the mass current and kinetic power distributions,
which are dominated by the properties of primarily high-mass objects,
and hence peak at high Mach numbers. The distribution peaks are more
pronounced when environmental effects are taken into account, as they
tend to move objects of both high and low masses towards the peak. 

Perhaps the most striking effect of accounting for the environmental
factor is the separation of a second, high Mach number peak in the
kinetic power distribution at low redshifts, due to an increasing
number of higher-mass structures concentrating inside
underdensities. This double-peaked behavior is present despite the
fact that the effect of filament heating of the accreted gas was not
included in our calculation. The latter process is expected to further
complicate the features of the kinetic power distribution. 

The integrated kinetic power processed by shocks peaks at a redshift of
$\sim 1$, and the effect of the local environment is to increase the
overall level of the processed energy at high redshifts by a factor of
$\sim 2$ as well as to move the peak of the kinetic power history
towards slightly higher redshifts. We found that accretion shocks
alone have processed by $z \sim 3$ energy comparable to that required 
to reionize the universe.

The integrated mass current history (i.e., the net baryonic 
mass processed through shocks) peaks at earlier epochs compared
to the kinetic power redshift, with the peak in both models occurring
at $z \sim 3$. By the current epoch, the baryon fraction shocked in
accretion shocks alone is between $40-50\%$. Since this material
represents baryons which can condense to form galaxies and stars, this
fraction represents an upper limit to the baryon fraction in galaxies.

The population study of cosmic accretion shocks presented in the
second part of this thesis is only one of many possible applications
of the double distribution of cosmic structures. With its capability to
treat both underdensities and overdensities, and collapsed as well as
diffuse regions, the double distribution can be used to build
analytical models for many problems which are currently attracting
much interest in the context of cosmological structure formation. The
filamentary structure of the universe and  the dependence of merger
histories on environmental factors are only two such problems that we
are currently pursuing using the tools developed in this thesis.
%
\appendix
%

\chapter{Limits of the Double Distribution}\label{ap_lim}

In this appendix, we examine the behavior of the double distribution in
the limiting cases $\beta \rightarrow \infty$ and $\beta \rightarrow
1$. 

\section{Behavior of  the Double Distribution in the limit \boldmath{$\beta
  \rightarrow \infty$}}

In order to find the behavior the double distribution as $\beta
\rightarrow \infty$, we recall that, because $S(m)$ 
decreases monotonically with $m$, its limit in the infinite $\beta$ regime will be
\begin{equation}
\lim _{\beta \rightarrow \infty} S(\beta m) =0 \,.
\end{equation}

Then, using the notation of the previous section, 
\begin{equation}
\lim_{\beta \rightarrow \infty} \frac{dn}{dmd\ed_{\ell}} (m, \ed_{\ell}, \beta,
a) = \lim_{S_2 \rightarrow 0} \frac{dn}{dmd\ed_{\ell}}(S_1, S_2, \ed_{\ell}, a, m)\,.
\end{equation}
The limit of a unit-area Gaussian when its width
vanishes is the Dirac delta-function $\delta_{\rm D}$, 
\begin{equation}
\lim _{\lambda \rightarrow 0 }\frac{1}{\sqrt{2\pi}\lambda} \exp\left[
-\frac{(x-x_0)^2}{2\lambda^2}\right] = \delta_{\rm D}(x-x_0)\,.
\end{equation}
Using this result, we get
\begin{eqnarray}
\lim_{S_2 \rightarrow 0} \frac{dn}{dmd\ed_{\ell}} &=& 
\frac{\rho_{\rm m}}{m}\left|\frac{dS_1}{dm}\right|
\frac{\ed_{\rm 0,c}-\ed_{\ell}}{2\pi} \times \nonumber \\
&&\lim_{S_2\rightarrow 0}\left\{
\exp\left[-\frac{(\ed_{\rm 0,c}-\ed_{\ell})^2}{2(S_1-S_2)}\right]
\frac{\exp\left[-\frac{\ed_{\ell}^2}{2S_2}\right]
-\exp\left[-\frac{(\ed_{\ell}-2\ed_{\rm 0,c})^2}{2S_2}\right]
}{S_2^{1/2}(S_1-S_2)^{3/2}} \right\}\nonumber \\
&=& \frac{\rho_{\rm m}}{m}\left|\frac{dS_1}{dm}\right|
\frac{\ed_{\rm 0,c}-\ed_{\ell}}{\sqrt{2\pi}} 
\frac{\exp\left[-\frac{(\ed_{\rm 0,c}-\ed_{\ell})^2}{2S_1}\right]}{S_1^{3/2}}
\times \nonumber \\
&& \left\{\lim_{S_2\rightarrow 0}\frac{\exp\left[-\frac{\ed_{\ell}^2}{2S_2}
\right]}{\sqrt{2\pi S_2}}
-\lim_{S_2\rightarrow 0}\frac{\exp\left[-\frac{(\ed_{\ell}-2\ed_{\rm 0,c})^2}{2S_2}
\right]}{\sqrt{2\pi S_2}}
\right\} \nonumber \\
&=& \frac{\rho_{\rm m}}{m}\left|\frac{dS_1}{dm}\right|
\frac{\ed_{\rm 0,c}-\ed_{\ell}}{\sqrt{2\pi}}
\frac{\exp\left[-\frac{(\ed_{\rm 0,c}-\ed_{\ell})^2}{2S_1}\right]}{S_1^{3/2}}
\left[\delta_{\rm D}(\ed_{\ell})-\delta_{\rm D}(\ed_{\ell}-2\ed_{\rm 0,c})
\right]\,.
\end{eqnarray}
However, the $\ed_{\ell}-$domain of the double distribution is between
$-\infty$ and $\ed_{\rm 0,c}$, and therefore the value $\ed_{\ell}=2\ed_{\rm 0,c}$ is
outside its domain. Hence the second Dirac delta-function is always zero, and
\begin{equation}
\lim_{\beta \rightarrow \infty} \frac{dn}{dmd\ed_{\ell}} =
\frac{\rho_{\rm m}}{m}\left|\frac{dS_1}{dm}\right|
\frac{\ed_{\rm 0,c}-\ed_{\ell}}{\sqrt{2\pi}}
\frac{\exp\left[-\frac{(\ed_{\rm 0,c}-\ed_{\ell})^2}{2S_1}\right]}{S_1^{3/2}}
\delta_{\rm D}(\ed_{\ell})
\,,
\end{equation}
proportional, as expected, to a Dirac delta-function centered at $\ed_{\ell}=0$.
\section{Behavior of the Double Distribution in the limit \boldmath{$\beta
  \rightarrow 1$}}
Denoting $S(\beta m)$ by $S_2$ and letting $\phi = S(m)/S(\beta m)$,
we seek the behavior of the double distribution in the limit $\beta
\rightarrow 1$ or $\phi \rightarrow 1$. Defining 
\begin{equation}
\mathcal{C} = \frac{\rho_{\rm m}}{m}\frac{\ed_{\rm 0,c}-\ed_{\ell}}{\sqrt{2\pi}}
\left|\frac{dS}{dm}\right| \frac{\exp\left[-\frac{\ed_{\ell}^2}{2S_2}\right]
-\exp\left[-\frac{(\ed_{\ell}-2\ed_{\rm 0,c})^2}{2S_2}\right]}{S_2^{1/2}}\,,
\end{equation}
we can write
\begin{eqnarray}
\lim_{\phi \rightarrow 1} \frac{dn}{dmd\ed_{\ell}} &=& 
\mathcal{C} \lim_{\phi \rightarrow 1}\frac{\exp\left[
-\frac{(\ed_{\rm 0,c}-\ed_{\ell})^2}{2S_2(\phi-1)}\right]}
{\sqrt{2\pi}S_2^{3/2}(\phi-1)^{3/2}} \nonumber \\
&=& \mathcal{C}\lim_{\phi \rightarrow 1}
\frac{(\phi-1)^{-3/2}}{\sqrt{2\pi}S_2^{3/2}\exp\left[\frac{(\ed_{\rm 0,c}-\ed_{\ell})^2}
{2S_2(\phi-1)}\right]}\nonumber \\
&\stackrel{\infty/\infty}{=}&\mathcal{C}\lim_{\phi \rightarrow 1}
\frac{-\frac{3}{2}(\phi-1)^{-5/2}}
{\exp\left[\frac{(\ed_{\rm 0,c}-\ed_{\ell})^2}
{2S_2(\phi-1)}\right]\left[-
\frac{\sqrt{2\pi}S_2^{3/2}(\ed_{\rm 0,c}-\ed_{\ell})^2}{2S_2(\phi-1)^{2}}
\right]}\nonumber \\
&=&\mathcal{C}\lim_{\phi \rightarrow 1}\frac{3
\exp\left[-\frac{(\ed_{\rm 0,c}-\ed_{\ell})^2}
{2S_2(\phi-1)}\right]}{\sqrt{2\pi}S_2^{1/2}(\phi-1)^{1/2}(\ed_{\rm 0,c}-\ed_{\ell})^2}
\nonumber \\
&=& \mathcal{C}\frac{3}{(\ed_{\rm 0,c}-\ed_{\ell})^2}\delta_{\rm D}
(\ed_{\rm 0,c}-\ed_{\ell})\,
\end{eqnarray}
proportional, as expected, to a Dirac delta-function around $\ed_{\rm 0,c}$.


\chapter{Derivation of the Press-Schechter Mass Function From the
  Double Distribution} \label{ap_int}
 Using $S_1$
to denote $S(m)$ and $S_2$ for $S(\beta m)$ we have:
\begin{eqnarray}
\int_{-\infty}^{\ed_{\rm 0,c}}d\ed \frac{dn}{dmd\ed}&=&
\frac{\rho_{\rm m}}{m}\left|\frac{dS_1}{dm}\right|
\frac{1}{2\pi S_2^{1/2}(S_1-S_2)^{3/2}}\times \\
&& \left\{
\int_{-\infty}^{\ed_{\rm 0,c}}d\ed
(\ed_{\rm 0,c}-\ed)
\exp\left[-\frac{\ed^2}{2S_2}\right]
 \exp\left[-\frac{(\ed_{\rm 0,c}-\ed)^2}{2(S_1-S_2)}\right]-
\right. \nonumber \\
&& \left. \int_{-\infty}^{\ed_{\rm 0,c}}d\ed
(\ed_{\rm 0,c}-\ed)
\exp\left[-\frac{(\ed-2\ed_{\rm 0,c})^2}{2S_2}\right]
 \exp\left[-\frac{(\ed_{\rm 0,c}-\ed)^2}{2(S_1-S_2)}\right]
\right\}\nonumber \\
&\stackrel{\ed'=\ed_{0,c}-\ed}{=}&
\frac{\rho_{\rm m}}{m}\left|\frac{dS_1}{dm}\right|
\frac{1}{2\pi S_2^{1/2}(S_1-S_2)^{3/2}}\times \\
&& \left\{
\int_{0}^{\infty}d\ed' \ed'
\exp\left[-\frac{(\ed_{\rm 0,c}-\ed')^2}{2S_2}\right]
 \exp\left[-\frac{\ed'^2}{2(S_1-S_2)}\right]-
\right. \nonumber \\
&& \left. \int_{0}^{\infty}d\ed'
\ed'
\exp\left[-\frac{(\ed_{\rm 0,c}+\ed')^2}{2S_2}\right]
 \exp\left[-\frac{\ed'^2}{2(S_1-S_2)}\right]
\right\}\nonumber \\
\end{eqnarray}
Then, performing the transformation 
 $\ed'\rightarrow -\ed'$ in the second integral, we get
\begin{eqnarray}
\int_{-\infty}^{\ed_{\rm 0,c}}d\ed \frac{dn}{dmd\ed}
&=&
\frac{\rho_{\rm m}}{m}\left|\frac{dS_1}{dm}\right|
\frac{1}{2\pi S_2^{1/2}(S_1-S_2)^{3/2}}\times \\
&& \left\{
\int_{0}^{\infty}d\ed' \ed'
\exp\left[-\frac{(\ed_{\rm 0,c}-\ed')^2}{2S_2}\right]
 \exp\left[-\frac{\ed'^2}{2(S_1-S_2)}\right]+
\right. \nonumber \\
&& \left. \int_{-\infty}^{0}d\ed'
\ed'
\exp\left[-\frac{(\ed_{\rm 0,c}-\ed')^2}{2S_2}\right]
 \exp\left[-\frac{\ed'^2}{2(S_1-S_2)}\right]
\right\}\nonumber \\
&=&
\frac{\rho_{\rm m}}{m}
\frac{1}{2\pi S_2^{1/2}(S_1-S_2)^{3/2}}
\left|\frac{dS_1}{dm}\right| \times \nonumber \\
&& \int_{-\infty}^{\infty}
\ed'd\ed'
\exp\left[-\frac{(\ed_{\rm 0,c}-\ed')^2}{2S_2}\right]
\exp\left[-\frac{\ed'^2}{2(S_1-S_2)}\right] \nonumber \\
&=& 
\frac{\rho_{\rm m}}{m}
\frac{\left|dS_1/dm \right|}{2\pi S_2^{1/2}(S_1-S_2)^{3/2}}
\ed_{\rm 0,c}\left(\!\!\frac{S_1-S_2}{S_1}\!\!\right)^{3/2}
\!\!\!\!
\sqrt{2\pi S_2}\exp \!
\left[-\frac{\ed_{\rm 0,c}^2}{2S_1}\right]\nonumber \\
&=& \sqrt{\frac{2}{\pi}}
\frac{\rho_{\rm m}}{m^2}\frac{\ed_{\rm 0,c}}{\sqrt{S_1}}
\left|\frac{d\ln \sqrt{S_1}}{d\ln m}\right|
\exp \left[-\frac{\ed_{\rm 0,c}^2}{2S_1}\right]
\end{eqnarray}
The final result is 
the Press-Schechter mass function formula, independently of
the value of $\beta$.

\chapter{Vacuum Integrals}\label{vac_ints}


\section{ The incomplete vacuum integral of the first kind 
\boldmath{$\mathcal{V}_1$}}
\subsection{Definition} 
We define the incomplete vacuum integral of the first kind as 
\begin{equation}\label{realdefv1}
\mathcal{V}_1(r,\mu) = 
\frac{3}{2}
\int_0^r\frac{\sqrt{x}dx}{\sqrt{(1-x)(-x^2-x+\mu)}}\,,
\end{equation}
with domain $0\le r\le 1$ and $\mu \ge 2$. 
\subsection{Properties}
Physically,
$\mathcal{V}_1(r,\mu)$ is proportional to the time required by a
perturbation of normalized curvature parameter
$\kappa/\omega^{1/3}=(\mu+1)/\mu^{2/3}$ to achieve a size $a_{\rm p} =
r a_{\rm p,ta}(\kappa/\omega^{1/3})$ {\em before turnaround}. Its
asymptotic behavior for $r\ll 1$ is 
\begin{equation}
\mathcal{V}_1(r,\mu) \stackrel{r\ll 1}{\approx} \frac{1}{\sqrt{\mu}}
r^{3/2}
\end{equation}
while for $\mu\gg 1$ it is 
\begin{equation}
\mathcal{V}_1(r,\mu) \stackrel{\mu \gg 1}{\approx} 
\frac{1}{\sqrt{2\mu}}\left[\frac{\pi}{2} - \sqrt{r(1-r)}-\sin ^{-1}\sqrt{1-r}
\right]
\end{equation}
In the case of the Eddington perturbation ($\mu =2$), 
we can derive a closed-form expression for $\mathcal{V}_1$: 
\begin{eqnarray}
\mathcal{V}_1(r,2) &=& \frac{3}{2}
\int_0^r\frac{\sqrt{x}dx}{(1-x)\sqrt{x+2}} \nonumber \\
&=& \frac{\sqrt{3}}{2}\left[\ln \frac{1+\sqrt{r}}{1-\sqrt{r}}
 -2\sqrt{3}\sinh^{-1}\sqrt{\frac{r}{2}}
\right. \nonumber \\
&& \left. + \ln \frac{2\sqrt{3}+\sqrt{3r}+3\sqrt{2+r}}
{2\sqrt{3}-\sqrt{3r}+3\sqrt{2+r}}
\right]\,.
\end{eqnarray} 
When $r=1$, the value of $\mathcal{V}_1(1,\mu)$ is the {\em complete} vacuum
integral of the first kind, which is a function of $\mu$ alone.
Physically, the complete vacuum integral of the first kind is
proportional to the time required for a perturbation of curvature
parametrized by $\mu$ to reach turnaround. The derivative of
$\mathcal{V}_1(1,\mu)$ appears in the calculation of the derivative
$\partial \ed_o/\partial \delta$, in the 3rd line of Table
\ref{mytable2}, and it is
\begin{equation}
\frac{d}{d\mu}\mathcal{V}_1(1,\mu) = -\frac{3}{4}\int_0^1
\frac{\sqrt{x}dx}{\sqrt{1-x}(-x^2-x+\mu)^{3/2}}\,.
\end{equation}

\section{The hyperbolic vacuum integral of the first kind
\boldmath{$\mathcal{H}_1$}}

\subsection{Definition}
We define the hyperbolic vacuum integral of the first kind as 
\begin{equation}
\mathcal{H}_1 (r,\varpi)=\frac{3}{2}
\int_0^r\frac{\sqrt{x}dx}{\sqrt{(1+x)(x^2-x+\varpi)}}\,,
\end{equation}
with domain $0\le r<\infty$ and
$\varpi > 1/4$. 

\subsection{Properties}
Physically, $\mathcal{H}_1 (r,\varpi)=$ is
proportional to the time  required by a
perturbation of normalized curvature parameter
$\kappa/\omega^{1/3}=(1-\varpi)/\varpi^{2/3}$ to achieve a size $a_{\rm p} =
r a_{\rm p,R}(\kappa/\omega^{1/3})$. Its asymptotic behavior for $r
\ll 1$ is
\begin{equation}
\mathcal{H}_1 (r,\varpi) \stackrel{r \ll 1}{\approx}
\frac{1}{\sqrt{\varpi}}r^{3/2}
\end{equation}
while for $r \gg 1$ it is 
\begin{equation}
\mathcal{H}_1 (r,\varpi) 
 \stackrel{r \gg 1}{\approx} C(\varpi) + \frac{3}{2}
\ln \left(2\sqrt{r^2-r+\varpi}+2r-1\right)
\end{equation}
where $C(\varpi)$ is a function dependent only on $\varpi$. 
In the case of a flat ($\varpi=1$) perturbation, $\mathcal{H}_1$ can
be integrated immediately to give
\begin{equation}
\mathcal{H}_1 (r,1)=
\frac{3}{2}\int_0^r\frac{\sqrt{x}dx}{\sqrt{x^3+1}}
= \sinh ^{-1}\sqrt{x^3}\,.
\end{equation}

\section{The incomplete vacuum integral of the second kind
\boldmath{$\mathcal{V}_2$}} 
\subsection{Definition}
We define the incomplete vacuum integral of the second kind as  
\begin{equation}
\mathcal{V}_2 (r,\mu)=
\frac{3}{4}\int_0^r \frac{x^{3/2}dx}{(1-x)^{3/2}(-x^2-x+\mu)^{3/2}}\,, 
\end{equation}
with domain same as for $\mathcal{V}_1(r,\mu)$. 

\subsection{Properties}
The incomplete vacuum integral of the second kind is
related to $\mathcal{V}_1(r,\mu)$ through 
\begin{equation}
\frac{\partial}{\partial (\kappa/\omega^{1/3})}\mathcal{V}_1(r,\mu)=
\mu^{2/3}\mathcal{V}_2(r,\mu)\,,
\end{equation}
with 
\begin{equation}
\kappa/\omega^{1/3} = (1+\mu)/\mu^{2/3}\,.
\end{equation}
In the case of the Eddington perturbation ($\mu =2$), we can derive closed-form
expressions for $\mathcal{V}_2$:
\begin{eqnarray}
\mathcal{V}_2(r,2) &=& 
\frac{3}{4}
\int_0^r\frac{x^{3/2}dx}{(1-x)^3(x+2)^{3/2}} \nonumber \\
&=& \frac{\sqrt{3}}{72}\left[
\frac{\sqrt{3r}(2-3r+4r^2)}{\sqrt{2+r}(1-r)^2}\right.\nonumber\\
&&\left.+\log\frac{1-r}{1+2r+\sqrt{3r(2+r)}}
\right]\,.
\end{eqnarray} 

\section{
The hyperbolic vacuum integral of the second kind
\boldmath{$\mathcal{H}_2$}}

\subsection{Definition}
We define the hyperbolic vacuum integral of the second kind as 
\begin{equation}
\mathcal{H}_2 (r,\varpi)=
\frac{3}{4}\int_0^r
\frac{x^{3/2}dx}{(1+x)^{3/2}(x^2-x+\varpi)^{3/2}}\,, 
\end{equation}
and its domain is that of $\mathcal{H}_1(r,\varpi)$. 

\subsection{Properties}
The hyperbolic vacuum integral of the second kind is related to 
$\mathcal{H}_1(r,\varpi)$ through
\begin{equation}
\frac{\partial}{\partial (\kappa/\omega^{1/3})}\mathcal{H}_1(r,\varpi)=
\varpi^{2/3}\mathcal{H}_2(r,\varpi)\,, 
\end{equation}
with 
\begin{equation}
\kappa/\omega^{1/3} = (1+\varpi)/\varpi^{2/3}\,.
\end{equation}
In the case of a flat ($\varpi=1$) perturbation, $\mathcal{H}_2$ takes
the form
\begin{equation}
\mathcal{H}_2(r, 1) = \frac{3}{2^{4/3}}\int_0^{2^{1/3}r}\left
(\frac{u}{u^3+2}\right)^{3/2}du\,,
\end{equation}
which is the integral entering the linear growth factor in the
$\Omega_{\rm m} + \Omega_\Lambda=1$ universe. Hence, the linear growth
factor function $A(x)$ can be written as
\begin{equation}
A(x) = \frac{2^{4/3}(x^3+2)^{1/2}}{3x^{3/2}}\mathcal{H}_2(2^{-1/3}x,1)\,.
\end{equation}


\chapter{Comparison to Bertschinger Similarity Solution}\label{apa}
In this appendix, we compare  the Mach number as derived from the 
temperature jump across the accretion shock surrounding a collapsed structure 
(presented in chapter 3) 
to the similarity solution derived by Bertschinger \cite{bert85}
for the case of a single, spherically symmetric collapsed structure
accreting matter in an otherwise homogeneous $\Omega_{\rm m}=1$ universe. According
to the Bertschinger solution, the accretion shock is positioned at a constant
fraction $\lambda_{\rm s}\approx 0.347$ of the radius $r_{\rm ta}$ of the matter
shell turning  around at a given cosmic time $t$, 
\begin{equation}
r_{\rm s}(m,t) = \lambda_{\rm s} r_{\rm ta}(m,t) 
= 0.347\left(\frac{9}{14}mGt^2\right)^{1/3}
\end{equation}
where $m$ is the mass of the structure at the specific time $t$.
The velocity of the infalling gas in the lab frame at the shock position is
\begin{equation}
v_{\rm g} = -1.43\frac{r_{\rm ta}(m,t)}{t}\,.
\end{equation}
The shock surface itself is
propagating outwards, with a velocity 
$v_{\rm s} = (8/9)\lambda_{\rm s}r_{\rm ta}/t$
\footnote{if the time dependence of $m$ is written
  out explicitly, the turnaround radius varies as $r_{\rm ta}(t) \propto t^{8/9}$ }.
Then, the absolute value of the gas velocity
{\it in the shock frame}, $v_1$, is 
\begin{eqnarray}
v_1 &=& |v_{\rm g}-v_{\rm s}| = \left|-1.43-\frac{8}{9}\lambda_s
\right|
\frac{r_{\rm ta}}{t} \\
&=& 1.74 \left(\frac{27GH_0}{14}\right)^{1/3}(1+z)^{1/2}m^{1/3}\,.
\end{eqnarray}
If now the pre-shock material has an adiabatic  sound speed $\cs$, the Mach number of the
accretion shock is
\begin{eqnarray}\label{fromber}
\mach(m,z,\cs)&=&\frac{v_1}{\cs} \nonumber \\ &=&
\frac{1.74}{\cs}
\left(\frac{27GH_0}{14}\right)^{1/3}(1+z)^{1/2}m^{1/3} \nonumber \\
&=& 92\left(\frac{15{\rm \,\,km\,s^{-1}}}{\cs}\right)(1+z)^{1/2}\nonumber \\
&&\times \left(\frac{m}{m_8}\right)^{1/3}\,.
\end{eqnarray}
Note that the Bertschinger solution was derived in the limit 
$\mach \rightarrow \infty$, and therefore it is only valid for 
$\mach \gg 1$.

We can compare Eq.~(\ref{fromber}) with the result we derived using the
temperature jump across the surface of the shock. Combining Eq.s 
(\ref{highmachlim}) (for the temperature jump in the high-$\mach$
limit) and (\ref{jumpvir}) we get
\begin{eqnarray}
\mach &\approx& 93 \left(\frac{\rm 15 \, \,  km\,s^{-1}}{c_s}\right)
(1+z)^{1/2}\left(\frac{f_c}{18\pi^2}\right)^{1/6} \nonumber \\
&&\times \left(\frac{m}{m_8}\right)^{1/3}\,,
\end{eqnarray}
in excellent agreement with Eq.~(\ref{fromber}). The small ($\sim 1 \%$)
deviation arises because, in obtaining $\mach$ from $T_2/T_1$, 
we have ignored any temperature structure
within the collapsed object, and have instead assumed that the virial
temperature of the structure is representative of the temperature
right behind the shock, while the Bertschinger analysis
calculates and takes into account the temperature structure
inside the shock surface.

\clearpage
%
\addcontentsline{toc}{chapter}{\bibname}
%
\singlespace
%
\bibliographystyle{dis}
%
\bibliography{dis}

\begin{thebibliography}{10}
\expandafter\ifx\csname url\endcsname\relax
  \def\url#1{\texttt{#1}}\fi
\expandafter\ifx\csname urlprefix\endcsname\relax\def\urlprefix{URL }\fi
\providecommand{\bibinfo}[2]{#2}
\providecommand{\eprint}[2][]{\url{#2}}

\bibitem{ps74}
\bibinfo{author}{W.~Press} and \bibinfo{author}{P.~Schechter},
  \bibinfo{journal}{ApJ} \textbf{\bibinfo{volume}{187}}, \bibinfo{pages}{425}
  (\bibinfo{year}{1974}).

\bibitem{b91}
\bibinfo{author}{J.~Bond}, \bibinfo{author}{S.~Cole},
  \bibinfo{author}{G.~Efstathiou}, and \bibinfo{author}{N.~Kaiser},
  \bibinfo{journal}{ApJ} \textbf{\bibinfo{volume}{379}}, \bibinfo{pages}{440}
  (\bibinfo{year}{1991}).

\bibitem{lc93}
\bibinfo{author}{C.~Lacey} and \bibinfo{author}{S.~Cole},
  \bibinfo{journal}{MNRAS} \textbf{\bibinfo{volume}{262}}, \bibinfo{pages}{627}
  (\bibinfo{year}{1993}).

\bibitem{ph90}
\bibinfo{author}{J.~Peacock} and \bibinfo{author}{A.~Heavens},
  \bibinfo{journal}{MNRAS} \textbf{\bibinfo{volume}{243}}, \bibinfo{pages}{133}
  (\bibinfo{year}{1990}).

\bibitem{j95}
\bibinfo{author}{K.~Jedamzik}, \bibinfo{journal}{ApJ}
  \textbf{\bibinfo{volume}{448}}, \bibinfo{pages}{1} (\bibinfo{year}{1995}).

\bibitem{bow91}
\bibinfo{author}{R.~Bower}, \bibinfo{journal}{MNRAS}
  \textbf{\bibinfo{volume}{248}}, \bibinfo{pages}{332} (\bibinfo{year}{1991}).

\bibitem{wef}
\bibinfo{author}{S.~White}, \bibinfo{author}{G.~Efstathiou}, and
  \bibinfo{author}{C.~Frenk}, \bibinfo{journal}{MNRAS}
  \textbf{\bibinfo{volume}{262}}, \bibinfo{pages}{1023} (\bibinfo{year}{1993}).

\bibitem{lc94}
\bibinfo{author}{C.~Lacey} and \bibinfo{author}{S.~Cole},
  \bibinfo{journal}{MNRAS} \textbf{\bibinfo{volume}{271}}, \bibinfo{pages}{676}
  (\bibinfo{year}{1994}).

\bibitem{ecf}
\bibinfo{author}{R.~Eke}, \bibinfo{author}{S.~Cole}, and
  \bibinfo{author}{C.~Frenk}, \bibinfo{journal}{MNRAS}
  \textbf{\bibinfo{volume}{282}}, \bibinfo{pages}{263} (\bibinfo{year}{1996}).

\bibitem{ls98}
\bibinfo{author}{J.~Lee} and \bibinfo{author}{S.~Shandarin},
  \bibinfo{journal}{ApJ} \textbf{\bibinfo{volume}{500}}, \bibinfo{pages}{14}
  (\bibinfo{year}{1998}).

\bibitem{smt01}
\bibinfo{author}{R.~Sheth}, \bibinfo{author}{H.~Mo}, and
  \bibinfo{author}{G.~Tormen}, \bibinfo{journal}{MNRAS}
  \textbf{\bibinfo{volume}{323}}, \bibinfo{pages}{1} (\bibinfo{year}{2001}).

\bibitem{st02}
\bibinfo{author}{R.~Sheth} and \bibinfo{author}{G.~Tormen},
  \bibinfo{journal}{MNRAS} \textbf{\bibinfo{volume}{329}}, \bibinfo{pages}{61}
  (\bibinfo{year}{2002}).

\bibitem{mw96}
\bibinfo{author}{H.~Mo} and \bibinfo{author}{S.~White},
  \bibinfo{journal}{MNRAS} \textbf{\bibinfo{volume}{282}}, \bibinfo{pages}{347}
  (\bibinfo{year}{1996}).

\bibitem{kns97}
\bibinfo{author}{G.~Kaufmann}, \bibinfo{author}{A.~Nusser}, and
  \bibinfo{author}{M.~Steinmetz}, \bibinfo{journal}{MNRAS}
  \textbf{\bibinfo{volume}{286}}, \bibinfo{pages}{795} (\bibinfo{year}{1997}).

\bibitem{s98}
\bibinfo{author}{R.~Sheth}, \bibinfo{journal}{MNRAS}
  \textbf{\bibinfo{volume}{300}}, \bibinfo{pages}{1057} (\bibinfo{year}{1998}).

\bibitem{tp98}
\bibinfo{author}{M.~Tegmark} and \bibinfo{author}{P.~Peebles},
  \bibinfo{journal}{ApJ} \textbf{\bibinfo{volume}{500}}, \bibinfo{pages}{L79}
  (\bibinfo{year}{1998}).

\bibitem{lk99}
\bibinfo{author}{G.~Lemson} and \bibinfo{author}{G.~Kaufmann},
  \bibinfo{journal}{MNRAS} \textbf{\bibinfo{volume}{302}}, \bibinfo{pages}{111}
  (\bibinfo{year}{1999}).

\bibitem{st99}
\bibinfo{author}{R.~Sheth} and \bibinfo{author}{G.~Tormen},
  \bibinfo{journal}{MNRAS} \textbf{\bibinfo{volume}{308}}, \bibinfo{pages}{119}
  (\bibinfo{year}{1999}).

\bibitem{sj00}
\bibinfo{author}{U.~Seljak}, \bibinfo{journal}{MNRAS}
  \textbf{\bibinfo{volume}{318}}, \bibinfo{pages}{203} (\bibinfo{year}{2000}).

\bibitem{cs02}
\bibinfo{author}{A.~Cooray} and \bibinfo{author}{R.~Sheth},
  \bibinfo{journal}{Phys.R.} \textbf{\bibinfo{volume}{372}}, \bibinfo{pages}{1}
  (\bibinfo{year}{2002}).

\bibitem{j98}
\bibinfo{author}{J.~Jing}, \bibinfo{journal}{ApJ}
  \textbf{\bibinfo{volume}{503}}, \bibinfo{pages}{L9} (\bibinfo{year}{1998}).

\bibitem{HO86}
\bibinfo{author}{D.~Hegyi} and \bibinfo{author}{K.~Olive},
  \bibinfo{journal}{ApJ} \textbf{\bibinfo{volume}{303}}, \bibinfo{pages}{56}
  (\bibinfo{year}{1986}).

\bibitem{Fuk98}
\bibinfo{author}{M.~Fukugita}, \bibinfo{author}{C.~Hogan}, and
  \bibinfo{author}{P.~Peebles}, \bibinfo{journal}{ApJ}
  \textbf{\bibinfo{volume}{503}}, \bibinfo{pages}{518} (\bibinfo{year}{1998}).

\bibitem{Sper03}
\bibinfo{author}{D.~S. {\it et al.}}, \bibinfo{journal}{ApJS}
  \textbf{\bibinfo{volume}{148}}, \bibinfo{pages}{175} (\bibinfo{year}{2003}).

\bibitem{Cyb03}
\bibinfo{author}{R.~Cyburt}, \bibinfo{author}{B.~.Fields}, and
  \bibinfo{author}{K.~Olive}, \bibinfo{journal}{PhLB}
  \textbf{\bibinfo{volume}{567}}, \bibinfo{pages}{227} (\bibinfo{year}{2003}).

\bibitem{Nic02}
\bibinfo{author}{F.~N. {\it et al.}}, \bibinfo{journal}{ApJ}
  \textbf{\bibinfo{volume}{573}}, \bibinfo{pages}{157} (\bibinfo{year}{2002}).

\bibitem{Mat03}
\bibinfo{author}{S.~Mathur}, \bibinfo{author}{D.~Weinberg}, and
  \bibinfo{author}{X.~Chen}, \bibinfo{journal}{ApJ}
  \textbf{\bibinfo{volume}{582}}, \bibinfo{pages}{82} (\bibinfo{year}{2003}).

\bibitem{finog03}
\bibinfo{author}{A.~Finoguenov}, \bibinfo{author}{U.~Briel}, and
  \bibinfo{author}{J.~Henry}, \bibinfo{journal}{A\&A}
  \textbf{\bibinfo{volume}{410}}, \bibinfo{pages}{777} (\bibinfo{year}{2003}).

\bibitem{nicastro}
\bibinfo{author}{F.~N. {\it et al.}}, \bibinfo{journal}{Nature}
  \textbf{\bibinfo{volume}{433}}, \bibinfo{pages}{495} (\bibinfo{year}{2005}).

\bibitem{Hel98}
\bibinfo{author}{U.~Hellsten}, \bibinfo{author}{N.~Gnedin}, and
  \bibinfo{author}{J.~Miralda-Escud\'{e}}, \bibinfo{journal}{ApJ}
  \textbf{\bibinfo{volume}{509}}, \bibinfo{pages}{56} (\bibinfo{year}{1998}).

\bibitem{PL98}
\bibinfo{author}{R.~Perna} and \bibinfo{author}{A.~Loeb},
  \bibinfo{journal}{ApJ} \textbf{\bibinfo{volume}{503}}, \bibinfo{pages}{L135}
  (\bibinfo{year}{1998}).

\bibitem{FBC02}
\bibinfo{author}{T.~Fang}, \bibinfo{author}{G.~Bryan}, and
  \bibinfo{author}{C.~Canizares}, \bibinfo{journal}{ApJ}
  \textbf{\bibinfo{volume}{564}}, \bibinfo{pages}{604} (\bibinfo{year}{2002}).

\bibitem{CO99}
\bibinfo{author}{R.~Cen} and \bibinfo{author}{J.~Ostriker},
  \bibinfo{journal}{ApJ} \textbf{\bibinfo{volume}{514}}, \bibinfo{pages}{1}
  (\bibinfo{year}{1999}).

\bibitem{D01}
\bibinfo{author}{R.~D. {\it et al.}}, \bibinfo{journal}{ApJ}
  \textbf{\bibinfo{volume}{552}}, \bibinfo{pages}{473} (\bibinfo{year}{2001}).

\bibitem{furlanetto}
\bibinfo{author}{S.~R. Furlanetto} and \bibinfo{author}{A.~Loeb},
  \bibinfo{journal}{ApJ} \textbf{\bibinfo{volume}{611}}, \bibinfo{pages}{642}
  (\bibinfo{year}{2004}).

\bibitem{KRCS05}
\bibinfo{author}{H.~Kang}, \bibinfo{author}{D.~Ryu}, \bibinfo{author}{R.~Cen},
  and \bibinfo{author}{D.~Song}, \bibinfo{journal}{ApJ}
  \textbf{\bibinfo{volume}{620}}, \bibinfo{pages}{21} (\bibinfo{year}{2005}).

\bibitem{min01}
\bibinfo{author}{F.~Miniati}, \bibinfo{author}{T.~Jones},
  \bibinfo{author}{H.~Kang}, and \bibinfo{author}{D.~Ryu},
  \bibinfo{journal}{ApJ} \textbf{\bibinfo{volume}{562}}, \bibinfo{pages}{233}
  (\bibinfo{year}{2001}).

\bibitem{min01b}
\bibinfo{author}{F.~Miniati}, \bibinfo{author}{D.~Ryu},
  \bibinfo{author}{H.~Kang}, and \bibinfo{author}{T.~Jones},
  \bibinfo{journal}{ApJ} \textbf{\bibinfo{volume}{559}}, \bibinfo{pages}{59}
  (\bibinfo{year}{2001}).

\bibitem{BSFG01}
\bibinfo{author}{G.~Brunetti}, \bibinfo{author}{G.~Setti},
  \bibinfo{author}{L.~Feretti}, and \bibinfo{author}{G.~Giovannini},
  \bibinfo{journal}{New Astron.} \textbf{\bibinfo{volume}{6}},
  \bibinfo{pages}{1} (\bibinfo{year}{2001}).

\bibitem{BerrD03}
\bibinfo{author}{R.~Berrington} and \bibinfo{author}{C.~Dermer},
  \bibinfo{journal}{ApJ} \textbf{\bibinfo{volume}{594}}, \bibinfo{pages}{709}
  (\bibinfo{year}{2003}).

\bibitem{GB03p}
\bibinfo{author}{S.~Gabici} and \bibinfo{author}{P.~Blasi},
  \bibinfo{journal}{ApJ} \textbf{\bibinfo{volume}{583}}, \bibinfo{pages}{695}
  (\bibinfo{year}{2003}).

\bibitem{BBCG04}
\bibinfo{author}{G.~Brunetti}, \bibinfo{author}{P.~Blasi},
  \bibinfo{author}{R.~Cassano}, and \bibinfo{author}{S.~Gabici},
  \bibinfo{journal}{MNRAS} \textbf{\bibinfo{volume}{350}},
  \bibinfo{pages}{1174} (\bibinfo{year}{2004}).

\bibitem{KJ05}
\bibinfo{author}{H.~Kang} and \bibinfo{author}{T.~Jones},
  \bibinfo{journal}{ApJ} \textbf{\bibinfo{volume}{620}}, \bibinfo{pages}{44}
  (\bibinfo{year}{2005}).

\bibitem{lw}
\bibinfo{author}{A.~Loeb} and \bibinfo{author}{E.~Waxman},
  \bibinfo{journal}{Nature} \textbf{\bibinfo{volume}{405}},
  \bibinfo{pages}{156} (\bibinfo{year}{2000}).

\bibitem{tk}
\bibinfo{author}{T.~Totani} and \bibinfo{author}{T.~Kitayama},
  \bibinfo{journal}{ApJ} \textbf{\bibinfo{volume}{545}}, \bibinfo{pages}{572}
  (\bibinfo{year}{2000}).

\bibitem{min02}
\bibinfo{author}{F.~Miniati}, \bibinfo{journal}{MNRAS}
  \textbf{\bibinfo{volume}{337}}, \bibinfo{pages}{199} (\bibinfo{year}{2002}).

\bibitem{SchMuk}
\bibinfo{author}{C.~Scharf} and \bibinfo{author}{R.~Mukherjee},
  \bibinfo{journal}{ApJ} \textbf{\bibinfo{volume}{580}}, \bibinfo{pages}{154}
  (\bibinfo{year}{2002}).

\bibitem{SuIn02}
\bibinfo{author}{T.~Suzuki} and \bibinfo{author}{S.~Inoue},
  \bibinfo{journal}{ApJ} \textbf{\bibinfo{volume}{573}}, \bibinfo{pages}{168}
  (\bibinfo{year}{2002}).

\bibitem{ti}
\bibinfo{author}{T.~Totani} and \bibinfo{author}{S.~Inoue},
  \bibinfo{journal}{APh} \textbf{\bibinfo{volume}{17}}, \bibinfo{pages}{79}
  (\bibinfo{year}{2002}).

\bibitem{GB03g}
\bibinfo{author}{S.~Gabici} and \bibinfo{author}{P.~Blasi},
  \bibinfo{journal}{APh} \textbf{\bibinfo{volume}{19}}, \bibinfo{pages}{679}
  (\bibinfo{year}{2003}).

\bibitem{kesh}
\bibinfo{author}{U.~Keshet}, \bibinfo{author}{E.~Waxman},
  \bibinfo{author}{A.~Loeb}, \bibinfo{author}{V.~Springel}, and
  \bibinfo{author}{L.~Hernquist}, \bibinfo{journal}{ApJ}
  \textbf{\bibinfo{volume}{585}}, \bibinfo{pages}{128} (\bibinfo{year}{2003}).

\bibitem{min03}
\bibinfo{author}{F.~Miniati}, \bibinfo{journal}{MNRAS}
  \textbf{\bibinfo{volume}{342}}, \bibinfo{pages}{1009} (\bibinfo{year}{2003}).

\bibitem{RPSM}
\bibinfo{author}{O.~Reimer}, \bibinfo{author}{M.~Pohl},
  \bibinfo{author}{P.~Sreekumar}, and \bibinfo{author}{J.~Mattox},
  \bibinfo{journal}{ApJ} \textbf{\bibinfo{volume}{588}}, \bibinfo{pages}{155}
  (\bibinfo{year}{2003}).

\bibitem{GB04}
\bibinfo{author}{S.~Gabici} and \bibinfo{author}{P.~Blasi},
  \bibinfo{journal}{APh} \textbf{\bibinfo{volume}{20}}, \bibinfo{pages}{579}
  (\bibinfo{year}{2004}).

\bibitem{ProdF04}
\bibinfo{author}{T.~Prodanovi\'{c}} and \bibinfo{author}{B.~Fields},
  \bibinfo{journal}{ApJ} \textbf{\bibinfo{volume}{616}}, \bibinfo{pages}{L115}
  (\bibinfo{year}{2004}).

\bibitem{ProdF04b}
\bibinfo{author}{T.~Prodanovi\'{c}} and \bibinfo{author}{B.~Fields},
  \bibinfo{journal}{ApJ} \textbf{\bibinfo{volume}{in press}}
  (\bibinfo{year}{2005}).

\bibitem{SuIn04}
\bibinfo{author}{T.~K. Suzuki} and \bibinfo{author}{S.~Inoue},
  \bibinfo{journal}{PASA} \textbf{\bibinfo{volume}{21}}, \bibinfo{pages}{148}
  (\bibinfo{year}{2004}).

\bibitem{KBH}
\bibinfo{author}{P.~Kuo}, \bibinfo{author}{S.~Bowyer}, and
  \bibinfo{author}{C.-Y. Hwang}, \bibinfo{journal}{ApJ}
  \textbf{\bibinfo{volume}{618}}, \bibinfo{pages}{675} (\bibinfo{year}{2005}).

\bibitem{bert85}
\bibinfo{author}{E.~Bertschinger}, \bibinfo{journal}{ApJS}
  \textbf{\bibinfo{volume}{58}}, \bibinfo{pages}{39} (\bibinfo{year}{185}).

\bibitem{ryu97}
\bibinfo{author}{D.~Ryu} and \bibinfo{author}{H.~Kang},
  \bibinfo{journal}{MNRAS} \textbf{\bibinfo{volume}{284}}, \bibinfo{pages}{416}
  (\bibinfo{year}{1997}).

\bibitem{min_shock}
\bibinfo{author}{F.~Miniati}, \bibinfo{author}{D.~Ryu},
  \bibinfo{author}{H.~Kang}, \bibinfo{author}{T.~Jones},
  \bibinfo{author}{R.~Cen}, and \bibinfo{author}{J.~Ostriker},
  \bibinfo{journal}{ApJ} \textbf{\bibinfo{volume}{542}}, \bibinfo{pages}{608}
  (\bibinfo{year}{2000}).

\bibitem{bertV}
\bibinfo{author}{E.~Bertschinger}, \bibinfo{journal}{ApJS}
  \textbf{\bibinfo{volume}{58}}, \bibinfo{pages}{1} (\bibinfo{year}{1985}).

\bibitem{RKJ03}
\bibinfo{author}{D.~Ryu}, \bibinfo{author}{H.~Kang},
  \bibinfo{author}{E.~Hallman}, and \bibinfo{author}{T.~Jones},
  \bibinfo{journal}{ApJ} \textbf{\bibinfo{volume}{593}}, \bibinfo{pages}{599}
  (\bibinfo{year}{2003}).

\bibitem{inoue05}
\bibinfo{author}{S.~Inoue} and \bibinfo{author}{M.~Nagashima}
  (\bibinfo{year}{2005}).

\bibitem{ch}
\bibinfo{author}{S.~Chandrasekhar}, \bibinfo{journal}{Rev. Mod. Phys.}
  \textbf{\bibinfo{volume}{15}}, \bibinfo{pages}{2} (\bibinfo{year}{1943}).

\bibitem{peeb84}
\bibinfo{author}{P.J.E.Peebles}, \bibinfo{journal}{ApJ}
  \textbf{\bibinfo{volume}{284}}, \bibinfo{pages}{439} (\bibinfo{year}{1984}).

\bibitem{lem1}
\bibinfo{author}{G.~Lema\^{i}tre}, \bibinfo{journal}{Compt. Rend.}
  \textbf{\bibinfo{volume}{196}}, \bibinfo{pages}{903} (\bibinfo{year}{1933}).

\bibitem{lem2}
\bibinfo{author}{G.~Lema\^{i}tre}, \bibinfo{journal}{Compt. Rend.}
  \textbf{\bibinfo{volume}{196}}, \bibinfo{pages}{1085} (\bibinfo{year}{1933}).

\bibitem{lah}
\bibinfo{author}{O.~Lahav}, \bibinfo{author}{P.~Lilje},
  \bibinfo{author}{J.~Primack}, and \bibinfo{author}{M.~Rees},
  \bibinfo{journal}{MNRAS} \textbf{\bibinfo{volume}{251}}, \bibinfo{pages}{128}
  (\bibinfo{year}{1991}).

\bibitem{peeb}
\bibinfo{author}{P.~Peebles}, \emph{\bibinfo{title}{The Large Scale Structure
  of the Universe}} (\bibinfo{publisher}{Princeton Univ. Press},
  \bibinfo{address}{Princeton, N.J.}, \bibinfo{year}{1980}).

\bibitem{b94}
\bibinfo{author}{F.~Bernardeau}, \bibinfo{journal}{A\&A}
  \textbf{\bibinfo{volume}{291}}, \bibinfo{pages}{697} (\bibinfo{year}{1994}).

\bibitem{bard86}
\bibinfo{author}{J.~Bardeen}, \bibinfo{author}{J.~Bond},
  \bibinfo{author}{N.~Kaiser}, and \bibinfo{author}{A.~Szalay},
  \bibinfo{journal}{ApJ} \textbf{\bibinfo{volume}{304}}, \bibinfo{pages}{15}
  (\bibinfo{year}{1986}).

\bibitem{mw00}
\bibinfo{author}{H.~Mo} and \bibinfo{author}{S.~White},
  \bibinfo{journal}{MNRAS} \textbf{\bibinfo{volume}{336}}, \bibinfo{pages}{112}
  (\bibinfo{year}{2002}).

\bibitem{ll}
\bibinfo{author}{L.~Landau} and \bibinfo{author}{E.~Lifshitz},
  \emph{\bibinfo{title}{Fluid Mechanics}}
  (\bibinfo{publisher}{Butterworth-Heinemann}, \bibinfo{address}{Boston},
  \bibinfo{year}{1999}).

\bibitem{pfd}
\bibinfo{author}{V.~Pavlidou} and \bibinfo{author}{B.~Fields},
  \bibinfo{journal}{PhysRevD} \textbf{\bibinfo{volume}{71}},
  \bibinfo{pages}{043510} (\bibinfo{year}{2005}).

\end{thebibliography}
%
\doublespace
%
\chapter*{Author's Biography}
\addcontentsline{toc}{chapter}{Biographical Note}

Vasiliki Pavlidou was born in Thessaloniki, Greece in 1977. She
received her B.S. in Physics from the Aristotle University of
Thessaloniki in 1999, and her M.S. in Astronomy from the University of
Illinois in 2001. She is married to Kostas Tassis, and they have had
several house plants, three of which are currently alive.

\end{document}